%% file: manuscript.tex
\documentclass[aps,pra,reprint,amsmath,amssymb,superscriptaddress,onecolumn,longbibliography,notitlepage]{revtex4-1}

\usepackage[hidelinks]{hyperref}

\usepackage{braket}
\usepackage{mathtools}
\usepackage{verbatim, xcolor}
\usepackage{siunitx}

\newcommand{\im}{\mathrm{i}}
\newcommand{\dif}{\mathrm{d}}
\newcommand{\e}[1]{e^{#1}}
\newcommand{\conj}{^\ast}
\newcommand{\inv}{^{-1}}
\newcommand{\dagg}{^\dagger}
\newcommand{\paren}[1]{\left(#1\right)}
\newcommand{\sbrak}[1]{\left[#1\right]}
\renewcommand{\set}[1]{\left\{#1\right\}}
\newcommand{\abs}[1]{\left|#1\right|}
\newcommand{\Paren}[1]{\bigl(#1\bigr)}
\newcommand{\Sbrak}[1]{\bigl[#1\bigr]}
\newcommand{\Vbrak}[1]{\bigl\langle#1\bigr\rangle}
\DeclareMathOperator{\sgn}{sgn}
\DeclareMathOperator{\mean}{mean}
\DeclareMathOperator{\RE}{Re}

\newcommand{\EJ}{E_\text{J}}
\newcommand{\EC}{E_\text{C}}
\newcommand{\jpo}{\text{JPO}}
\newcommand{\dc}{\text{dc}}
\newcommand{\ac}{\text{ac}}
\newcommand{\bus}{\text{bus}}
\newcommand{\slow}{\text{slow}}
\newcommand{\fast}{\text{fast}}
\newcommand{\nom}{\text{nom}}
\newcommand{\temax}{\text{max}}
\newcommand{\J}{\text{J}}

\begin{document}
\title{A quantum annealer with fully programmable all-to-all coupling via Floquet engineering}
\author{Tatsuhiro~Onodera}
\thanks{Equal contribution.}
\affiliation{E.\,L. Ginzton Laboratory, Stanford University, Stanford, CA 94305, USA}
\author{Edwin~Ng}
\thanks{Equal contribution.}
\affiliation{E.\,L. Ginzton Laboratory, Stanford University, Stanford, CA 94305, USA}
\author{Peter~L.~McMahon}
\email{pmcmahon@cornell.edu}
\affiliation{E.\,L. Ginzton Laboratory, Stanford University, Stanford, CA 94305, USA}
\affiliation{School of Applied and Engineering Physics, Cornell University, Ithaca, NY 14853, USA}

\input{main.tex}

\clearpage

\input{references.tex}
\clearpage

\begin{center}
  \textbf{\large A quantum annealer with fully programmable all-to-all coupling via Floquet engineering: Supplementary Materials}\\[1cm]
\end{center}

\renewcommand{\thepage}{S\arabic{page}}
\renewcommand{\thesection}{S\arabic{section}}
\renewcommand{\thetable}{S\arabic{table}}
\renewcommand{\thefigure}{S\arabic{figure}}
\renewcommand{\theequation}{S\arabic{equation}}
\renewcommand{\thefigure}{S\arabic{figure}}

\numberwithin{equation}{section}

\setcounter{section}{0}
\setcounter{equation}{0}
\setcounter{figure}{0}
\setcounter{table}{0}
\setcounter{page}{1}

\renewcommand{\bibnumfmt}[1]{[#1]}
\renewcommand{\citenumfont}[1]{#1}

\input{supplementary.tex}

\end{document}

%% file: main.tex
\begin{abstract}
Quantum annealing is a promising approach to heuristically solving difficult combinatorial optimization problems. However, the connectivity limitations in current devices lead to an exponential degradation of performance on general problems. We propose an architecture for a quantum annealer that achieves full connectivity and full programmability while using a number of physical resources only linear in the number of spins. We do so by application of carefully engineered periodic modulations of oscillator-based qubits, resulting in a Floquet Hamiltonian in which all the interactions are tunable; this flexibility comes at a cost of the coupling strengths between spins being smaller than they would be had the spins been directly coupled. Our proposal is well-suited to implementation with superconducting circuits, and we give analytical and numerical evidence that fully connected, fully programmable quantum annealers with $1000$ qubits could be constructed with Josephson parametric oscillators having cavity-photon lifetimes of $ \SI{\sim 100}{\micro\second}$, and other system-parameter values that are routinely achieved with current technology. Our approach could also have impact beyond quantum annealing, since it readily extends to bosonic quantum simulators and would allow the study of models with arbitrary connectivity between lattice sites.
\end{abstract}

\maketitle

\section{Introduction}
Quantum annealers are computational devices designed for solving combinatorial optimization problems, most typically Ising optimization problems \cite{Kadowaki1998,Farhi2000a, Boixo2014}. An Ising problem is specified by connections among $N$ spins on a graph, as well as local fields on each spin. One of the foremost challenges in the experimental realization of quantum annealers is the requirement that quantum annealers be able to represent densely connected Ising problems with minimal overhead in the number of qubits (and other physical components) used \cite{Perdomo-Ortiz2017,Katzgraber2018,Hamerly2019, Hauke2019}. If a quantum annealer is not able to directly represent a particular problem because the problem graph has higher connectivity than the physical annealer does, then one incurs a penalty in the number of qubits needed to represent the Ising problem. For example, the largest fully connected Ising problem that can be represented in the D-Wave 2000-qubit quantum annealer is one that has $N=64$ spins \cite{Hamerly2019}; this limitation arises because the connectivity in this particular quantum annealer is very sparse (the connectivity graph has maximum degree six).

Superconducting circuits are one of the most prominent technologies for realizing quantum information processing devices, including quantum annealers, and form the basis for many of the major projects to construct experimental quantum annealers \cite{Johnson2011, Weber2018, Chen2017}. However, when qubit connectivity is achieved via physical pairwise couplers (as is the case for the efforts described in Refs.~\cite{Johnson2011, Weber2018, Chen2017}), there is a substantial engineering impediment to realizing full connectivity: each qubit would need $N-1$ physical couplers, and arranging such couplers spatially has proven to be impractical for large $N$. On the other hand, bus architectures have been demonstrated for superconducting-circuit qubits in the context of circuit-model quantum computing \cite{Majer2007,DiCarlo2009,Mariantoni2011}, and bus-mediated interactions naturally provide all-to-all coupling \cite{Majer2007,Song2017}, with the use of one physical coupler per qubit. In this paper, we address the challenge of realizing full programmability of these all-to-all couplings, in the context of developing a quantum annealer.

In classical neuromorphic computing, a scheme allowing full programmability has been proposed in an all-to-all coupled system of Kuramoto oscillators via periodically modulated external inputs \cite{Hoppensteadt1999}. In a completely separate community and context, nonlinear-oscillator-based qubits, whose operation relies on the continuous-variable nature of the oscillators, have also been established as a promising building block for the realization of superconducting-circuit quantum annealers, owing partially to their resilience to photon loss \cite{Goto2016b, Nigg2017b, Puri2017a}. In this work, we take inspiration from both these lines of research to show that the all-to-all, bus-mediated couplings in a system of quantum nonlinear oscillators can be made fully programmable by careful design of the modulation of each oscillator's instantaneous frequency. We utilize the mathematical tools of Floquet theory \cite{Bukov2015b, Eckardt2015} to engineer the desired interactions between oscillators and establish that the functionality of our dynamically-coupled system is equivalent to that of a statically-coupled system with pairwise physical couplers.

Our scheme applies generically to a variety of nonlinear oscillators, including those realized in platforms besides superconducting circuits, such as optics \cite{Wang2013, Marandi2014b, McMahon2016, Inagaki2016} or nanomechanics \cite{Arndt2014, Lifshitz2003}. However, for concreteness, we focus on Kerr parametric oscillators \cite{Goto2016a, Goto2016b, Puri2017b} in which the Kerr nonlinearity is provided by a Josephson junction---i.e., a Josephson parametric oscillator (JPO) \cite{Krantz2016, Puri2017b, Nigg2017b, Frattini2018, Wang2019}. JPOs have been utilized in two other schemes for achieving programmable couplings:  a proposed realization \cite{Puri2017a} of the LHZ architecture \cite{Lechner2015} (which requires $N(N-1)/2$ physical qubits to represent $N$ spins), and an inductive-shunt scheme \cite{Nigg2017b} (which only provides $\mathcal O(N \log N)$ programmable parameters, out of a total of $\mathcal{O}(N^2)$ in general).

In summary, the previously known approaches to building a quantum annealer with JPOs are, variously, incompatible with dense connectivity due to engineering limitations; requiring of a large overhead in the number of qubits and/or couplers (i.e., scaling with $N^2$); or lacking full programmability. In contrast, our proposal uses a bus to provide all-to-all connectivity, and a dynamical approach to coupling that enables full programmability while requiring only a number of oscillators and couplers linear in $N$ to solve Ising problems with $N$ spins.

\section{Results}
In this paper, we study the design of a quantum annealer whose purpose is to solve the Ising optimization problem, defined as finding the $N$-spin configuration $\sigma_i \in \left\{ -1,+1 \right\}$ ($i=1,\ldots,N$) that minimizes the classical spin energy $E(\sigma) \coloneqq \sum_{j \neq i} C_{ij} \sigma_i \sigma_j$, where $C$ is a symmetric real matrix. A choice of $C$ specifies a problem instance to be solved, and $C$ can be interpreted as the adjacency matrix of a graph whose vertices are spins and whose edges represent spin-spin interactions. In general, $C$ can have $\mathcal{O}(N^2)$ non-zero entries. It is desirable for a quantum annealer to be fully programmable, such that there are no restrictions on the structure of $C$, and that the annealer not use more than $N$ oscillators to represent a given $N$-spin problem, nor use more than $\propto N$ other physical components. In this paper, we show how this can be achieved using nonlinear oscillators in a bus architecture together with dynamically realized couplings designed via Floquet engineering.

\begin{figure}
  \centering
  \includegraphics[width=1.0\textwidth]{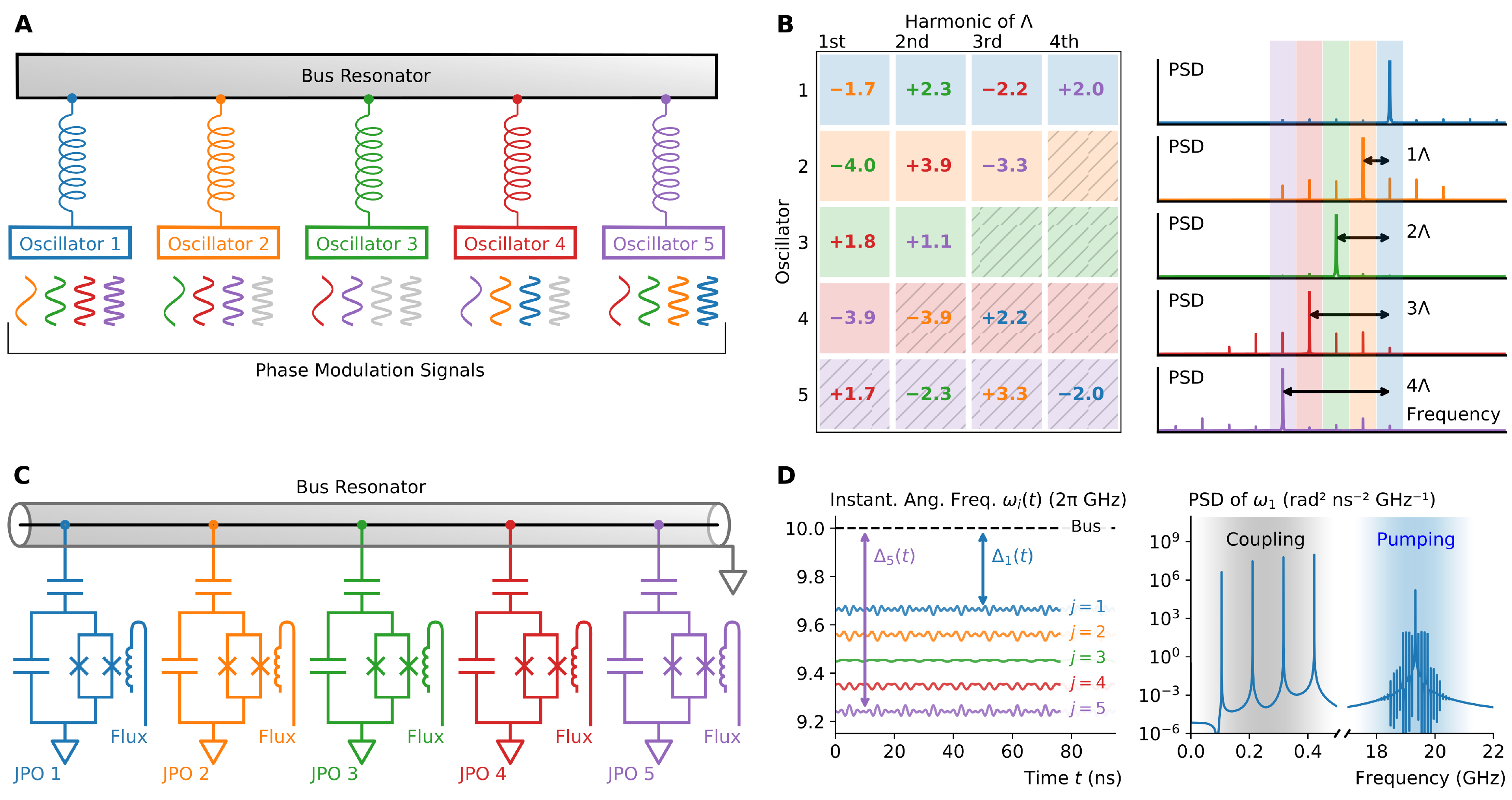}
  \caption{\textbf{Achieving fully programmable all-to-all coupling in oscillator networks via phase modulation.} (\textbf{A}~and~\textbf{B})~An implementation-independent overview of the architecture, showing an example with $N=5$ oscillators each coupled to a common bus resonator and each driven by a set of phase modulation (PM) signals. As depicted in (A) each oscillator is modulated by up to $N-1$ harmonics with frequencies $k\Lambda$ for integers $k<N$. The amplitude of each harmonic for each oscillator is given by an element in a phase-modulation matrix $F$, an example of which is shown on the left side of (B) (times a factor of \num{100} for clarity). The color of the PM signal on each oscillator indicates with which other oscillator the PM causes an interaction: for example, the $k=1$ harmonic PM applied to Oscillator 1 (shown in orange) causes it to couple with Oscillator 2, and the $k=2$ harmonic PM applied to Oscillator 1 (shown in green) causes it to couple with Oscillator 3. When a signal potentially mediates coupling to two other oscillators (e.g., $k=2$ for Oscillator 3), we color by the oscillator with the lower center frequency (larger oscillator index $i$); the signals colored gray are not applied as they would not create any useful couplings. As shown in the power spectral density (PSD) of the oscillators' canonical positions on the right side of (B), the oscillators have center frequencies spaced by $\Lambda$; thus each PM signal at frequency $k\Lambda$ creates a primary sideband which overlaps in frequency with the center frequency of the neighbor $k$ oscillators away. Taken together, the effect of these sideband interactions is to generate a desired effective Ising coupling matrix (see the Supplementary Materials for the matrix used in this example). (\textbf{C})~A superconducting-circuits implementation of the dynamical-coupling architecture. A generic oscillator may be realized by a Josephson parametric oscillator (JPO), whose frequency can be modulated by a flux line, while the bus may be realized by a microwave resonator. (\textbf{D})~Quantitative overview of the scheme for a JPO system, using the modulations shown in (B) and depicting the instantaneous angular frequencies $\omega_i(t)$ of the JPOs, both in the time domain (left) and as a PSD of $\omega_1$ in the frequency domain (right). The PM signals manifest as modulations in the instantaneous frequency of each JPO, or, equivalently, in the detuning $\Delta_i(t)$ between the bus and Oscillator $i$. In the PSD, the PM signal appears in a low-frequency band (\SI{<0.5}{GHz}), while the parametric drive to the JPO (at twice the JPO's frequency) occurs as usual in a high-frequency band \SI{\sim20}{GHz} \cite{Nigg2017b, Puri2017b}. The parameters used for this calculation are obtained from the bus-coupled JPO model described in the Supplementary Materials.}
  \label{fig:schematic}
\end{figure}

Figure \ref{fig:schematic}A shows an overview of our proposed architecture. The $N$ nonlinear oscillators of the quantum annealer are coupled to a common (bus) resonator. If the center frequencies of the oscillators are sufficiently far-detuned from the bus resonance, then the bus mediates a photon-exchange interaction between any pair of oscillators. In particular, denoting the annihilation operator for the $i$th oscillator as $\hat a_i$, the bus mediates interactions that contribute terms of the form $\hat a_i^\dagger \hat a_j + \hat a_i \hat a_j^\dagger$ to the system Hamiltonian \cite{Majer2007} which generically couples every oscillator to every other oscillator. Thus with $N$ nonlinear oscillators coupled to a bus, we can implement all-to-all coupling. However, the couplings up to this point are not programmable. The principal result in this paper is that we can engineer complete programmability of all the couplings by phase-modulating the oscillators in a specific way.

In our scheme, the oscillators are, to a good approximation, detuned from each other by multiples of a fundamental frequency $\Lambda$, such that the $k$th-nearest neighbor of any given oscillator is detuned from it by $k\Lambda$. Despite the presence of the bus, for $\Lambda$ sufficiently large the oscillators are effectively uncoupled in the absence of modulation (by the rotating-wave approximation). To effect dynamical coupling, each oscillator is controlled with a phase modulation (PM) signal containing harmonics of $\Lambda$, such that the resulting sidebands of each oscillator overlap in frequency with the center frequencies of the other oscillators. More precisely, the $i$th oscillator is phase-modulated by $\delta\phi_i(t) \coloneqq -\sum_{k=1}^{N-1} F_i^{(k)} \sin(k\Lambda t)$, which causes its instantaneous frequency $\omega_i(t)$ to pick up a time-varying component $\dot{\delta\phi}_i(t) = -\sum_{k=1}^{N-1} F_i^{(k)} k\Lambda \cos(k\Lambda t)$. Here, the coefficients $F^{(k)}_i$ encode the strength of the $k$th-harmonic component in the PM of oscillator $i$, and they can be summarized by a matrix $F$ of dimension $N\times(N-1)$ whose $k$th column consists of the elements $F^{(k)}_1, \ldots, F^{(k)}_N$. Intuitively, the strengths of the dynamically induced couplings are determined by the strengths of the sidebands, which in turn are controlled by the elements of $F$. Thus, the task of programming these couplings reduces to making an appropriate choice for $F$; we will present a method for choosing these coefficients shortly. Figure \ref{fig:schematic}B is a cartoon depicting the realization of this PM scheme for a representative \num{5}-spin problem instance; the coefficients $F^{(k)}_i$ are shown on the left, while on the right are the power spectral densities (PSDs) of each oscillator's canonical position $x_i(t) \coloneqq \cos\bigl(\int_0^t \omega_i(t^\prime) \mathop{\textrm{d}t^\prime}\bigr)$. Each PSD shows a strong peak at its respective oscillator's center frequency, together with sidebands of various amplitudes (controlled by $F$) that overlap with the center frequencies of the other oscillators.

While our method is independent of the exact type of nonlinear oscillator used, we specialize our discussion in this paper to a superconducting-circuits realization. For concreteness, we consider Josephson parametric oscillators (JPOs) \cite{Krantz2016,Puri2017b,Nigg2017b, Wang2019}, although other superconducting-circuit-based oscillators are plausible choices as well \cite{Leghtas2015, Mirrahimi2014c}. In the absence of coupling, the eigenstates of the JPO are 0-phase and $\pi$-phase coherent states \cite{Puri2018, Goto2016a}; these two eigenstates can be used to encode, respectively, the spin-up and spin-down configurations of an Ising spin \cite{Goto2016b}, and superpositions of these eigenstates (i.e., Schr\"odinger cat states) have been experimentally demonstrated \cite{Wang2019}. Figure \ref{fig:schematic}C shows $N$ JPOs coupled to a superconducting resonator, which acts as the bus in this platform. Energy is supplied to each JPO by a flux line carrying a time-varying current. Conventionally, the dc component of the current determines the center frequency of the JPO, and an ac component at twice that frequency provides the parametric drive. In our scheme, the center frequency of each JPO is additionally modulated in time by $\dot{\delta\phi}(t)$, which is achieved by an additional modulation of the flux-line current corresponding to $\dot{\delta\phi}$ \cite{Roy2016,Reagor2018}. (In an alternative technological realization using a system of optical oscillators, this PM could potentially be applied by a physical phase modulator.)

For this JPO realization of a nonlinear-oscillator-based quantum annealer, Fig.~\ref{fig:schematic}D shows a more quantitative picture of our dynamical-coupling scheme, with plots of the instantaneous angular frequencies $\omega_i(t)$ both in the time domain (for all the oscillators) and in the frequency domain (for the first oscillator only). First, on the time-domain plot, we see that the effect of the PM is to create oscillations in the instantaneous detuning $\Delta_i(t)$ between the fixed bus resonance frequency and $\omega_i(t)$; in the absence of these modulations (or on time-averaging), the oscillator frequencies are approximately evenly spaced. As expected, the PM consists of four harmonics; looking at the frequency-domain plot, we see that these harmonics lie in a ``coupling band'' with frequency content ${}\ll \SI{1}{GHz}$. These plots also illustrate two more technical features specific to our construction (see the Supplementary Materials for more information). First, the deviations in oscillator frequency from the center (i.e., the modulation depths) are small compared to the mean values of the bus-oscillator detunings; this qualitative feature ensures that the native couplings $J_{ij}$ are approximately time-independent despite the PM. Second, there is a significant gap between the coupling band and the ``pumping band'' in which the parametric drive is realized; because of this separation of time scales, we can first design the flux-line currents to support the PM needed for dynamical coupling, while the parametric drive simply follows that modulation as needed. Since the ac part of the flux-line current is directly proportional to the ac part of $\omega_i(t)$ (see the Supplementary Materials), we see that the entire control signal only requires approximately \SI{20}{GHz} of bandwidth at most, which is realizable with current microwave technology.

We have thus far not explained how to choose the modulation coefficients $F$ for a given problem instance $C$. An intuitive, but incorrect, choice is to simply set $F^{(k)}_i \propto C_{i,i+k}$, since generating a sideband at frequency $k\Lambda$ on oscillator $i$ intuitively causes it to interact with oscillator $i+k$. However, this is not entirely accurate, as the symmetrically generated sideband at $-k\Lambda$ (caused by the same $F^{(k)}_i$ coefficient) also leads to the same interaction with oscillator $i-k$; for arbitrary $C$, these two interactions may need to differ in general. Furthermore, we must also consider higher-order sidebands as well: even if we were to only phase modulate oscillator $i$ at frequency $k\Lambda$, weaker sidebands at $\pm 2k\Lambda$, $\pm 3k\Lambda$, and so on are also generated, causing interactions with oscillators $i\pm2k$, $i\pm3k$, and so on, respectively. Thus, there is a nontrivial relation between $F$ and the desired couplings $C$, and $F$ needs to be chosen in a way such that the various contributions of $F$ to the effective couplings among oscillators combine appropriately to give the desired couplings $C$. To do this formally, we apply the mathematical tools of Floquet theory \cite{Eckardt2015, Bukov2015b}.

First, to more explicitly define the problem we are trying to solve, a quantum annealer based on Josephson parametric oscillators can be described by a (rotating-frame) Hamiltonian of the form \cite{Goto2016b, Puri2017b}
\begin{equation} \label{eq:H-static}
\hat{H}_\textrm{static} \coloneqq -\sum_i \left( \delta_i \hat a^\dagger_i \hat a_i + \frac{\chi}{2} \hat a_i^{\dagger2} \hat a_i^2 \right) + \sum_i\frac{r(t)}{2}\left( \hat a_i^2 +  {\hat a_i}^{\dagger 2} \right) - \lambda_\textrm{C} \sum_{j\neq i} C_{ij} \hat a_i^\dagger \hat a_j,
\end{equation}
where $\chi$ is the Kerr nonlinear rate, $\delta_i$ is the detuning between oscillator $i$ and the half-harmonic of its parametric drive, and $r(t)$ is the (slowly time-varying) amplitude of the parametric drives; in this work we choose $r(t)$ to be a linear ramp in time for the duration of the computation (see the Supplementary Materials for more details). Here, $\lambda_\text{C}$ is a problem-strength parameter dictating the strength of the oscillator-oscillator couplings $C_{ij}$ (which are normalized in this work to $\max_{j\neq i} |C_{ij}| = 1$). One way to realize such a Hamiltonian is to have $\mathcal O(N^2)$ physical pairwise couplers, which can be programmed given a desired $C_{ij}$; these programmed couplings can then be held static throughout the annealing process while the parametric drive is varied.

By contrast, our goal is to realize this annealing Hamiltonian via dynamical control of the effective couplings among the oscillators. As a result, we start instead with the (rotating-frame) Hamiltonian (see the Supplementary Materials for a derivation)
\begin{equation} \label{eq:H-dynamical}
\hat{H}_\text{dynamical} \coloneqq -\sum_i \left(\delta_i \hat a^\dagger_i \hat a_i + \frac{\chi}{2} \hat a_i^{\dagger2} \hat a_i^2\right) + \sum_i\frac{r(t)}{2}\left( \hat a_i^2 +  {\hat a_i}^{\dagger 2} \right) - \sum_{j \neq i} J_{ij} \exp\left[-\mathrm{i}\Bigl(\Lambda(i-j)t +\delta\phi_j(t)-\delta\phi_i(t)\Bigr)\right] \hat a_i^\dagger \hat a_j,
\end{equation}
which is the same as $\hat H_\text{static}$ with the exception of the final coupling term. In this case, $J_{ij}$ are the bus-mediated coupling rates natively present in the system but which are not fully programmable in general. By detuning the oscillators relative to one another by multiples of $\Lambda$ and applying PM to the oscillators according to $\delta\phi_i(t)$, the result is the ``native'' coupling term $\hat H_\text{native} \coloneqq - \sum_{j \neq i} J_{ij} \exp\left[-\mathrm{i}\Bigl(\Lambda(i-j)t +\delta\phi_j(t)-\delta\phi_i(t)\Bigr)\right] \hat a_i^\dagger \hat a_j$. The time-varying phase factor is due to the PM of the $\Lambda$-detuned oscillators, and control over its time dependence forms the core of our dynamical-coupling scheme.

If we denote the coupling term in the static annealing Hamiltonian \eqref{eq:H-static} as $\hat H_\text{target} \coloneqq -\lambda_\text{C} \sum_{j \neq i} C_{ij} \hat a_i^\dagger \hat a_j$, then the goal is to achieve $\hat H_\text{native} \approx \hat H_\text{target}$, under some appropriate sense of the approximation. As previously mentioned, we choose the modulations to be $-\sum_{k=1}^{N-1} F_i^{(k)} \sin(k\Lambda t)$, which means that $\hat H_\text{native}$ is periodic with frequency $\Lambda$. If $\Lambda$ is much larger than all other system timescales, then the results of Floquet theory allow us to make the Floquet approximation $\hat{H}_\text{native} \approx \lambda_\text{C} \sum_{j \neq i} {C_\text{eff}}_{ij} \hat a_i^\dagger \hat a_j$ where
\begin{equation} \label{eq:C-eff}
{C_\text{eff}}_{ij} \coloneqq \int_0^{2\pi/\Lambda} \! \frac{J_{ij}}{\lambda_\text{C}} \cos\biggl[(i-j)\Lambda t + \sum_{k=1}^{N-1} \left( F^{(k)}_i - F^{(k)}_j \right) \sin(k\Lambda t) \biggr] \frac{\mathrm{d} t}{2\pi/\Lambda}
\end{equation}
describes the effective couplings between oscillators $i$ and $j$ due to all the sideband interactions. (More formally, this approximation is the leading-order term in the Floquet-Magnus expansion of $\hat H_\text{native}$ in $1/\Lambda$.) Thus, all that remains is to choose the coefficients in $F$ such that ${C_\text{eff}}_{ij} \approx C_{ij}$. As discussed in the Supplementary Materials, while it is possible to solve for $F$ by direct numerical nonlinear optimization, there are a number of ways to make this precomputation step more tractable and robust. In particular, one can consider a second-order Taylor expansion of ${C_\text{eff}}_{ij}$ (intuitively, by considering effective interactions only up to the second-order sidebands) and obtain a system of quadratic equations that can be numerically solved.

\begin{figure}[h!]
  \centering
  \includegraphics[width=1.0\textwidth]{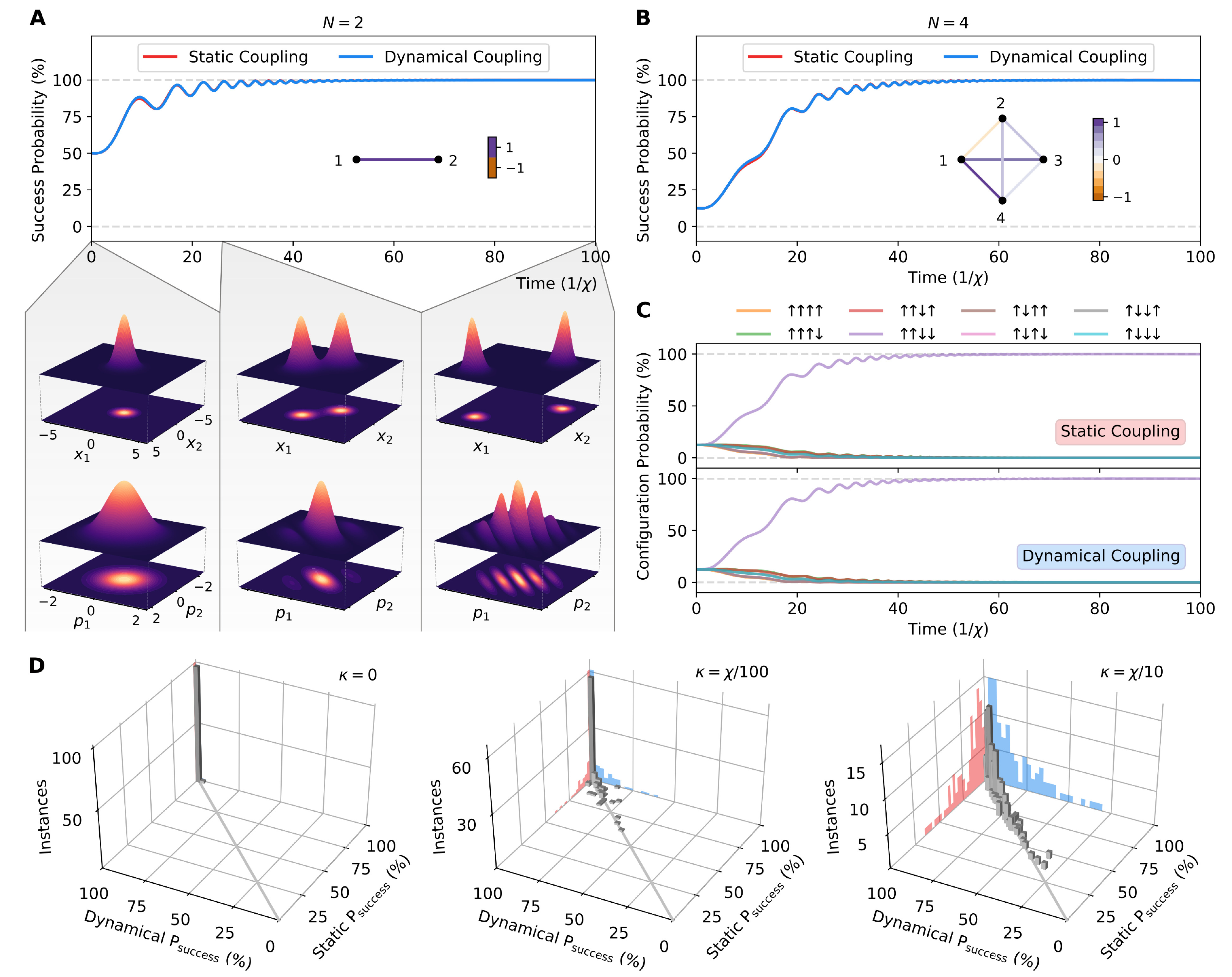}
  \caption{\textbf{Quantum simulations.} (\textbf{A} and \textbf{B}) Evolution of the success probability of finding a ground state of the inset Ising problem instances, for both a system whose couplings are implemented by static pairwise physical couplers (static coupling) and a system using this work's dynamical-coupling approach (dynamical coupling), for the case of no dissipation. Shown to the bottom of (A) are the joint quantum states of the oscillators (represented as probability distributions $|\psi(x_1,x_2)|^2$ and $|\psi(p_1,p_2)|^2$ in canonical position and momentum coordinates, respectively) at three different times for the dynamically-coupled system. Initially the system is in a vacuum state (with equal probability on each of the four possible spin configurations); by the end of the evolution, the system is in a coherent superposition $\frac{1}{\sqrt{2}} \left( \ket{\uparrow\downarrow} + \ket{\downarrow\uparrow} \right)$ (in the qubit subspace), so the probability to measure a ground state is $\approx 100\%$. (\textbf{C}) The probabilities of obtaining each of the possible spin configurations (i.e., configuration probability) as a function of the evolution time, for the problem instance and simulations shown in (B). Since the problem does not have any external fields, each spin configuration shown is degenerate to flipping every spin; for brevity, we have added together the probabilities corresponding to these degenerate configurations. (\textbf{D}) Correlation matrices of success probabilities $P_\text{success}$ (evaluated at the end of the computation) between the statically-coupled and dynamically-coupled systems, for 100 instances of the SK7 problem class. (For the trajectory simulation data used to generate these correlation matrices, see the Supplementary Materials.) Here, we also introduce a nonzero field decay rate $\kappa$ (relative to the timescale $\chi$), to show that the correlations are high even in the presence of photon losses. Projected to the sides in blue and red are the corresponding histograms for the dynamically-coupled and statically-coupled systems, respectively. For simulation parameters and methodology, see the Supplementary Materials.}
  \label{fig:quantum}
\end{figure}

\begin{figure}[b!]
  \centering
  \includegraphics[width=\textwidth]{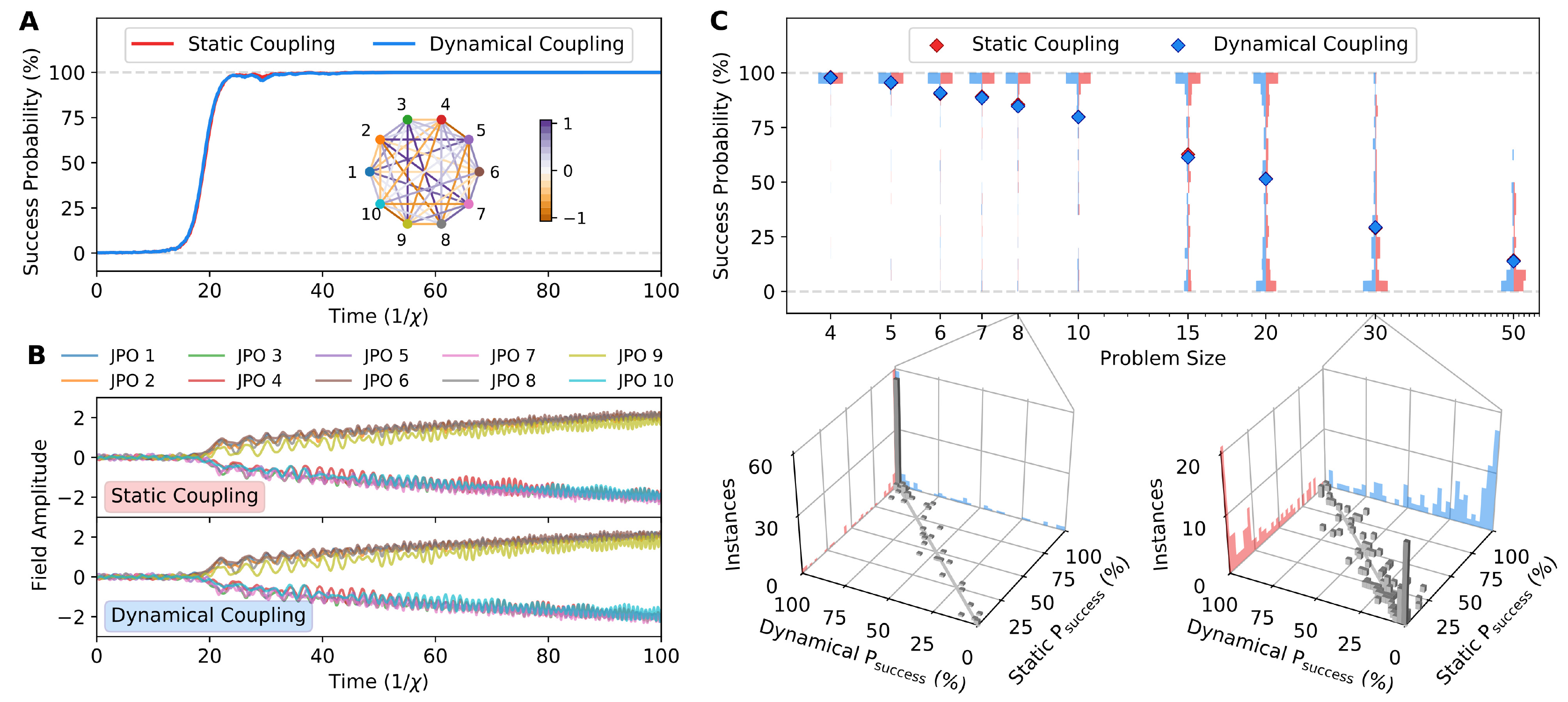}
  \caption{\textbf{Simulations of classical equations of motion.} (\textbf{A}) Evolution of the success probability of finding a ground state of the inset 10-spin Ising problem instance, for both a system whose couplings are implemented by static pairwise physical couplers (static coupling) and a system using this work's dynamical-coupling approach (dynamical coupling). To estimate the probability, we sample random initial conditions according to a Gaussian distribution (see Supplementary Materials, Ref.~\cite{Goto2016b}). (\textbf{B}) Trajectories of the classical field amplitudes of each of the 10 oscillators, for one of the samples constituting the simulation in (A). (\textbf{C}) Histograms of the success probability $P_\text{success}$ for various problem sizes $N$, each estimated using 100 instances of the SK7 problem class. The diamonds indicate the average success probability, while the horizontal bars show the histogram. For $N=8$ and $N=30$, the correlation matrices between statically- and dynamically-coupled systems are shown at the bottom. (For the correlation matrices for all other problem sizes, see the Supplementary Materials.) Projected to the sides of the correlation matrices, in blue and red, are the corresponding histograms for the dynamically-coupled and statically-coupled systems, respectively. For simulation parameters and methodology, see the Supplementary Materials.}
  \label{fig:classical}
\end{figure}

To demonstrate the effectiveness of our approach, we perform numerical simulations of both the statically-coupled system governed by $\hat H_\text{static}$ as well as the dynamically-coupled system governed by $\hat H_\text{dynamical}$ (with an appropriate choice of $F$ given $C$), and we show that they achieve nearly indistinguishable results, as one would expect if one has achieved $\hat{H}_\textrm{dynamical} \approx \hat{H}_\textrm{static}$. Figure \ref{fig:quantum}A shows the results of simulating the quantum evolution of both systems as they perform quantum annealing on a two-spin problem, where the coupling between the spins is antiferromagnetic. We can see that both the statically-coupled system and the dynamically-coupled system obtain the correct final state, which is a superposition $\frac{1}{\sqrt{2}} \left( \ket{ \uparrow \downarrow } + \ket{ \downarrow \uparrow } \right)$ in the qubit basis, and would produce one of the correct ground-state spin configurations $\uparrow \downarrow$ or $\downarrow \uparrow$ upon measurement. Moreover, we see that the evolution of the success probability to obtain a ground state is nearly identical for both architectures at all times, suggesting that the dynamically-coupled system is closely mimicking the behavior of the statically-coupled system, as desired.

Figure~\ref{fig:quantum}B shows the same evolution of the success probability for a particular $N=4$ problem instance, chosen from the class of finite-range, integer-valued Sherrington-Kirkpatrick (SK) instances studied in a foundational quantum-annealing benchmark work \cite{Rønnow2014}; in particular, we utilize range-7 graphs (SK7) (see the Supplementary Materials for details about instance generation). We again see that the evolutions of the success probabilities match very closely. Aside from the success probability, Fig.~\ref{fig:quantum}C shows the projections of the quantum state onto the spin configurations (of which there are \num{16} for $N = 4$), which demonstrates that the two architectures produce nearly indistinguishable evolutions in the projections as well.

In Fig.~\ref{fig:quantum}D, we show the results from $N=4$ simulations for \num{100} different randomly generated problem instances (again drawn from the SK7 problem class). We simulate both the statically-coupled and dynamically-coupled quantum annealers and show the correlations between their respective success probabilities. To explore the effects of decoherence due to photon loss, we also examine these correlations for three different values of cavity-field decay rate $\kappa$ (which is related to the cavity-photon lifetime as $T_\textrm{cav} \coloneqq (2\kappa)^{-1}$). We see that even in the cases of non-zero loss ($\kappa > 0$), both annealers are still able to find ground states with high success probability, which demonstrates the loss-resilience results found in previous work~\cite{Puri2017a, Nigg2017b}. Just as importantly, the correlations remain strong in the presence of photon losses. As a result, the loss resilience of the statically-coupled architecture carries over to the dynamically-coupled architecture, and even for those instances where the statically-coupled system suffers in success probability due to photon loss events, the dynamically-coupled system follows its behavior as expected.

While it would be desirable to validate our theoretical results with quantum simulations having more than $N=4$ spins, full quantum simulations are prohibitively expensive for $N>4$. Fortunately, for an oscillator-based quantum-optical system, there is a natural set of classical equations of motion (EOMs) which one can derive from the quantum model; formally, this consists of replacing the annihilation operators $\hat{a}_i$ with coherent-state amplitudes $\alpha_i \in \mathbb{C}$ in the quantum Heisenberg EOMs. Such a set of classical EOMs does not fully capture the dynamics of the system in the quantum regime, but we can nevertheless simulate these classical EOMs for both the statically-coupled and dynamically-coupled annealers and compare their dynamical behavior. If the two systems correspond well, we gain further confidence that our theoretical results can lead to the desired performance on a dynamically-coupled quantum annealer. Figure~\ref{fig:classical} shows the results of simulating these classical EOMs for both architectures on SK7 problems. Figure~\ref{fig:classical}A shows the evolutions of the success probabilities for a particular $N=10$ problem instance, while Fig.~\ref{fig:classical}B shows the evolutions of the field amplitudes $\alpha_i(t)$. We see that both the success probabilities and the oscillator amplitudes of the two architectures match well. Figure~\ref{fig:classical}C shows the distribution of success probabilities for different problem sizes $N$ up to $N=50$, using 100 problem instances for $N \leq 30$, and 30 instances for $N=50$; the corresponding correlation plots for $N=8$ and $N=30$ are also shown to the bottom of the figure (additional correlation plots can be found in the Supplementary Materials). There is good agreement between the simulation results of the statically- and dynamically-coupled systems: the success probability as a function of $N$ scales in the same way, and the correlation plots show similar performance on an instance-by-instance basis.

Having shown that the dynamically-coupled system indeed reproduces the behavior of the desired quantum annealer, we now discuss an important tradeoff inherent in our PM scheme for dynamical coupling. Examining the expression \eqref{eq:C-eff} for $C_\text{eff}$, we see that a key parameter is the ratio between the native couplings $J_{ij}$ and the desired problem-strength parameter $\lambda_\text{C}$. For simplicity, we take in this work an approach where the bus-resonator interactions are designed in such a way that the native couplings are approximately constant: $J_{ij} \approx \lambda_\text{J} > 0$ for $j \neq i$ (see the Supplementary Materials for more details). Under this framework, we can identify the ``dynamical-coupling parameter'' $\eta \coloneqq \lambda_\text{C}/\lambda_\text{J}$, so called because intuitively, $\eta$ captures the ``cost'' of the dynamical-coupling scheme. More precisely (see the Supplementary Materials for details), for any given problem, there is a fixed, required value for $\lambda_\text{C}/\chi$ to ensure successful annealing (regardless of whether this is obtained by physical pairwise couplers or through our dynamical-coupling scheme). As a result, a small value for $\eta$ implies a large required value for $\lambda_\text{J}/\chi$. However, as we discuss later, the absolute scale of $\lambda_\text{J}$ is generally hardware-limited, either by the physically realizable bus-oscillator coupling rates or the amount of available bandwidth in the system. Assuming we operate at one of these hardware limits on $\lambda_\text{J}$, a small value of $\eta$ thus implies a correspondingly small value for the nonlinear rate $\chi$, which is unfavorable in the presence of a fixed cavity-photon lifetime. (A full account of the hierarchy of timescales and the chain of parameter requirements are given in the Supplementary Materials.) Therefore, it is desirable to use as high a value of $\eta$ as possible.

\begin{figure}[t]
  \centering
  \includegraphics[width=\textwidth]{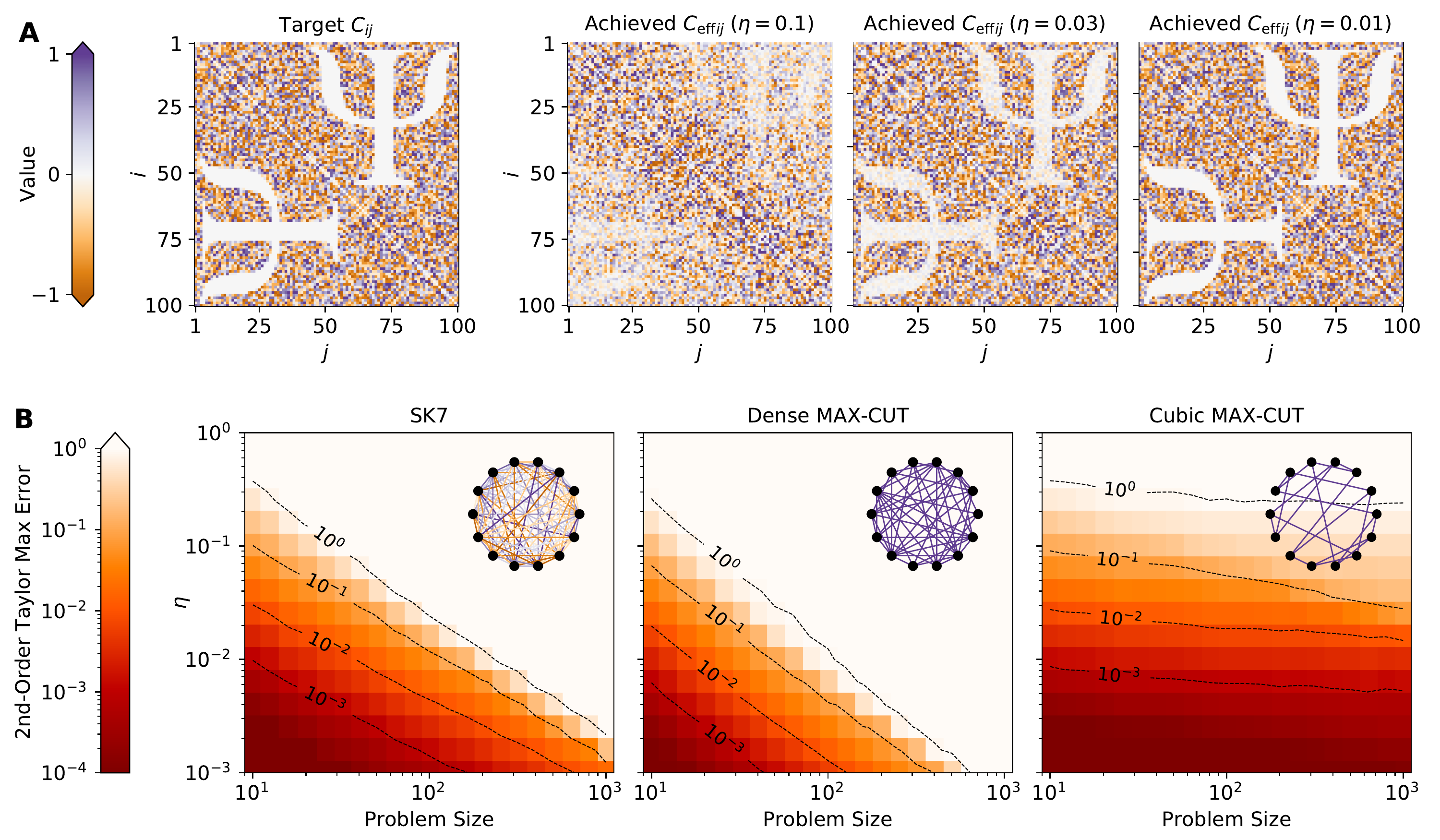}
  \caption{\textbf{Accuracy of the effective couplings.} (\textbf{A}) Comparison of the effective coupling matrices $C_\text{eff}$ (target Ising problem matrix $C$ shown to the left) for different settings of the dynamical-coupling parameter $\eta$, which sets the ratio between the effective coupling strength and the strength of the native coupling provided via the bus. As $\eta$ is decreased, the accuracy of the $C_\text{eff}$ increases. (\textbf{B}) The maximum entry-wise error in the effective coupling matrix $C_\text{eff}$, averaged over 100 problem instances for each size $N$, as a function of the dynamical-coupling parameter $\eta$. We show the scaling for three different classes of Ising problems, with illustrative examples of instances shown as insets. For all panels, the modulation matrix $F$ used to produce $C_\text{eff}$ is found by a numerical optimizer that attempts to maximize the accuracy by using a second-order Taylor approximation to \eqref{eq:C-eff}, with native couplings $J_{ij}$ constructed to be approximately uniform (see the Supplementary Materials).}
  \label{fig:error}
\end{figure}

On the other hand, a clear requirement for dynamical coupling to work well is the ability to obtain $C_\text{eff} \approx C$ (so that $\hat H_\text{dynamical} \approx \hat H_\text{static}$) for any desired $C$. When $\eta$ becomes large, however, the prefactor $J_{ij}/\lambda_\text{C} \approx 1/\eta$ in \eqref{eq:C-eff} becomes small. As discussed in the Supplementary Materials, this effect can become detrimental to our ability to obtain $C_\text{eff} \approx C$: intuitively, the maximum magnitude of the elements of $C$ is by construction unity, but for a sufficiently small prefactor for the integral (i.e., sufficiently small $1/\eta$), it becomes impossible to find a suitable set of coefficients $F^{(k)}_i$ that could allow the integral over the cosine to compensate. This phenomenon is demonstrated in Fig.~\ref{fig:error}. Given a particular target coupling matrix $C$, Fig.~\ref{fig:error}A shows, over a range of values for $\eta$, the $C_\text{eff}$ matrices corresponding to the optimal $F$ matrix found by our (second-order-Taylor-expansion-based) numerical routine (see the Supplementary Materials). We see that when $\eta$ is chosen to be too large, the correspondence between $C$ and $C_\text{eff}$ is rather poor. However, by decreasing $\eta$, one can achieve improved accuracy. More quantitatively, Fig.~\ref{fig:error}B shows the maximum element-wise error in the $C_\text{eff}$ matrix, as a function of $\eta$ and problem size $N$ (averaged over an ensemble of 100 instances for each $N$). To show how the resulting error depends on the structure of $C$, we consider three problem classes: the SK7 class discussed above, the class of unweighted MAX-CUT problems with 50\% edge density, and the class of unweighted MAX-CUT problems on cubic graphs (see the Supplementary Materials for details about instance generation). As expected from our intuition about the role of $\eta$, the particular scaling of the error with $\eta$ depends on the type of problems considered: for SK7, the required $\eta$ goes approximately as $\eta \propto 1/N$, while for Dense MAX-CUT, $\eta \propto 1/(N \log N)$; in contrast with these two dense problem classes, cubic MAX-CUT problems require $\eta \propto 1/\log N$. From this figure, we see that, assuming we can tolerate an upper limit for the error in $C_\text{eff}$ (say, of \SI{3}{\percent}), there is a maximum value for $\eta$ that we are able to use for any given $N$. By staying below this maximal value, we can ensure feasible solutions for the modulation coefficients $F_i^{(k)}$ which generate effective couplings to within the acceptable error. In the Supplementary Materials, we derive for each problem class an explicit, functional form of $\eta$ with respect to problem size $N$ such that we are guaranteed an error of at most \SI{3}{\percent}, but which is also not so conservative that the resulting nonlinear rate $\chi$ is unnecessarily low.

Finally, to study the technological feasibility of our dynamical-coupling scheme, it is important to understand how this tradeoff between the accuracy of the effective couplings and the nonlinear rate---in combination with the requirements for the Floquet approximation---translate into concrete scaling requirements for hardware figures of merit as a function of problem size. As previously mentioned, two relevant hardware limitations are the maximum coupling rate $g_\text{max}$ between each oscillator and the bus, as well as the maximum realizable detuning $\Delta_\text{max}$ between the bus and each oscillator. In the presence of these two limitations, we find (see the Supplementary Materials for a full analysis) that there is a maximum allowable nonlinear rate $\chi_\text{max}$ for dynamical coupling to work.

\begin{figure}
  \centering
  \includegraphics[width=0.73\textwidth]{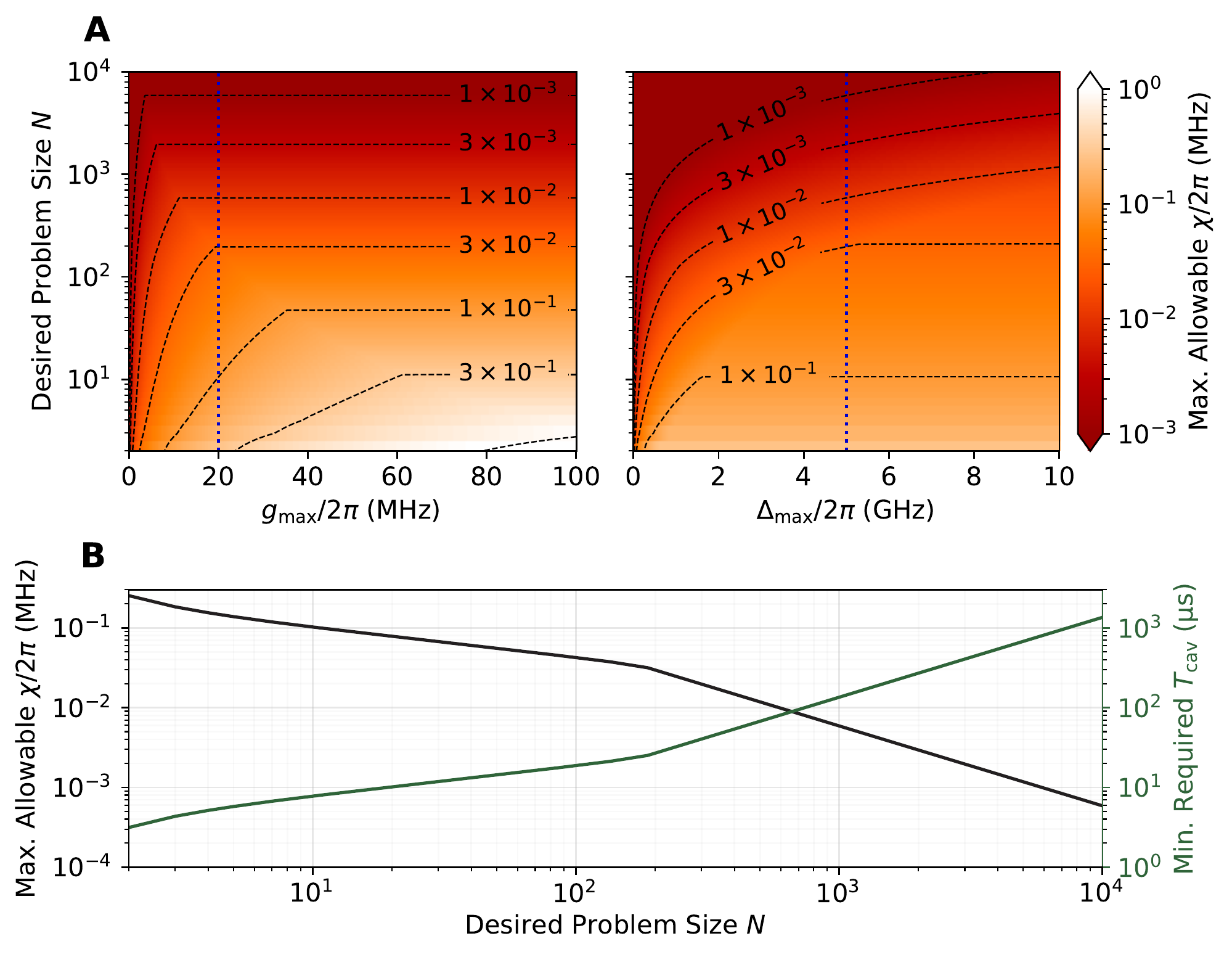}
  \caption{\textbf{Scaling requirements for SK7 problems.} Due to the tradeoff between accuracy of the effective couplings and the required strength of native couplings, hardware limitations on the achievable bus-oscillator detunings and bus-oscillator couplings impose a maximum allowable nonlinear rate $\chi$ in the system. (\textbf{A}) The maximum allowable nonlinear rate as a function of the desired problem size $N$ and the hardware-limited bus-oscillator coupling rate $g_\text{max}$ (left, with $\Delta_\text{max}/2\pi = \SI{5}{GHz}$) and the hardware-limited bus-oscillator detuning $\Delta_\text{max}$ (right, with $g_\text{max}/2\pi = \SI{20}{MHz}$). Dashed vertical lines in blue indicate the slice of these plots shown in (B) below. (\textbf{B}) The black line (left axis) shows the maximum allowable nonlinear rate as a function of problem size for fixed  $\Delta_\text{max}/2\pi = \SI{5}{GHz}$ and $g_\text{max}/2\pi = \SI{20}{MHz}$. The green line (right axis) shows the consequence of the maximum allowable nonlinear rate for the minimum required cavity-photon lifetime $T_\text{cav}$, assuming that the quantum annealer requires $\chi \leq \kappa/10$. See the Supplementary Materials for the corresponding scaling requirements for MAX-CUT problem classes.}
  \label{fig:parameter}
\end{figure}

In Fig.~\ref{fig:parameter}A, we show $\chi_\text{max}$ as a function of problem size $N$, upon varying the hardware limits $g_\text{max}$ and $\Delta_\text{max}$ (for the SK7 problem class; see the Supplementary Materials for MAX-CUT). We observe that at fixed values of $g_\text{max}$ and $\Delta_\text{max}$, $\chi_\text{max}$ decreases as $N$ increases. To more concretely interpret the implications of the requirement $\chi \leq \chi_\text{max}$, we note that $\kappa \ll \chi$ is a necessary condition for the annealer to operate in the quantum, rather than the dissipative, regime. Thus, the requirement of $\chi \leq \chi_\text{max}$ translates directly to a minimum required cavity-photon lifetime $T_\textrm{cav} = (2\kappa)^{-1}$. Taking $\kappa = \chi/10$ for concreteness, we plot the minimum required cavity-photon lifetime $T_\textrm{cav}$ on the right axis of Fig.~\ref{fig:parameter}B as a function of $N$, for $g_\textrm{max}/2\pi = \SI{20}{MHz}$ and $\Delta_\textrm{max}/2\pi = \SI{5}{GHz}$. This plot indicates that $T_\text{cav} \approx \SI{100}{\micro\second}$ is required to achieve $N=1000$. The values for $g_\text{max}$ and $\Delta_\text{max}$ are readily achievable experimentally; regarding the $T_\text{cav}$ requirement, transmon qubits with $T_1$ times beyond \SI{500}{\micro\second} have been measured \cite{Gambetta2019}. We also mention that, in the spirit of Refs.~\cite{Johnson2011, Boixo2014}, one could also ignore this latter requirement on $T_\text{cav}$ and build a system with the best $T_\text{cav}$ one can achieve---even if it is smaller than the required $T_\text{cav}$ to satisfy $\kappa = \chi/10$---and experimentally explore the performance of a highly dissipative quantum annealer. For a small experimental demonstration that still obeys the $\kappa = \chi/10$ requirement, $T_\textrm{cav} \sim \SI{10}{\micro\second}$ should be sufficient to realize an $N=10$ version of the system; this is a very realistic parameter regime for current experiments.

\section{Discussion}
There are two fundamental tradeoffs that one is making by adopting our dynamical-coupling architecture versus a static-coupling architecture: firstly, it is necessary to perform some classical precomputation to obtain the phase modulation coefficients (i.e., the matrix $F$), and secondly, each oscillator's cavity-photon lifetime in a dynamical-coupling architecture will likely need to be longer than it would in a static-coupling architecture (this is a consequence of the required native coupling strength $\lambda_\text{J}/\chi$, and hence the required  cavity-photon lifetime, increasing with the desired accuracy of the achieved couplings ${C_\text{eff}}_{ij}$). In exchange for accepting these two downsides, one is able to build a fully programmable, fully connected $N$-spin quantum annealer with a number of qubits and couplers that only scales as $N$, as opposed to current proposals for statically-coupled quantum annealers, which require a number of qubits and/or couplers that scales as $N^2$. 

Our method for computing $F$---by minimizing a second-order Taylor approximation of effective coupling error---requires only $\mathcal{O}(N^3)$ time, which is efficient, and remarkably so given that the Ising problem matrix contains $\mathcal{O}(N^2)$ entries in general, hence the runtime could at best be $\mathcal{O}(N^2)$. Moreover, in our numerical experiments, the wall-clock times when running our implementation of the method on a single-core processor were approximately \SI{2}{min} for $N=1000$, so the precomputation time should not be a significant practical concern, given the difficulty of solving hard instances of Ising problems. With regards to the required cavity-photon lifetime, we have provided (in the Supplementary Materials) a prescription for the parameters in the dynamical-coupling scheme such that the error in the realized couplings will be at most $\SI{\sim3}{\percent}$, and we have computed the resulting lifetime requirement as a function of $N$. Our scaling analysis shows that our dynamical-coupling approach could be realized for $N=1000$ with existing superconducting-circuits technology, provided that JPOs with cavity-photon lifetimes of $\SI{\sim 100}{\micro \second}$ can be engineered.

In the current approach from D-Wave Systems \cite{Johnson2011,Bunyk2014,Hamerly2019}, as well as the proposed approaches in Refs.~\cite{Puri2017a, Lechner2015}, realizing a 1000-spin quantum annealer for fully connected problems would require on the order of $10^6$ qubits. This stands in contrast to our proposed architecture, which would require only 1000 oscillators to achieve the same number of spins. Beyond the engineering expense of implementing a relatively larger number of qubits, quantum annealers using problem embedding can also suffer from an additional exponential degradation in success probability \cite{Hamerly2019}. The dynamical-coupling approach uses the minimal number of physical qubits possible, and hence avoids this additional penalty in performance.

Outside the context of quantum annealing for solving optimization problems, our work has a strong connection with Floquet engineering of quantum simulators and quantum spin chains \cite{ Oka2019, Moessner2017, Goldman2016, Kyriienko2018, Gorg2018}. A quantum annealer, in addition to being an optimization machine, is also a realization of a quantum simulator of the transverse-field Ising model \cite{Gardas2018}. We anticipate that our techniques for dynamical control of the couplings in a spin system will also enable the realization of novel simulation capabilities for more general spin Hamiltonians. Furthermore, our technique and derivations are not restricted to simulators of spin systems, but apply directly to generic bosonic simulators, and may allow more complex Hamiltonians to be engineered than are currently realized with pairwise physical couplers in Bose-Hubbard simulators built with superconducting circuits \cite{Houck2012, Roushan2017, Roushan2017a, Ma2019, Yan2019}. 

\begin{acknowledgments}
We thank Johannes Majer, Leigh Martin, Tim Menke, Kevin O'Brien, William Oliver, Shruti Puri, Chris Quintana, and Shyam Shankar for helpful discussions, and Daniel Wennberg and Ryotatsu Yanagimoto for comments on a draft of the manuscript. This research was partially funded by NSF award PHY-1648807 (T.O., E.N.) and Stanford University (the Nano- and Quantum Science and Engineering Postdoctoral Fellowship; P.L.M.). \textit{Author contributions:} P.L.M. conceived and supervised the project. T.O. and E.N. derived the analytical models, performed the numerical simulations, and produced the figures. P.L.M., T.O., and E.N. wrote the manuscript. \textit{Competing interests:} P.L.M., T.O. and E.N. are listed as inventors on a U.S. provisional patent application (No.\ 62/814750) related to the architecture described in this paper.
\end{acknowledgments}

%% file: supplementary.tex
This document is structured as follows:
\begin{itemize}
\item In Section~\ref{sec:dynamical-couplings}, we describe our design principles, based on quantum Floquet theory, for programming arbitrary oscillator-oscillator couplings via dynamical control of a set of oscillators with a fixed set of all-to-all native couplings.  We cast the problem as a generic, hardware-independent design problem of specifying phase-modulation indices for each oscillator.  When the modulations required are small, we provide approximate solutions to the design problem in the form of analytic solutions at first order as well as an efficient numerical routine at second order. We also perform an analysis of the accuracy of the effective couplings and identify the dynamical-coupling parameter as a pivotal quantity in controlling the errors. Finally, we provide some guidelines for choosing the Floquet frequency in order to ensure that the Floquet approximation holds.

\item In Section~\ref{sec:QA-KPO}, we review the use of nonlinear Kerr parametric oscillators (KPOs) in quantum annealing to solve Ising spin problems.  We present an archetypal model of a statically-coupled KPO network as introduced by Ref.~\cite{Goto2016b}, whose performance and solution mechanism we aim to approximate with our dynamical-coupling scheme.  We also elaborate on some of the technical operational requirements of the statically-coupled annealer with respect to cavity field decay rate and problem-instance edge density, which we incorporate into our dynamically-coupled annealer.

\item In Section~\ref{sec:bus-KPO}, we present a hardware model for the realization of the KPO in the form of a Josephson parametric oscillator (JPO).  We derive the relationship between the flux control signals applied to the JPO and resulting instantaneous frequency of the oscillator; the flux control signals are the control variables for achieving dynamical coupling.  We also derive a model for an all-to-all-coupled network of JPOs using a common bus resonator, which provides the fixed native couplings that our dynamical-coupling scheme transforms into programmable effective couplings.  We formulate and address the design problem that arises in performing dynamical coupling: what control signal should one apply to the flux lines in order to realize the dynamical modulations prescribed by our dynamical-coupling scheme given in Section~\ref{sec:dynamical-couplings}?

\item In Section~\ref{sec:native-coupling}, we analyze the structure of the native coupling matrix and how it is affected by both the dynamical modulations (which can lead to temporal variation in the couplings) as well as the bus-oscillator detunings of the oscillators (which affect the uniformity of the native couplings). Using this analysis, we give a prescription for choosing the bus-oscillator couplings and nominal detunings of each oscillator, in a way that ensures that the native coupling matrix possesses desirable properties such as approximate time-independence and uniformity while obeying all other constraints imposed by the analyses of previous sections.  In particular, these properties dramatically simplify the procedure for solving the control design problem formulated in Section~\ref{sec:bus-KPO}.

\item In Section~\ref{sec:scaling-considerations}, we bring together the results of the above sections to give, for any desired Ising problem matrix, an explicit prescription for choosing the system parameters and modulations in order to realize dynamical coupling in a bus-mediated JPO network, while still respecting the constraints and approximations made in the above sections. With this prescription in hand, we then analyze the implications for scalability in the context of hardware limitations on the oscillator-bus detuning (i.e., system bandwidth) and the oscillator-bus coupling. We find that the main implication is that a maximum allowable nonlinear rate is imposed on the system, which translates into a minimum required cavity photon lifetime that generally increases with problem size.

\item In Section~\ref{sec:methodology}, we describe the open-quantum system formalism we use to study the dynamics of the quantum annealer in our simulations. We also detail how we compute the metrics we use in our main manuscript figures, such as the success probability and spin configuration probabilities. We derive a classical model which is amenable to simulations of larger problem sizes in order to gain confidence that dynamical coupling can be achieved for more than a handful of oscillators.

\item In Section~\ref{sec:Ising-problem-classes}, we define more formally the problem classes (SK7, Dense MAX-CUT, and Cubic MAX-CUT) we use in this work to study the performance of the dynamical-coupling scheme. We also briefly describe how we generate instances drawn from these problem classes.

\item Finally, in Section~\ref{sec:data-dump}, we provide additional data from our simulations which may be relevant or useful to interpreting our main results. In particular, we present the trajectories that resulted from simulating all 100 problem instances we used in our quantum simulations, as well as the correlation matrices for every problem size we considered in our classical simulations.
\end{itemize}

Finally, some notes on our conventions:
\begin{enumerate}
    \item Throughout this work, we set $\hbar = 1$, so that energy (i.e., our Hamiltonians) are in units of frequency.
    \item Furthermore, whenever the limits of a summation are not explicitly stated, it is implied that the index runs over the set $\set{1,\ldots,N}$, where $N$ is the problem size.
    \item Finally, whenever we use the notation $x \sim y$, we mean the heuristic statement ``$x$ is on the order of $y$''.
\end{enumerate}

\newpage
\section{Design principles for dynamical couplings} \label{sec:dynamical-couplings}
In the context of engineering dynamical couplings, we are primarily concerned with the problem of designing a set of control variables to cause a Hamiltonian possessing native couplings to approximate some other desired target Hamiltonian exhibiting static couplings of a particular form.  More concretely, we suppose we start with a native Hamiltonian having the oscillatory form
\begin{equation} \label{eq:H-native}
    \hat H_\text{native} = -\sum_{i \neq j} J_{ij}(t) \exp\sbrak{-\im\Paren{\phi_j(t) - \phi_i(t)}} \hat a_i\dagg \hat a_j,
\end{equation}
where $J_{ij}(t)$ (satisfying $J_{ij}=J_{ji}$) is the magnitude of the native coupling between modes $\hat a_i$ and $\hat a_j$, and $\phi_i(t)$ are scalar functions of time.  This Hamiltonian can be interpreted as the excitation-exchange energy between oscillators with nominal native couplings $J_{ij}(t)$ and where $\phi_i(t)$ is the total accrued phase of oscillator $i$ in the Hamiltonian's frame.  We show in Section~\ref{sec:bus-KPO} one realization of such a Hamiltonian in a system of nonlinear oscillators, in which time-dependent control of the functions $\phi_i(t)$ can be designed.

Our goal of engineering dynamical couplings is to achieve
\begin{equation} \label{eq:coupling-problem}
\hat H_\text{native} \approx \hat H_\text{target} \coloneqq \lambda_\text{C} \sum_{i\neq j} C_{ij} \hat a_i\dagg \hat a_j,
\end{equation}
where $C_{ij}$ (satisfying $C_{ij} = C_{ji}$ and $\max_{ij} |C_{ij}| = 1$) is the desired target coupling between modes $\hat a_i$ and $\hat a_j$, while the problem-strength parameter $\lambda_\text{C} \in \mathbb{R}^+$ controls the overall strength of the target couplings.  In general, if we only have limited control over $J_{ij}(t)$ and $\phi_i(t)$, it is impossible to satisfy such a relationship (i.e., obtain $\hat H_\text{native} = \hat H_\text{target}$) for all times. To solve this problem, we utilize quantum Floquet theory \cite{Bukov2015b}: if $\hat H_\text{native}$ is a periodic Hamiltonian with period $2\pi/\Lambda$, where $\Lambda \in \mathbb{R}^+$ is the Floquet frequency, then it can be approximated by its time average over one period:
\begin{equation} \label{eq:H-eff}
    \hat H_\text{eff} \coloneqq \int_0^{2\pi/\Lambda} \hat H_\text{native}(t) \, \frac{\dif t}{2\pi/\Lambda},
\end{equation}
provided that $\Lambda$ is sufficiently large (i.e., larger than any other energy scale in the system Hamiltonian) \cite{Bukov2015b}.  The evolution of the system under $\hat H_\text{eff}$ matches the evolution under $\hat H_\text{native}$ at periodic times $2\pi k/T$, where $k \in \mathbb Z$.  For this reason, the dynamics under the ``Floquet approximation'' $\hat H_\text{native} \approx \hat H_\text{eff}$ is a stroboscopic approximation to the original dynamics.  (Formally this is a leading-order approximation in the Magnus expansion of $\hat H_\text{native}$ \cite{Bukov2015b}.)  We discuss in this section how we use the Floquet approximation can be used to solve the design problem \eqref{eq:coupling-problem}.

\subsection{Floquet design problem}
In principle, we can use any form of $\phi_i(t)$ that causes $\hat H_\text{native}$ to satisfy the Floquet approximation while achieving \eqref{eq:coupling-problem}.  For concreteness, however, we choose the following ansatz for the form of $\phi_i(t)$:
\begin{subequations} \label{eq:phi-ansatz}
\begin{align}
    \phi_i(t) &= \omega_0 t - (i-1)\Lambda t + \delta\phi_i(t), \quad\text{with} \\
    \delta\phi_i(t) &\coloneqq -\sum_{k=1}^{N-1} F^{(k)}_i \sin(k \Lambda t).
\end{align}
\end{subequations}
This ansatz allows us to reduce the design problem for $\phi_i(t)$ into a design problem for the (scalar) Floquet design parameters $F^{(k)}_i \in \mathbb{R}$ for $i \in \set{1, \ldots, N}, k \in \set{1, \ldots, N-1}$. (In the main manuscript $F^{(k)}_i$ is alternative named phase modulation coefficients.)  Note that we also assume that $J_{ij}(t)$ can be made periodic with period $2\pi/\Lambda$. (Preferably, $J_{ij}(t)$ is approximately constant on the timescale of $\Lambda$.)

Interpretationally, if the energy shifts due to other terms of the Hamiltonian and $\dot{\delta\phi}_i(t)$  are small, then the mean frequency of the $i$th oscillator is approximately $\omega_0 - (i-1)\Lambda$, corresponding to evenly-spaced detunings by the Floquet frequency $\Lambda$.  The functions $\delta\phi_i(t)$ then represent additional dynamical phase modulation applied to each oscillator, which are used to engineer the effective interactions, with parametric controls represented by the scalars $F^{(k)}_j$.

Provided that the $\phi_i(t)$ have the form prescribed in \eqref{eq:phi-ansatz}, then the Floquet approximation tells us that
\begin{subequations}
\begin{equation}
    \hat H_\text{native} \approx \sum_{i \neq j} I_{ij} \hat a_i\dagg \hat a_j,
\end{equation}
where the Floquet integral is
\begin{equation} \label{eq:floquet-integral}
    I_{ij} \coloneqq \int_0^{2\pi/\Lambda} \!\! J_{ij}(t) \cos\biggl[(i-j)\Lambda t + \sum_{k=1}^{N-1} \paren{F^{(k)}_i - F^{(k)}_j} \sin(k\Lambda t) \biggr] \frac{\dif t}{2\pi/\Lambda}.
\end{equation}
\end{subequations}
Thus, instead of trying to solve \eqref{eq:coupling-problem} directly for all times, we can instead try to satisfy
\begin{equation} \label{eq:floquet-equation}
    I_{ij} = \lambda_\text{C} C_{ij}.
\end{equation}
This is the Floquet design equation, which results in \eqref{eq:coupling-problem} holding true in a time-averaged sense.  We note that the time averaging does not preclude the possibility of designing a slow or adiabatic variation in, for example, $\lambda_\text{C}$ (i.e., that it be a function $\lambda_\text{C}(t)$ of time), since any variation on timescales much longer than $2\pi/\Lambda$ is approximately constant in the Floquet integral; furthermore, such adiabatic variations need not be periodic in $\Lambda$.

The design parameters $F^{(k)}_i$ can be interpreted in a signal-processing sense as phase modulation indices since they cause the phase of the cosine function in \eqref{eq:floquet-integral} to modulate in time.  Thus, our scheme for achieving dynamical coupling can be equivalently viewed in a signal-processing framework as a phase-modulated coupling scheme.

\subsection{Small-modulation approximation} \label{sec:small-modulation}
One way to solve the Floquet design equation in general involves inserting the appropriate expression for $J_{ij}$ and performing an optimization or root-finding search for $F^{(k)}_j$ by iterated evaluation of the integral.  However, under certain special assumptions and/or restrictions, it is possible to simplify the problem further.

A number of assumptions can be made to simplify the Floquet equation.  Here, we consider one such special case where 1) the time dependence of $J_{ij}(t)$ is negligible; and 2) the magnitude of the phase modulation terms satisfy
\begin{equation} \label{eq:varphi}
    \varphi_{ij} \coloneqq \sum_{k=1}^{N-1} \paren{F_i^{(k)}-F_j^{(k)}}\sin(k\Lambda t) \sim \zeta,
\end{equation}
where $\zeta \ll 1$.  Then we can expand \eqref{eq:floquet-integral} to first order in $\zeta$ (i.e., around $\varphi_{ij} = 0$) and \eqref{eq:floquet-equation} becomes, to first order,
\begin{equation} \label{eq:floquet-equation-taylor1}
    -\frac 1 2 J_{ij} \paren{F^{(i-j)}_i - F^{(i-j)}_j} = \lambda_\text{C} C_{ij},
\end{equation}
for $i > j$.  Thus, the problem reduces to a simple set of linear equations in $F^{(k)}_i$, which is readily solvable.  We call this case the time-independent, small-modulation approximation.  This equation likewise has an interpretation in terms of signal processing.  In the rotating frame of the coupling interaction, the $i$th oscillator has carrier frequency $i\Lambda$, and so phase modulation at frequency $(i-j)\Lambda$ produces sidebands on oscillator $i$ that overlap with oscillator $j$ at carrier frequency $j\Lambda$.

There are a number of general solutions to \eqref{eq:floquet-equation-taylor1}, due to the symmetries of the equations.  One solution prescribes that the $i$th oscillator be driven by $i-1$ different frequencies, according to the recursive relation
\begin{equation}
    F^{(k)}_{k+j} = -\frac{2\lambda_\text{C}}{J^{(k)}_{j}} C_{j}^{(k)} + F^{(k)}_j,
\end{equation}
with a base case of $F_{i}^{(k)} = 0$ for $1 \leq i \leq N$ and $i \leq k \leq N-1$.  Here and elsewhere, $C^{(k)}_i \coloneqq C_{i,i+k}$ and $J^{(k)}_i \coloneqq J_{i,i+k}$ denotes the $i$th element of the $k$th diagonal of $C$ and $J$, respectively.

Another solution, which emphasizes symmetry in the design parameters, is given by
\begin{align} \label{eq:F-taylor1-symmetric}
  F^{(k)}_i = \begin{cases}
    \sum_{m=0}^{\lfloor(N-k-i)/k\rfloor} \lambda_\text{C} C^{(k)}_{i+mk}/J^{(k)}_{i+mk}, & i \in \set{1, \ldots, \min(k, N-k)} \\
    -F^{(k)}_{[(i+k-1) \bmod k] + 1}, & i \in \set{\max(N-k,k)+1, \ldots, N} \\
    F^{(k)}_{i-k} - 2 \lambda_\text{C}   C^{(k)}_{i-k}/J^{(k)}_{i-k}, & i \in \set{k+1, \ldots, N-k} \\
    0, & \text{otherwise}
  \end{cases},
\end{align}
As discussed later in Section~\ref{sec:F-structure}, this choice of solution centers the phase modulation strength on the middle oscillator $(i = N/2)$, which makes the dynamical-coupling scheme more robust to noise.  (See Section~\ref{sec:F-structure} for further analysis of this solution.)  Unless otherwise noted, this is the symmetry choice we will henceforth utilize for the first-order analytic solution.

While the first-order Taylor approximation provides an analytic form for the design parameters, we can further improve the approximation by expanding the Floquet integral to second order in $\zeta$.  At second order, \eqref{eq:floquet-equation} becomes
\begin{align}
  \frac{\lambda_\text{C} C_{ij}}{J_{ij}} &= -\frac 1 2\paren{F_i^{(i-j)} - F_j^{(i-j)}} -\frac 1 8 \sum_{l=1}^{N-1-(i-j)} \paren{F_i^{(l)}-F_j^{(l)}}\paren{F_i^{(i-j+l)}-F_j^{(i-j+l)}} \label{eq:floquet-equation-taylor2} \\
  &\quad -\frac 1 8 \sum_{l=i-j+1}^{N-1} \paren{F_i^{(l)}-F_j^{(l)}} \paren{F_i^{(l-i+j)}-F_j^{(l-i+j)}} + \frac 1 8 \sum_{l=1}^{i-j-1} \paren{F_i^{(l)}-F_j^{(l)}} \paren{F_i^{(i-j-l)}+F_j^{(i-j-l)}} \nonumber.
\end{align}

In contrast to the first-order expansion \eqref{eq:floquet-equation-taylor1}, it is not possible in general to analytically solve this set of equations.  However, as an algebraic system of second-order equations in $F_j^{(k)}$, it is still substantially easier to solve numerically than the full Floquet problem \eqref{eq:floquet-equation}, and the residual errors are almost negligible even for realistic modulation depths (which stands in contrast to the situation for the first-order expansion).  Further expansions in $\zeta$ are also possible, at the cost of algebraic complexity.

\subsection{Numerical optimization for Floquet design parameters} \label{sec:optimizeF}
As previously mentioned, the Floquet equation can be solved either by direct root finding or numerical optimization (specifically, minimization) over $F_i^{(k)}$ of the objective function
\begin{equation} \label{eq:optimizeF}
    \coloneqq \sum_{i\neq j}\paren{ \lambda_\text{C} C_{ij} - \int_0^{2\pi} J_{ij} \cos\sbrak{(i-j)t + \sum_{k=1}^{N-1} \Paren{F^{(k)}_i - F^{(k)}_j}\sin k t} \frac{\dif t}{2\pi}}^2,
\end{equation}
given two matrices $\lambda_\text{C} C$ and $J$ assumed to be time-independent here for simplicity, which allows us to perform a change of variables $t \mapsto t/\Lambda$ to scale out the Floquet frequency $\Lambda$.  While direct root finding is generally prohibitively expensive, optimization of \eqref{eq:optimizeF} usually works well for small $N$.  However, this optimization occurs over $\mathcal O(N^2)$ variables (the number of $F^{(k)}_i$ entries), which degrades the quality of solutions for larger $N$ (at least, for local methods and/or fixed optimization time).

To get around this issue, our preferred method for solving the Floquet equation is a ``column-by-column'' approach.  This is inspired by the empirical observation that, given any particular set of design parameters $F_i^{(k)}$, modifying the values of the $k$th ``column'' $\Paren{F_1^{(k)}, F_2^{(k)}, \ldots, F_N^{(k)}}$ primarily (though not exclusively) produces changes in the $k$th \emph{diagonal} of the Floquet integral $\Paren{I_{1,1+k}, I_{2,2+k}, \ldots, I_{N-k,N}}$ (and, symmetrically, in the $-k$th diagonal).  That is, instead of doing a single optimization over $\mathcal O(N^2)$ variables, we can opt to perform $\mathcal O(N)$ different optimizations, each over $\mathcal O(N)$ variables.  Of course, because the various values of $k$ are in fact coupled, this approach is unlikely to produce the global optimum.  Nevertheless, we have found that error values of $\abs{C_{ij} - I_{ij}/\lambda_\text{C}} \sim \num{e-4}$ are readily achievable for problem sizes $N \lesssim 100$, and it is reasonable to expect this behavior to continue for larger $N$ as well.

To be more precise, while optimizing column $k$, we utilize the more restricted objective function
\begin{subequations} \label{eq:optimizeF-column}
\begin{equation} \label{eq:optimizeF-column-obj}
  \sum_{j=1}^{N-k} \paren{\frac{\lambda_\text{C} C^{(k)}_j}{J^{(k)}_j} - \int_0^{2\pi} \cos \sbrak{\paren{F_{j+k}^{(k)}-F_j^{(k)}}\sin kt + u^{(k)}_j(t)}\frac{\dif t}{2\pi}}^2,
\end{equation}
where the ``cached'' functions are
\begin{equation}
  u^{(k)}_j(t) \coloneqq kt + \sum_{l \neq k} \paren{F_{j+k}^{(l)}-F_j^{(l)}}\sin lt,
\end{equation}
\end{subequations}
which contain all the values of $F_i^{(l)}$ that are not currently being optimized over and are thus static from the perspective of the $k$th-column optimization.  (To see the connection between \eqref{eq:optimizeF-column} and \eqref{eq:optimizeF}, consider $k \coloneqq i-j$, so that $i = j+k$.)  Thus, we need only update $u^{(k)}_j$ right before starting optimization over the $k$th column.

In this approach, we see that the objective function \eqref{eq:optimizeF-column-obj} costs $\mathcal O(N^2)$ to evaluate: a factor of $N$ for the summation and a factor of $N$ for the integral, which needs $\mathcal O(N)$ samples in time.  The updating of the cache functions $u^{(k)}_j(t)$ costs $\mathcal O(N^3)$: a factor of $N$ for the summation, a factor of $N$ for updating the value at every sample time $t$, and a final factor of $N$ since there are $\mathcal O(N)$ different $u^{(k)}_j$ for each fixed $k$.  Comparing \eqref{eq:optimizeF} and \eqref{eq:optimizeF-column}, we improve the cost of the objective function by $\mathcal O(N)$, which may be significant depending on how the cost of the optimizer scales with the cost of the objective function, compared to the cost of the cache update.

As a technical point, we also note that in \eqref{eq:optimizeF-column-obj}, there are only $N-k$ elements in the summation, but as written, there are $N$ different values of the Floquet design parameters $F_i^{(k)}$; this redundancy can be eliminated by a choice for the symmetry of the design parameters.  For example, in the form chosen for \eqref{eq:F-taylor1-symmetric}, only $\Paren{F_1^{(k)}, \ldots, F_{N-k}^{(k)}}$ are unique, while all other values in that column are either zero or a function of those $N-k$ values.

Similarly, this column-by-column approach can also be adapted to perform numerical optimization in solving the second-order equations \eqref{eq:floquet-equation-taylor2}.  In this case, the objective function is
\begin{subequations} \label{eq:optimizeF-column-taylor2}
\begin{equation} \label{eq:optimizeF-column-taylor2-obj}
    \sum_{j=1}^{N-k} \paren{\frac{\lambda_\text{C} C^{(k)}_j}{J^{(k)}_j} + \frac 1 4\paren{F^{(k)}_{j+k}-F^{(k)}_j}\sbrak{2 + \theta^{N-1}_{2k} \paren{F^{(2k)}_{j+k} - F^{(2k)}_j}} + u^{(k)}_j}^2,
\end{equation}
where the binary-valued step function $\theta^N_M = 1$ iff $N \geq M$.  The cached variables are now
\begin{align}
    u^{(k)}_j &\coloneqq \frac 1 8 \sum_{l=1}^{N-1-k} \paren{1-\delta_{k,l}} \paren{F^{(l)}_{j+k} - F^{(l)}_j}\paren{F^{(k+l)}_{j+k} - F^{(k+l)}_j} \\
    &\quad {} +\frac 1 8 \sum_{l=k+1}^{N-1} \paren{1-\delta_{2k,l}} \paren{F^{(l)}_{j+k} - F^{(l)}_j}\paren{F^{(l-k)}_{j+k} - F^{(l-k)}_j} -\frac 1 8 \sum_{l=1}^{k-1} \paren{F^{(l)}_{j+k} - {(l)}_j}\paren{F^{(k-l)}_{j+k} - F^{(k-l)}_j}. \nonumber
\end{align}
\end{subequations}
Again, these cached variables only involve values of $F^{(l)}_i$ that are not currently being optimized over and are thus static from the perspective of the $k$th-column optimization.

In this second-order approach, the objective function \eqref{eq:optimizeF-column-taylor2-obj} costs only $\mathcal O(N)$ to evaluate, while updating the cached variables costs $\mathcal O(N^2)$.  Thus, the second-order approach is more efficient than the full-order approach \eqref{eq:optimizeF-column} by $\mathcal O(N)$, at the cost of having some residual error (which itself is already better than the residual error in the first-order analytic solution).

If the optimization is performed using a gradient-based algorithm, the second-order approach also has the advantage of having a simple analytic gradient.  The gradient of \eqref{eq:optimizeF-column-taylor2} in the direction of $F^{(k)}_i$ is given by
\begin{subequations} \begin{gather} \label{eq:gradient-taylor2}
    \sum_{j=1}^{N-k} v^{(k)}_j \paren{\frac{\partial F^{(k)}_{j+k}}{\partial F^{(k)}_i} - \frac{\partial F^{(k)}_{j}}{\partial F^{(k)}_i}}, \text{where} \\
    v^{(k)}_j \coloneqq \frac 1 2 \paren{\frac{\lambda_\text{C} C^{(k)}_j}{J^{(k)}_j} + \frac 1 4\paren{F^{(k)}_{j+k}-F^{(k)}_j}\sbrak{2 + \theta^{N-1}_{2k} \paren{F^{(2k)}_{j+k} - F^{(2k)}_j}} + u^{(k)}_j} \sbrak{2 + \theta^{N-1}_{2k} \paren{F^{(2k)}_{j+k} - F^{(2k)}_j}}.
\end{gather} \end{subequations}
The partial derivatives in \eqref{eq:gradient-taylor2} are not explicitly evaluated because their values depend on the choice of symmetries in $F^{(k)}_j$; in an upper-triangular form, for example, the first term would be zero and the second a Kronecker delta; for a symmetry choice like in \eqref{eq:F-taylor1-symmetric}, the matrix of partial derivatives would be a sparse matrix.  Thus, despite the sum in \eqref{eq:gradient-taylor2}, the evaluation of a single element of the gradient has $\mathcal O(1)$ cost, as long as the matrix of partial derivatives has $\mathcal O(1)$ nonzero entries per column (i.e., per fixed $k$).

In summary, provided that we choose a parametrization resulting in a sparse matrix of partial derivatives in $F^{(l)}_i$, the cost of optimizing the Floquet design parameters using a second-order approach is $\mathcal O\Paren{N\cdot(N^2+MN)}$, where $M$ is the number of gradient and objective function evaluations needed to converge at each fixed $k$.  The outside factor of $N$ is needed because we visit each value of $k$ (i.e., column) a constant number of times, while the factor of $N^2$ is the cost of the caching.  If $M = \mathcal O(N)$, then the total cost is $\mathcal O(N^3)$.

\begin{figure}
    \centering
    \includegraphics[width=0.45\textwidth]{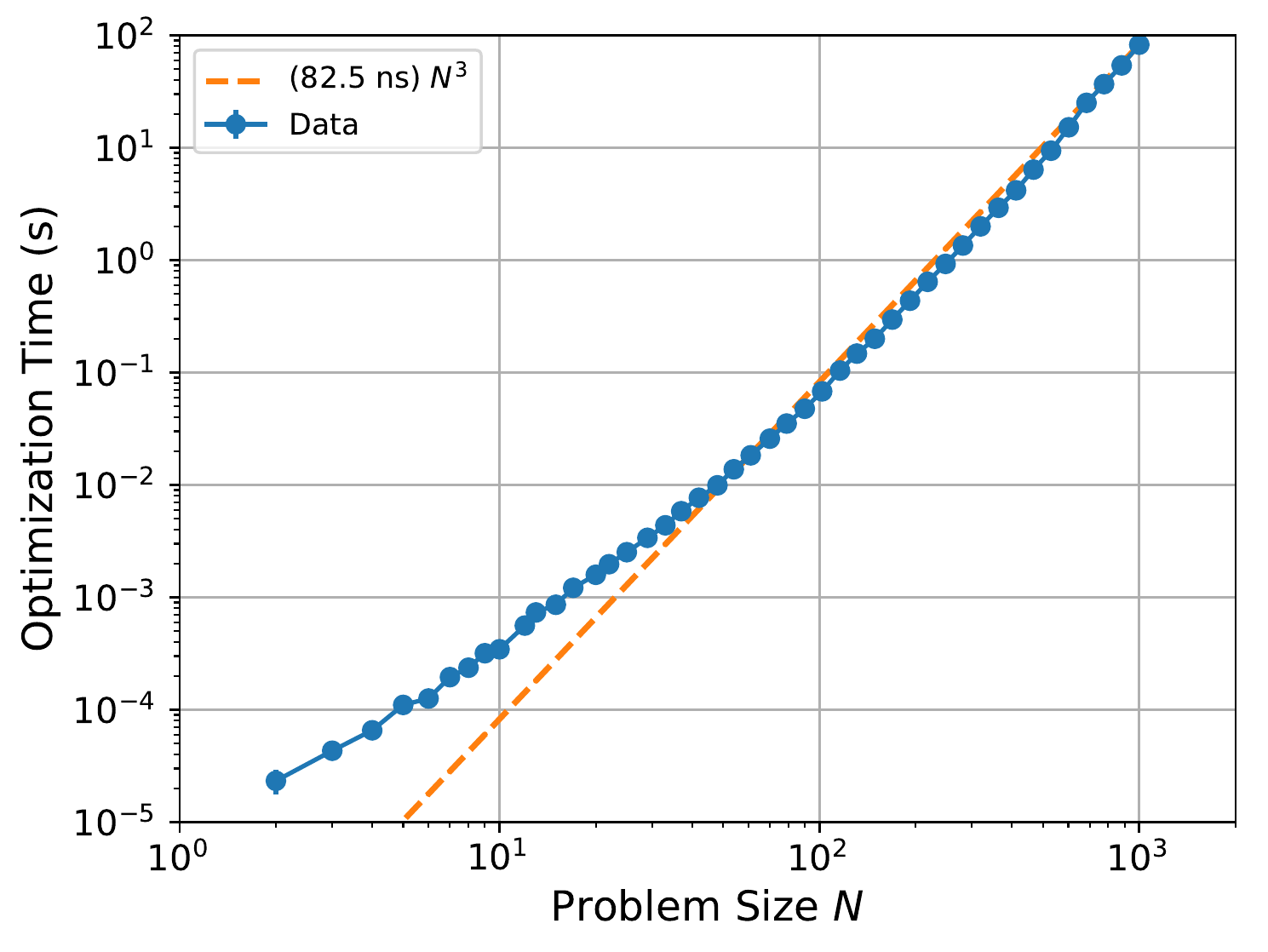}
    \caption{\textbf{Time taken to solve Floquet design problem.} Time taken to solve the Floquet design problem using a second-order, ``column-by-column'' approach as described in Section~\ref{sec:optimizeF} and implemented in Julia.  The optimization time is the time required to compute the $F^{(k)}_j$ Floquet design parameters and is shown, for each problem size $N$, as an average over ten SK7 problem instances.  For the procedure used to select the structure of the native couplings $J_{ij}$, see Section~\ref{sec:parameter-summary}, and for details on the optimization routine, see Section~\ref{sec:optimization-algorithm}.  This calculation was performed with a single Julia process (i.e., using a single core) running on an Intel i7-5600U CPU at a nominal clock speed of \SI{2.60}{GHz}.  Note that error bars (standard error on the mean) are shown but are not visible on all points.}
    \label{fig:benchmark}
\end{figure}

To validate this analysis, we implemented and ran a program using the above optimization method for solving the second-order Floquet design problem.  To test the scaling of the optimization procedure, we run it on random instances of SK7 across different problem sizes. The results of this numerical experiment are summarized in Figure~\ref{fig:benchmark}, from which it is clear that 1) the design problem can be solved with only moderate processing time, even for a problem size of $N = \num{e3}$; and 2) the scaling at large $N$ goes as $\mathcal O(N^3)$, as was argued in the discussion above.

\subsubsection{Additional details on numerical optimization routine} \label{sec:optimization-algorithm}
For the purposes of performing simulations and other numerical studies in this work, we implemented a straightforward version of the optimization routines described above.  Our program is written in the Julia language (version 0.6.4), and the numerical optimization is performed by the Julia package Optim.jl (version 0.14.1) \cite{mogensen2018optim}.  We mention here some additional details particular to our methodology, which may be relevant to the interpretation of our results.

We utilize the column-by-column approach as described in \eqref{eq:optimizeF-column} or \eqref{eq:optimizeF-column-taylor2} (for a second-order Taylor approximation to the former), since it is more scalable than the joint approach described by \eqref{eq:optimizeF}.  In this approach, there are $N-1$ objective functions, one for each value of $k$ in \eqref{eq:optimizeF-column} and \eqref{eq:optimizeF-column-taylor2}.  In each of these objective functions, the optimization variables are $\Paren{F^{(k)}_1, \ldots, F^{(k)}_{N-k}}$; the other values of $F^{(k)}_j$ are determined by symmetry, using \eqref{eq:F-taylor1-symmetric}.  Thus, if the Floquet design parameters $F^{(k)}_j$ were collected as a matrix, then the $k$th objective function involves optimizing over the $k$th ``column'' of that matrix.  The goal is to iteratively optimize each of these columns, using the first-order analytic solution \eqref{eq:F-taylor1-symmetric} as the starting point.

For each column being optimized, we make a call to the optimization routine provided by Optim.jl, using the conjugate gradient method with at most 250 iterations (i.e., steps) \cite{mogensen2018optim}.  The value of the attained objective function minimum is then analyzed.  If this minimum is below a certain tolerance level, then we deem that column to be ``converged'' and remove it from the set of columns that require optimizing further.  This tolerance level is chosen for this work to be $(\num{e-3}\eta)^2(N-k)$, where $\eta$ is the dynamical-coupling parameter (see Sections~\ref{sec:F-structure},~\ref{sec:eta},~and~\ref{sec:parameter-summary}).  Interpretationally, this choice of the tolerance implies that the average element-wise residual error in solving the Floquet design equation \eqref{eq:floquet-equation} is approximately \num{e-3}, at least for the case of approximately uniform native couplings $J_{ij} \approx \lambda_\text{J}$.  (However, we note that for the second-order optimizer, there is also a residual error due to the Taylor approximation as well, which increases with $\lambda_\text{C}/\lambda_\text{J} = \eta$ and can in fact dominate the residual error.)

Once every column to be optimized has been visited, the program then recomputes the objective function values for each column that has converged.  Since the various columns are coupled, it is possible that the objective function value for, say, $k = 1$, previously below the tolerance, now exceeds the tolerance due to the optimization over, say, $k = 5$.  Thus, if any column that has converged now exceeds the tolerance, it is added back into the set of columns requiring additional optimization.  This new set of columns to optimize then becomes the starting point for another ``pass'', and this whole procedure repeats until either 1) the number of such passes exceeds some maximum (set to \num{25} in this work), or if there are no longer any columns left to optimize.  Furthermore, a column which fails to improve its objective function value by more than a fixed amount (set to \SI{1}{\percent} in this work) on two consecutive passes is also removed (permanently) from the set of columns to optimize.  Finally, at the end of the routine, if any of the objective function values are worse than those of the first-order analytic solution \eqref{eq:F-taylor1-symmetric} (specifically, exceeding the latter by more than \SI{1}{\percent} of the tolerance level), then the optimization results are rejected and the latter is returned.

\subsection{Structure of the Floquet design parameters} \label{sec:F-structure}
\begin{figure}[t!]
    \includegraphics[width=1.0\linewidth]{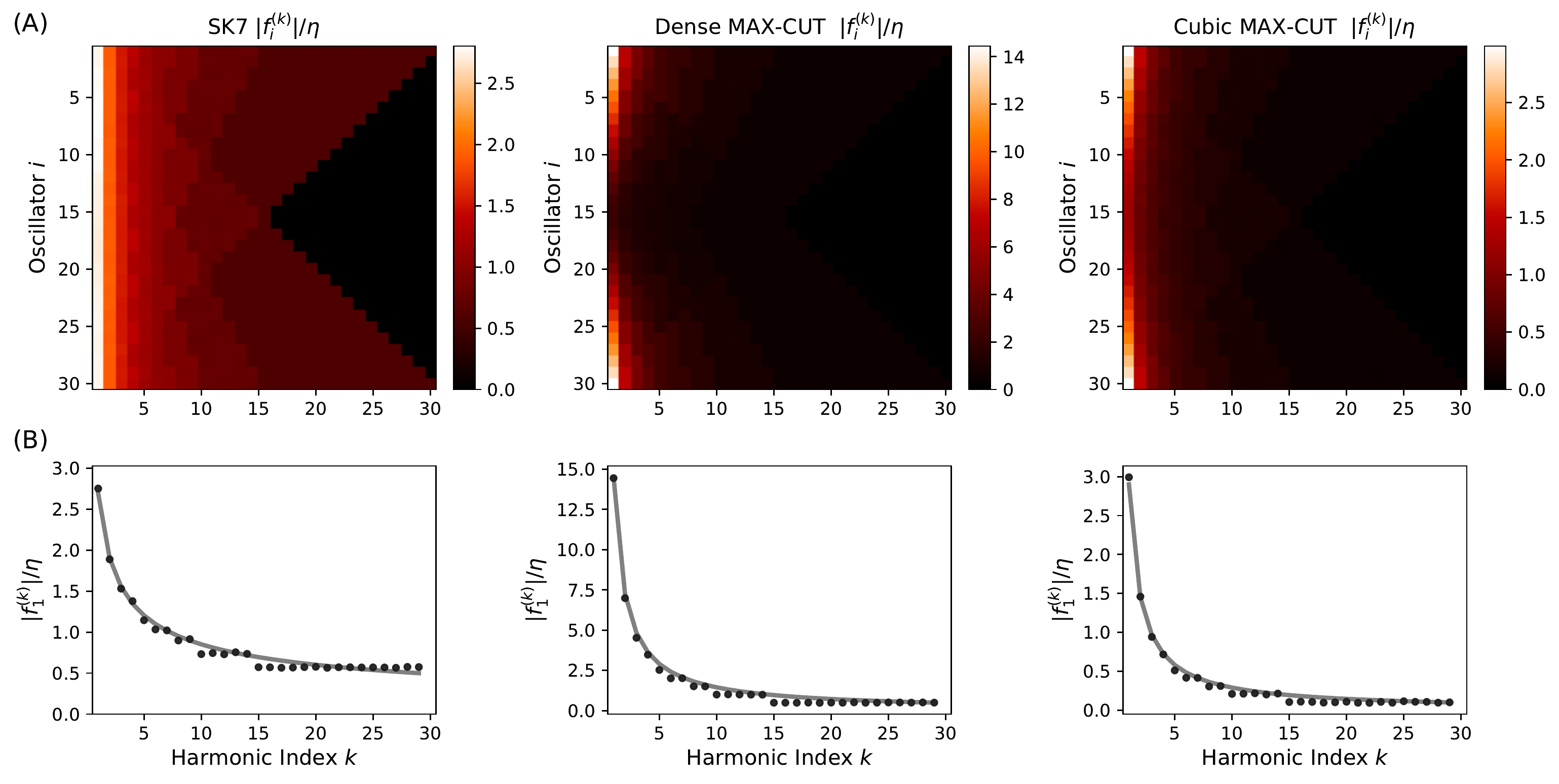}
    \caption{\textbf{Structure of the Floquet design parameters.} Magnitude of the Floquet design parameters $f^{(k)}_j$ in the first-order small-modulation approximation (see Section~\ref{sec:small-modulation}) and assuming a constant, uniform native couplings $J_{ij} = \lambda_\text{J}$, for a problem size of $N = 30$.  Each point is computed as a mean over \num{5000} problem instances.  (A): Heat map showing the mean magnitude as a function of oscillator index $i$ and the harmonic index $k$.  (B): The mean magnitude for the first oscillator, which is equivalent to the first row of each image in (A). The black dots are values produced from numerically averaging over many samples of problem instances while the gray solid line represents the analytic expressions given by \eqref{eq:F1k-analytic}. }
    \label{fig:F-structure}
\end{figure}

Since the Floquet design parameters dictate the physical behavior of the oscillators undergoing phase modulation, it is useful to understand the general structure of $F^{(k)}_j$.  Here, we provide some qualitative analysis which yields relationships that will be useful later on in the analysis of errors and scaling in hardware-specific contexts.  More specifically, we analyze the small-modulation, first-order solution \eqref{eq:F-taylor1-symmetric}: in most of the situations we are concerned with, the numerical solutions do not drastically change the qualitative structure at first order.

The small-modulation solution \eqref{eq:F-taylor1-symmetric} depends on both the target couplings $C_{ij}$ as well as the native couplings $J_{ij}$.  Let us denote by $f^{(k)}_j$ the first-order small-modulation solution to the design problem with $J_{ij}$ taken to be a uniform and time-independent (i.e., $J_{ij}(t) = \lambda_\text{J}$ for $i\neq j$ and $\lambda_\text{J} \in \mathbb R^+$).  This case anticipates approximations we later apply in which the hardware-provided native couplings are, in fact, quite close to this form (see Section~\ref{sec:native-coupling}).

We see that the dynamical-coupling ratio
\begin{equation}
    \eta \coloneqq \frac{\lambda_\text{C}}{\lambda_\text{J}}
\end{equation}
naturally appears in the expression for \eqref{eq:F-taylor1-symmetric}, and it fully captures the dependence of $f^{(k)}_j$ on $\lambda_\text{C}$ and $\lambda_\text{J}$.  However, the utility of the ratio $\eta$ extends beyond just analyzing $f^{(k)}_j$; as we will see, $\eta$ plays a critical role in determining both the performance (see Section~\ref{sec:eta}) as well as the scaling (see Section~\ref{sec:scaling-considerations}) of the dynamical-coupling scheme.  Intuitively, it captures how much of the native coupling strength (represented by $\lambda_\text{J}$) is utilized to generate the effective target coupling strength (represented by $\lambda_\text{C}$).

We present in Figure~\ref{fig:F-structure} the average magnitude of $f^{(k)}_j$ for each of the representative problem classes.  In Figure~\ref{fig:F-structure}\,A, we see that, quite generally, the oscillators at the two extremes of large and small detunings (i.e., nearer to the first and last rows) experience greater overall phase modulation than the oscillators closer to the middle.  We also see that the magnitude of the phase modulation generally decreases as the harmonic of the modulation (i.e., the row index) increases.  In addition, Figure~\ref{fig:F-structure}\,B, highlights the fact that the modulation strength for SK7 problems decays more slowly compared to the MAX-CUT problems.

The latter behavior can be understood by observing that \eqref{eq:F-taylor1-symmetric} in this case simplifies to
\begin{equation} \label{eq:F1k}
f_1^{(k)} =  \eta \sum_{m=1}^{\lfloor (N-1)/k \rfloor } C_{1+(m-1)k}^{(k)}.
\end{equation}
Assuming that $\lfloor (N-1)/k \rfloor \approx (N-1)/k$ and making use of the statistical properties of each problem class, we can derive that
\begin{equation} \label{eq:F1k-analytic}
    \mean |{f_{1}}^{(k)}| \approx \begin{cases}
        \eta/2 \cdot (N-1)^{1/2} \cdot k^{-1/2} & \text{for SK7} \\
        \eta/2 \cdot (N-1) \cdot k\inv & \text{for Dense MAX-CUT} \\
        3\eta \cdot (N-1)/N \cdot k\inv & \text{for Cubic MAX-CUT}
    \end{cases}\;.
\end{equation}
We see that different problem classes have different scaling with respect to $N$ and $k$. In the case of SK7, the target couplings $C_{ij}$ have random alternations in sign, which causes cancellations in the sum of \eqref{eq:F1k}, so that the mean magnitude decays more slowly with modulation frequency. On the other hand, for sparse problem classes, there are asymptotically fewer terms in \eqref{eq:F1k}, leading to a different scaling.  This functional form is shown in Figure~\ref{fig:F-structure}\,B, which indicates good agreement with empirical results, on average.

\subsection{Effective coupling error and the dynamical-coupling parameter} \label{sec:eta}
For the dynamical-coupling scheme to accurately approximate the target Hamiltonian, it is important for the effective couplings under time-averaging to approximate the target couplings accurately.  Given a set of desired target couplings $\lambda_\text{C} C_{ij}$ and a (time-independent) set of native couplings $J_{ij}$, suppose we have solved the Floquet design equation \eqref{eq:floquet-equation} for the design parameters $F^{(k)}_j$ (using, for example, the methods of Section~\ref{sec:optimizeF}).  We then define
\begin{equation}
    {C_\text{eff}}_{ij} \coloneqq \int_0^{2\pi} \frac{J_{ij}}{\lambda_\text{C}} \cos\sbrak{(i-j)t + \sum_{k=1}^{N-1} \Paren{F^{(k)}_i - F^{(k)}_j}\sin kt} \frac{\dif t}{2\pi}
\end{equation}
to be the effective couplings produced by that set of $F^{(k)}_j$.  Of course, if the Floquet design equation is solved well, $C_\text{eff}$ should be close to $C$.  Nevertheless, it is clear that $C_\text{eff}$ can be different from $C$, especially if, for example, we use a ``column-by-column'' optimization approach or derive our objective functions from a small-modulation approximation.

For this work, we define the error of the effective couplings to be
\begin{align}
    \mathcal E_\text{C} \coloneqq \max_{i\neq j} \abs{{C_\text{eff}}_{ij} - C_{ij}}.
\end{align}
This error metric of taking the maximum over all entries is conservative (one could choose to compute an element-wise average error instead, for example) and characterizes the largest element-wise deviation of the effective couplings from the target ones.  We aim here to characterize the error $\mathcal{E}_\text{C}$ and to connect the scaling of the error with the dynamical-coupling parameter $\eta$.

In Section~\ref{sec:small-modulation}, we showed how the Floquet design equation can be simplified by an expansion in the modulation depth $\zeta$.  Intuitively, we expect the numerical algorithms presented in Section~\ref{sec:optimizeF} to perform better and produce lower error when the expansion parameter $\zeta$ is small.  By examining \eqref{eq:varphi}, $\zeta$ is limited by the largest $\varphi_{ij}$.  Furthermore, we have observed in Section~\ref{sec:F-structure} that oscillators with the smallest and largest detunings ($i=1,N$) are driven most strongly (so that $\varphi_{1N}$ is largest), so we can reason that
\begin{equation} \label{eq:zeta-bound}
        \zeta \lesssim \max_{t} \varphi_{1N}(t) = \max_t \sum_{k=1}^{N-1} 2F_1^{(k)}\sin(k\Lambda t) \leq 2\sum_{k=1}^{N-1} \left|F_1^{(k)}\right|,
\end{equation}
where we have used $F_{N}^{(k)} = -F_1^{(k)}$ on the equality.  We now further assume, as argued in Section~\ref{sec:F-structure}, that the structure of $F^{(k)}_j$ is similar to the structure of $f^{(k)}_j$, derived in the first-order, constant-$J_{ij}$ approximation.  Then using \eqref{eq:F1k-analytic}, we arrive at the conclusion that $\zeta$ is, to a good approximation, \emph{linearly proportional} to the dynamical-coupling parameter $\eta$, at least on average over problem instances.  This in turn suggests that the error of the effective couplings reduces as $\eta$ is made smaller.  This is physically intuitive, as a small imposed ratio between the target coupling strength and native coupling strength implies more flexibility in engineering the system to approximate the desired target couplings.

Taken together, this discussion motivates the assertion that $\eta$ is, in addition to $\Lambda$, one of the key dynamical-coupling parameters that determines the performance of the scheme.  At the same time, given a fixed desired $\lambda_\text{C}/\chi$ in the target Hamiltonian, a small value of $\eta$ implies a large value of $\lambda_\text{J}/\chi = (\lambda_\text{C}/\chi)/\eta$, which in turn implies a small value of the nonlinear rate $\chi$; this latter condition imposes a more stringent requirement on the field decay rate (see Section~\ref{sec:nonlinear-rate} for more details).  Thus, a natural tradeoff is to choose the largest $\eta$ that achieves an acceptable error.

It is reasonable to expect that, for any given problem instance, there is a precise value of $\eta$ such that the Floquet design equation \eqref{eq:floquet-equation} fails to have a solution.  While this ``threshold'' $\eta$ varies across problem instances, we have found empirically (e.g., through numerical optimization) that, even on ensemble average over a problem class, there tends to be a sharp boundary between those values of $\eta$ with acceptable errors after optimization and those for which the optimizer cannot improve the error relative to the first-order analytic solution (i.e., there is a dramatic change in the error once $\eta$ exceeds a certain threshold value).  This boundary value of $\eta$ is a function of $N$, and we can understand its functional form by the following argument.  We first substitute the analytic formulas for the mean magnitude of the Floquet design parameters $f^{(k)}_j$ from \eqref{eq:F1k-analytic} into $F^{(k)}_j$ in \eqref{eq:zeta-bound}.  After some simplifying approximations (such as $N-1 \approx N$), this results in
\begin{equation}
    \zeta \lesssim \begin{cases}
        2\eta \cdot N & \text{for SK7} \\
        \eta \cdot N(1+\log N) & \text{for Dense MAX-CUT} \\
        6\eta \cdot (1+\log N) & \text{for Cubic MAX-CUT}
    \end{cases}.
\end{equation}
Heuristically, when $\zeta$ exceeds a certain modulation depth, the Floquet design equation tends to have no solution.  If we denote this nominal value by $\zeta_\temax$, then we can solve for $\eta$ in the above to conclude that if we want a modulation depth smaller than $\zeta_\temax$, we should take $\eta \leq \eta_\text{bound}$, given by
\begin{equation} \label{eq:eta-bound}
    \eta_{\text{bound}} = \begin{cases}
        \zeta_{\text{max}}/2 \cdot N\inv & \text{for SK7} \\
        \zeta_{\text{max}} \cdot \Paren{N(1+\log N)}\inv & \text{for Dense MAX-CUT} \\
        \zeta_{\text{max}}/6 \cdot (1+\log N)\inv & \text{for Cubic MAX-CUT}
    \end{cases}.
\end{equation}

\begin{figure}
    \centering
    \includegraphics[width=0.9\textwidth]{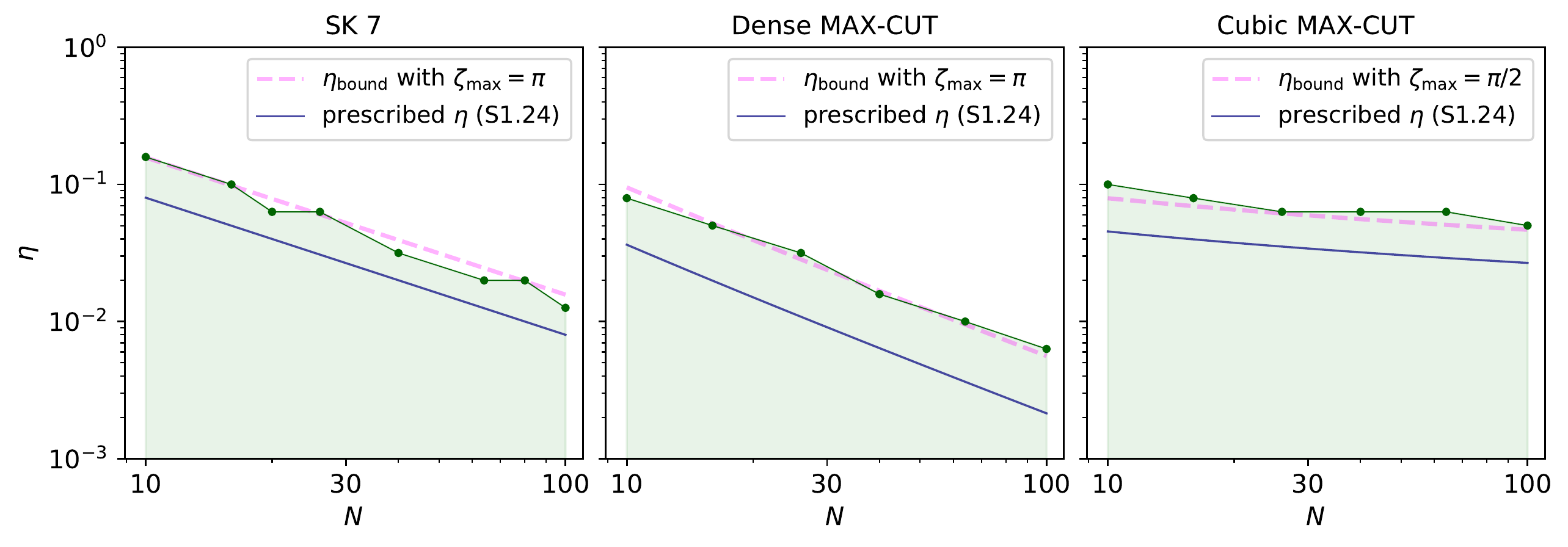}
    \caption{\textbf{Upper-boundary values of dynamical-coupling parameter.} The region depicted in green indicates the range of the dynamical-coupling parameter $\eta$ in which the numerical solution to the Floquet design equation---based on a full-order, ``column-by-column'' approach as described in Section~\ref{sec:optimizeF}---achieves low error $\mathcal E_\text{C}$, as a function of problem size $N$. Specifically, we choose $\mathcal E_\text{C} \leq \SI{3}{\percent}$, although the qualitative features are insensitive to this choice due to the sharpness of the error transition. The error is computed as a mean over ten problem instances, and a mesh grid of 8 points (in $N$) and 31 points (in $\eta$) is used to generate the boundary curve.  The pink lines show the heuristically derived functional forms for $\eta_{\text{bound}}$ from \eqref{eq:eta-bound}.  The blue lines indicate the choice of $\eta$ as a function of $N$ used in this work, as given by \eqref{eq:eta}.  For the choice of the native couplings $J_{ij}$, refer to Section~\ref{sec:parameter-summary}; for details on the numerical optimization routine, see Section~\ref{sec:optimizeF}.}
    \label{fig:eta-numerical}
\end{figure}

Figure~\ref{fig:eta-numerical} shows (in green) the range of values of $\eta$ for which our numerical optimizer finds acceptable errors.  Alongside this data (in pink) are the functional forms for $\eta_\text{bound}$ derived above.  We see that the functional forms are reproduced well across the different problem classes, while the offset can also be made to agree by setting $\zeta_{\text{max}}$ to reasonable values (of order unity).

The upshot of this discussion is that the highest possible value of $\eta$ as a function of $N$ (and hence the best-case scaling of the dynamical-coupling scheme as a whole) is constrained by $\eta_\text{bound}$.  In practice, however, finding the true solution of the Floquet design equation---or even being able to use the full-order numerical solver---is only possible for small problem sizes.  As mentioned in Section~\ref{sec:optimizeF}, when the problem size is large, we generally turn to a second-order Taylor approximation approach for the optimizer.  In this case, the error of the resulting solution is no longer sharp, as the quality of the second-order approximation \eqref{eq:floquet-equation-taylor2} (which is used to construct the objective function) also depends on $\eta$ (through $\zeta$).  In this case, we need to qualify our discussion by taking into account the choice of acceptable error.  For simplicity, we consider $\mathcal E_\text{C} \leq \SI{3}{\percent}$ as an acceptable error in this work.

\begin{figure}
    \includegraphics[width=0.9\linewidth]{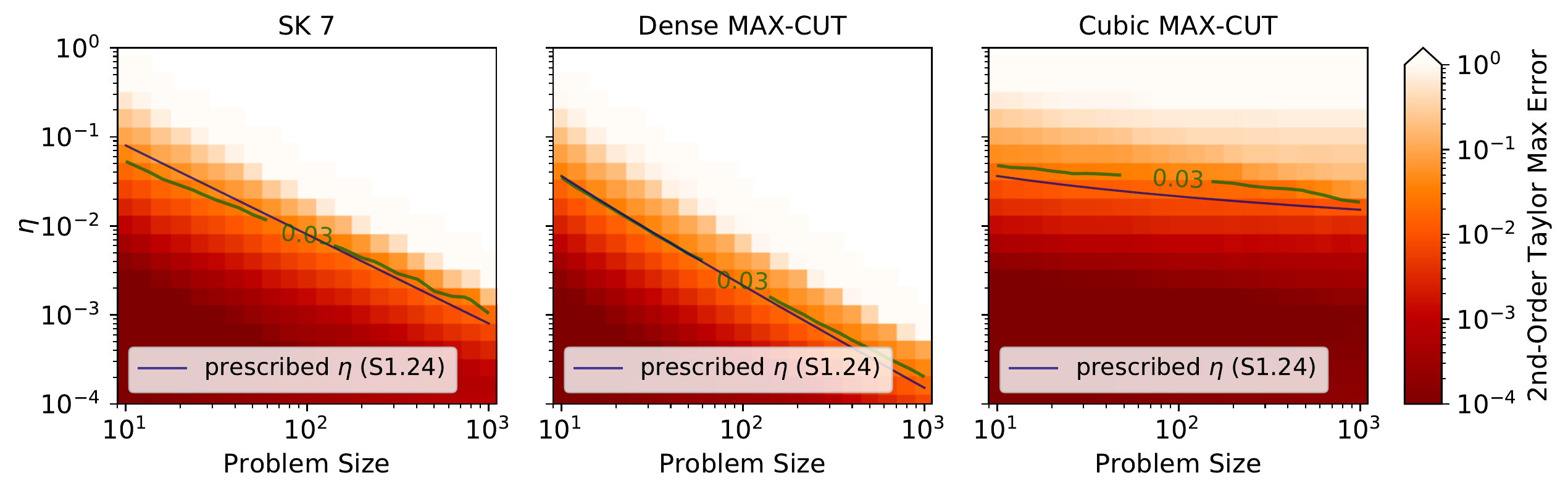}
    \caption{\textbf{Error scaling of the second-order numerical optimizer.} Heat map of the error $\mathcal E_\text{C}$ resulting from the numerical solution to the Floquet design equation using a second-order, ``column-by-column'' approach as described in Section~\ref{sec:optimizeF}, plotted as a function of the dynamical-coupling parameter $\eta$ and problem size $N$.  The level curve for an acceptable error of $\mathcal E_\text{C} = \SI{3}{\percent}$ is shown in green.  The error is computed as a mean over 100 problem instances for each point.  The blue lines indicate the choice of $\eta$ as a function of $N$ used in this work, as given by \eqref{eq:eta}.  For the choice of the native couplings $J_{ij}$, refer to Section~\ref{sec:parameter-summary}; for details on the numerical optimization routine, see Section~\ref{sec:optimizeF}.  Note that to save on computation, we omit the region where $\eta$ has safely exceeded $\eta_\text{bound}$ (namely, the upper-right triangular region in the left and center plots, around where the colormap saturates); by extrapolation, we take the error in this region to exceed 1.}
    \label{fig:eta-taylor2}
\end{figure}

In Figure~\ref{fig:eta-taylor2}, we show the errors which are empirically attained by the second-order numerical optimizer.  The green curve shows the values of $\eta$ that achieve the acceptable error of \SI{3}{\percent}, below which the error is generally smaller.  As expected, the \SI{3}{\percent} error boundary is qualitatively similar in shape to $\eta_{\text{bound}}$ (cf., Figure~\ref{fig:eta-numerical}), except for an overall offset due to the Taylor expansion.  Thus, even though the second-order approach produces larger error overall, the acceptable-error cutoff for $\eta$ scales in much the same way as the fundamental upper-bound $\eta_\text{bound}$ discussed above.

Motivated by these observations, we choose $\eta$ in this work by taking our theoretically derived fundamental bounds of $\eta$ (i.e., $\eta_\text{bound})$ and multiplying them by an empirically determined multiplicative factor:
\begin{equation} \label{eq:eta}
    \eta = \begin{cases}
        0.8 \cdot N\inv & \text{for SK7} \\
        1.2 \cdot N\inv(1+\log N)\inv & \text{for Dense MAX-CUT} \\
        0.12 \cdot (1+\log N)\inv & \text{for Cubic MAX-CUT}
    \end{cases}.
\end{equation}
This choice of $\eta$ is plotted in blue in both Figures \ref{fig:eta-numerical} and \ref{fig:eta-taylor2}.  Because of the scaling depicted in Figure~\ref{fig:eta-taylor2}, we have reasonable confidence that even if a second-order approach is used for solving the Floquet design problem, we are within a constant factor of the optimal scaling possible, as implied by Figure~\ref{fig:eta-numerical}.  As such, we use this prescription for $\eta$ in all our simulations in this work as well as for the scaling analysis performed in Section~\ref{sec:nonlinear-rate}.

As a technical point, we note that in Figure~\ref{fig:eta-taylor2}, there is a crossover between the blue and green lines, indicating that at smaller $N$, \eqref{eq:eta} is not quite enough to ensure $\mathcal E_\text{C} \leq \SI{3}{\percent}$.  However, we can circumvent this by simply using the full-order solver for $N \leq 100$ (which is efficient at small $N$), and only switch to the second-order approach for $N > 100$.  Of course, this technicality does not affect any of our scaling arguments.

\subsection{Requirements on Floquet frequency} \label{sec:Lambda}
Finally, we discuss the Floquet frequency $\Lambda$, which sets the periodicity of used in the Floquet approximation to be $2\pi/\Lambda$.  As mentioned above, the natively-coupling Hamiltonian \eqref{eq:H-native} is approximated by the time-average effective Hamiltonian \eqref{eq:H-eff} only if $\Lambda$ is much greater than all other system rates.  If this assumption is not met, then it is possible that the physics of the dynamically-coupled system fails to approximate that of the target statically-coupled system, despite the Floquet equation \eqref{eq:floquet-equation} being met (i.e., despite ${C_\text{eff}}_{ij} \approx C_{ij}$).  More precisely, we require
\begin{equation} \label{eq:floquet-requirement}
    \Lambda = A \times \max\paren{\text{all system rates}},
\end{equation}
where $A \gg 1$ is a large multiplicative factor.  As will be discussed in later sections---see especially Sections~\ref{sec:QA-KPO}, ~\ref{sec:native-coupling},and ~\ref{sec:nonlinear-rate} (and as presented in the main manuscript), the most relevant rates for our system are $\chi$, $\delta_i$, $r_\text{max}$, and $\lambda_\text{J}$.

\begin{figure}
    \centering
    \includegraphics[width=0.6\textwidth]{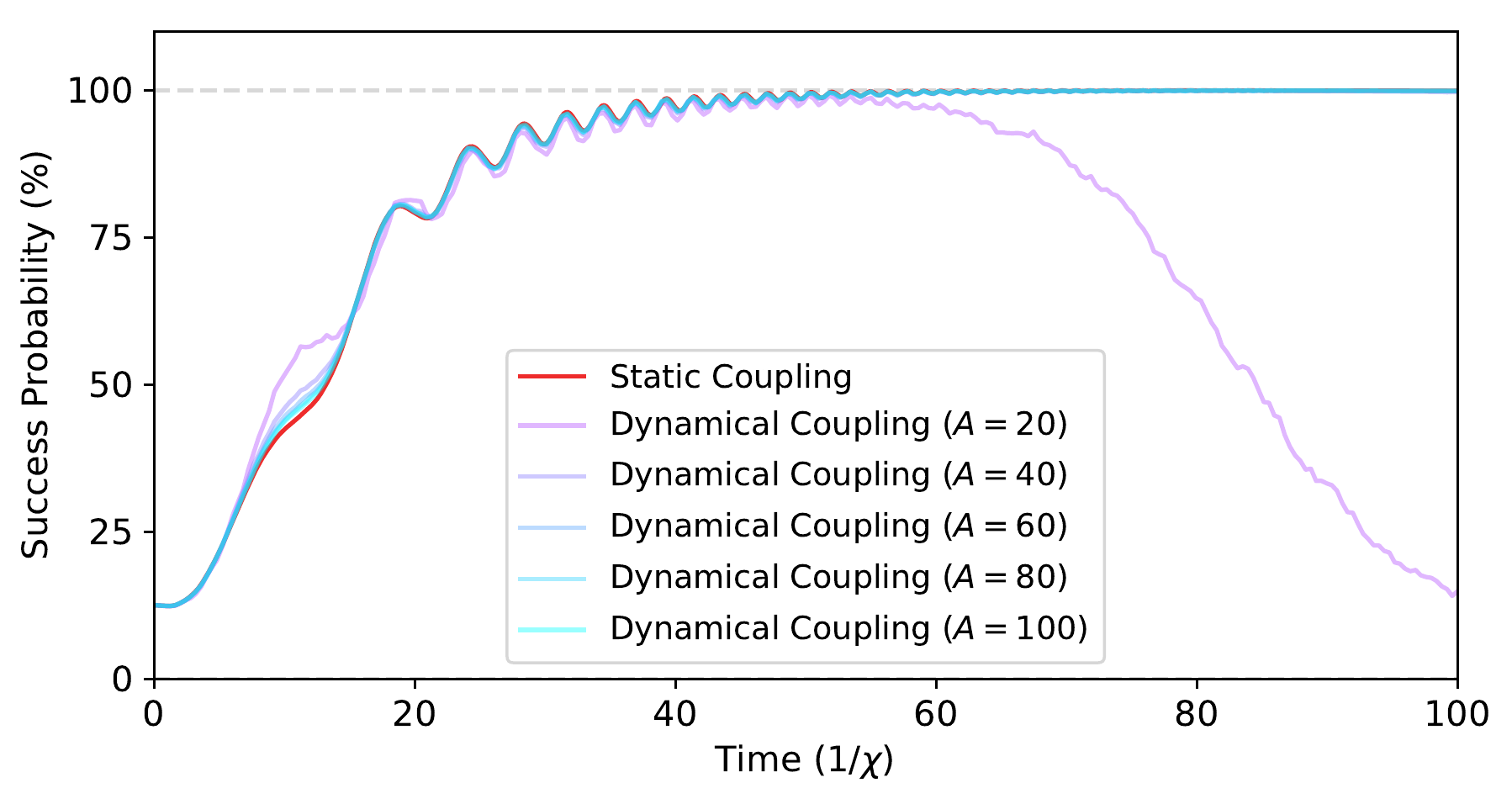}
    \caption{\textbf{Timescale requirements for Floquet approximation.} Comparison between the behavior of the statically-coupled target system (red, in back), vs.\ that of the corresponding dynamically-coupled system using a range value for the Floquet frequency $\Lambda$ (parameterized by the factor $A$, as defined in \eqref{eq:floquet-requirement}).  The metric plotted here is the success probability (see Section~\ref{sec:quantum}) on a particular problem instance (specifically, the instance used for Figure~2\,B in the main manuscript).  Here, the physical model for the oscillators is a bus-coupled system of JPOs (see Section~\ref{sec:bus-KPO}), under appropriate approximations (see Section~\ref{sec:native-coupling}) and with parameters chosen according to Section~\ref{sec:parameter-summary}.  For quantum simulation methods, see Section~\ref{sec:quantum}.}
    \label{fig:Lambda}
\end{figure}

In order to determine the value of $A$ that suffices, we simulate both the target statically-coupled system and the corresponding dynamically-coupled system using different values of $A$ and examine the respective evolution of a key performance metric, the success probability (see Section~\ref{sec:quantum}).  In Figure~\ref{fig:Lambda}, we show one example of how the dynamically-coupled system progressively matches the evolution of the target statically-coupled system (at least, on timescales greater than $2\pi/\Lambda$) as $A$ is increased.  For this particular problem instance and system parameters, we see that the value of $A$ should be larger than \num{40} for good correspondence.

From such empirical observations, we conservatively use in this work a value of $A=\num{100}$ for quantum simulations.  For simulations of the classical EOMs, similar empirical investigations show that a larger value of  $A=\num{200}$ is needed for comparable convergence.

\newpage
\section{Quantum annealing with Kerr parametric oscillators} \label{sec:static-quantum-annealing}
\label{sec:QA-KPO}
In this section, we review some operational principles for solving Ising problems with networks of Kerr parametric oscillators (KPOs) via quantum annealing.  A number of approaches towards this end have been proposed \cite{Goto2016b,Nigg2017b,Puri2017a}, but for simplicity, we focus on the directly- and statically-coupled model used in Ref.~\cite{Goto2016b}, as this is the model that our dynamical-coupling scheme aims to approximate.

A generic KPO network is represented by a Hamiltonian of the general form
\begin{equation} \label{eq:kpo-network}
    -\sum_i \delta_i \hat a\dagg_i \hat a_i - \sum_i \frac{\chi}{2} \hat a_i^{\dagger2} \hat a_i^2 + \sum_i\paren{\frac{r}{2} \hat a_i^2 + \text{H.c.}} - \lambda_\text{C} \sum_{i\neq j} C_{ij} \hat a_i\dagg \hat a_j,
\end{equation}
where $\chi > 0$ is the Kerr nonlinear rate, $r$ is the amplitude of the parametric pump drive, $\delta_i$ is the detuning of the $i$th oscillator from the half-harmonic of the pump, $\lambda_\text{C} C_{ij}$ is the strength of the coupling between oscillator $i$ and $j$.  Generally, the symmetric coupling matrix $C_{ij}$ is chosen to be that of a (bias-free) Ising problem whose ground state is desired.  Finally, although implicit in \eqref{eq:kpo-network}, all these parameters can potentially be time-dependent, varying as needed by the annealing procedure.

In the absence of coupling and detuning, the eigenstates of each KPO are the 0-phase and $\pi$-phase coherent states $\ket{+r/\chi}$ and $\ket{-r/\chi}$, which can be utilized in a quantum annealer to encode, respectively, the spin-up and spin-down configurations of an Ising spin system.  In a quantum annealer based on this encoding, the state of the KPO network after the annealing procedure is ideally the multimode Schr\"{o}dinger cat state
\begin{align}
    \label{eq:multimode-cat}
    \bigotimes_{i=1}^N \ket{\sigma_i r/\chi}  + \bigotimes_{i=1}^N \ket{-\sigma_i r/\chi},
\end{align}
where $\sigma$ is the ground-state configuration of the Ising problem of interest; and it comprises $\sigma = (\sigma_1, \ldots, \sigma_N)$ with $\sigma_i\in\set{-1, 1}$.  Given this state, measurement of the sign of the in-phase field amplitudes of the KPOs produces solutions to the Ising problem.

We can consider two approaches in which such a KPO network can be used for performing quantum annealing and obtaining the final state \eqref{eq:multimode-cat}.  One way is to take a purely qubit-based approach \cite{Farhi2000a}, as the KPO has been shown to exhibit a qubit subspace (spanned by $\ket{\pm r/\chi}$) in which \eqref{eq:kpo-network} can be used to realize universal control \cite{Goto2016a,Puri2017b}.  In this approach, the individual KPOs are each first prepared in the Pauli-$X$ $+1$-eigenstate, which corresponds to the cat state $\ket{r/\chi} + \ket{-r/\chi}$; as described in \cite{Puri2017b}, this can be done by engineering the evolution of the pump parameter $r$ appropriately.  After the preparation step, the annealing procedure then requires realizing and controlling two Hamiltonian terms in the qubit basis.  The first is the so-called driver Hamiltonian $\sum_i \hat X_i $, which can be realized in the qubit subspace by proper choice of the detunings $\delta_i$ \cite{Goto2016a, Puri2017b}. The second is the problem Hamiltonian $\sum_{i\neq j} C_{ij} \hat Z_i \hat Z_j$, which is realized in the qubit subspace by the coupling term in \eqref{eq:kpo-network}.  The key point here is that the state \eqref{eq:multimode-cat}---when interpreted in the qubit subspace---is the ground state of this latter Hamiltonian, so that the end result is the state \eqref{eq:multimode-cat}, provided the assumptions of the quantum adiabatic theorem are met (i.e., in the context of standard adiabatic quantum computation).  In this approach, the pump is held fixed to stabilize the two coherent states $\ket{\pm r/\chi}$ and ensure that the system remains in the qubit subspace, while the coupling term is gradually turned on by adiabatically time-varying $\lambda_\text{C}$.

For our work, however, we focus on a more recent continuous-variable approach, as exemplified by Ref.~\cite{Goto2016b} but also utilized in Refs.~\cite{Nigg2017b,Puri2017a}.  This approach has the advantage of possessing a straightforward classical limit, allowing us to numerically study scaling to larger problem sizes.  In contrast to the purely qubit-based approach, the initial state of the quantum annealer in this continuous-variable approach is the vacuum state.  Then, the quantum annealing procedure is realized by time-varying the parametric pump drive according to some function $r(t)$, where $r(0) = 0$.  A useful choice for $r(t)$ takes the form of a linear ramp with slope $r_\text{max}/T_\text{ramp}$ clamped at $r_\text{max}$:
\begin{equation}
    r(t) = r_\text{max} \min\paren{t / T_\text{ramp}, \, 1}.
\end{equation}
The detunings $\delta_i$ are static but independently tuned for each oscillator such that the highest-eigenvalue eigenstate of \eqref{eq:kpo-network} at $r = 0$ is the vacuum, and one satisfactory choice is \cite{Goto2016b}
\begin{equation}\label{eq:goto-detunings}
    \delta_i = \lambda_\text{C} \sum_j \abs{C_{ij}}.
\end{equation}
This choice ensures that the initial vacuum state of the oscillator network is adiabatically connected to the multimode cat state \eqref{eq:multimode-cat} at the end of the computation, thus implementing quantum annealing \cite{Goto2016b}.  In this approach, the coupling term $\lambda_\text{C} \sum_{i,j} C_{ij} {\hat a_i}\dagg \hat a_j$ is also held static throughout the procedure, with some suitable choice of the problem-dependent strength parameter $\lambda_\text{C}$; we discuss this tuning parameter further in Section~\ref{sec:static-parameters} below.

Of course, these networks of KPOs are not only described by the Hamiltonian \eqref{eq:kpo-network}, as they are also coupled to the environment. Following the approach in \cite{Puri2017a, Nigg2017b}, we model this dissipation due to the environment with Lindblad operators \cite{Wiseman2010}. In particular, we treat the photon loss for the $i$th oscillator with the Lindblad operator
\begin{align}
    \label{eq:lindblad}
    \hat L_i = \sqrt{2 \kappa} \hat a_i
\end{align}
where $\kappa$ is the field decay rate associated with this dissipation process.

\subsection{Parameter considerations for annealing} \label{sec:static-parameters}
The performance of a continuous-variable quantum annealer can be quite sensitive to the choice of parameters and annealing schedules used in the scheme.  Focusing on the continuous-variable annealing scheme with a clamped-linear pump schedule as in Ref.~\cite{Goto2016b}, we discuss in this subsection the prescriptions we follow for selecting the ramp time $T_\text{ramp}$, the maximum pump rate $r_\text{max}$, and the problem-strength parameter $\lambda_\text{C}$.  These prescriptions are then directly mapped onto the corresponding parameters in our dynamical-coupling scheme.

Among the three parameters, $\lambda_\text{C}$ is the most subtle, as its optimal choice depends on the problem size $N$ as well as the problem structure.  To first order, Ref.~\cite{Goto2016b} shows that the KPO system is an effective quantum annealer under the condition that
\begin{equation} \label{eq:first-order-condition}
    \abs{-\sum_i \frac{\chi}{2} \hat a_i^{\dagger2} \hat a_i^2 + \sum_i\paren{\frac{r(t)}{2} \hat a_i^2 + \text{H.c.}}} \gg \Biggl|\sum_i \delta_i \hat a\dagg_i \hat a_i + \lambda_\text{C} \sum_{i\neq j} C_{ij} \hat a_i\dagg \hat a_j\Biggr|.
\end{equation}
Thus, in order for the annealer to be effective, we should scale the value of $\lambda_\text{C}$ to cancel out the scaling in the sum over $C_{ij}$, to ensure the right-hand side remains bounded as required.

\begin{figure}[t!]
    \centering
    \includegraphics[width=0.84\textwidth]{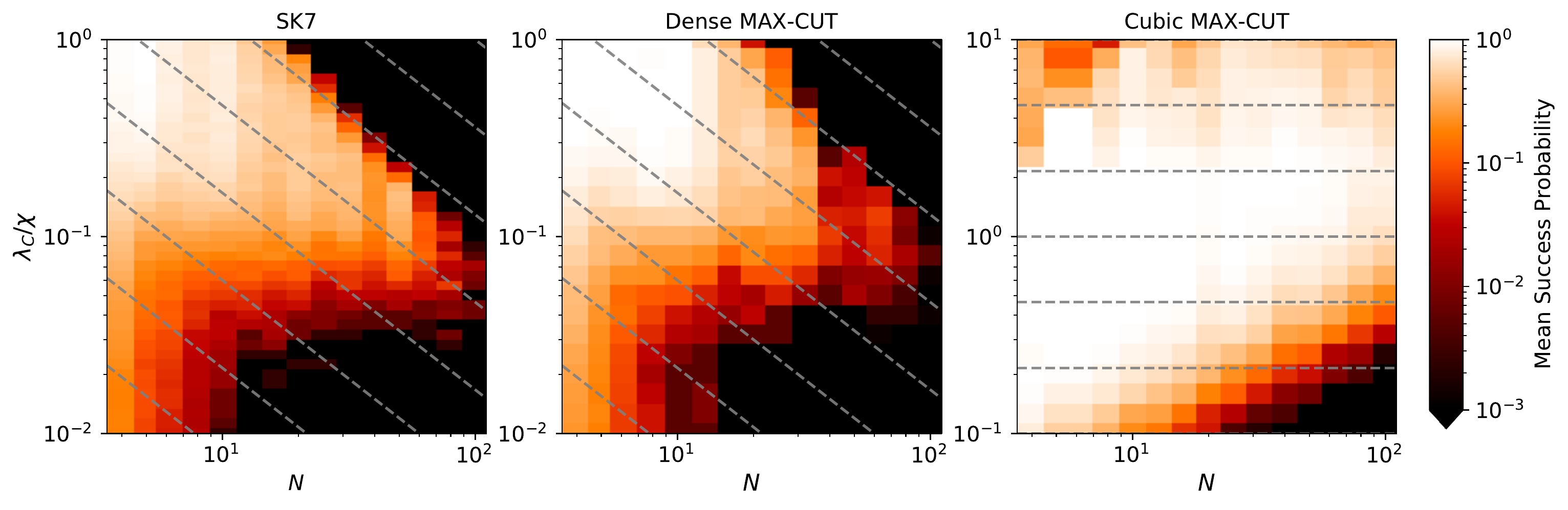}
    \caption{\textbf{Dependence of KPO network annealer on problem-strength parameter.} Success probability of the classical EOMs for statically-coupled KPO networks \label{eq:eom-static} as a function of the problem-strength parameter $\lambda_\text{C}/\chi$ and problem size $N$, for the three problem classes. The success probability is computed as a mean over at least ten problem instances for each class.  The gray dashed lines show curves that are linearly proportional to the functional forms given in \eqref{eq:lambdaC}. For the approach to simulating classical EOMs, see Section~\ref{sec:classical}.}
    \label{fig:lambdaC}
\end{figure}

We present in Figure~\ref{fig:lambdaC} the success probability, computed via simulations of the classical EOMs that are described in Section~\ref{sec:classical}, of various problem classes as we vary $\lambda_\text{C}$ and $N$. For simplicity the loss is set to zero $\kappa=0$ for these simulations. The result shows that dense problem classes (SK7, Dense MAX-CUT) require $\lambda_\text{C}/\chi \propto 1/N$, while sparse problem classes (Cubic MAX-CUT) require a constant $\lambda_\text{C}$.  This numerical investigation is consistent with \eqref{eq:first-order-condition} in that the optimal choice for the problem-strength parameter is $\lambda_\text{C}/\chi \sim r_\text{max}\Paren{\sum_{ij}|C_{ij}|}\inv$. Based on these results, unless stated otherwise, we recommend the following forms for $\lambda_\text{C}$:
\begin{equation}
    \label{eq:lambdaC}
    \frac{\lambda_\text{C}}{\chi} = \begin{cases}
        4/N & \text{for SK7} \\
        8/N & \text{for Dense MAX-CUT} \\
        2 & \text{for Cubic MAX-CUT}
    \end{cases}.
\end{equation}

\begin{figure}[b!]
    \centering
    \includegraphics[width=0.82\textwidth]{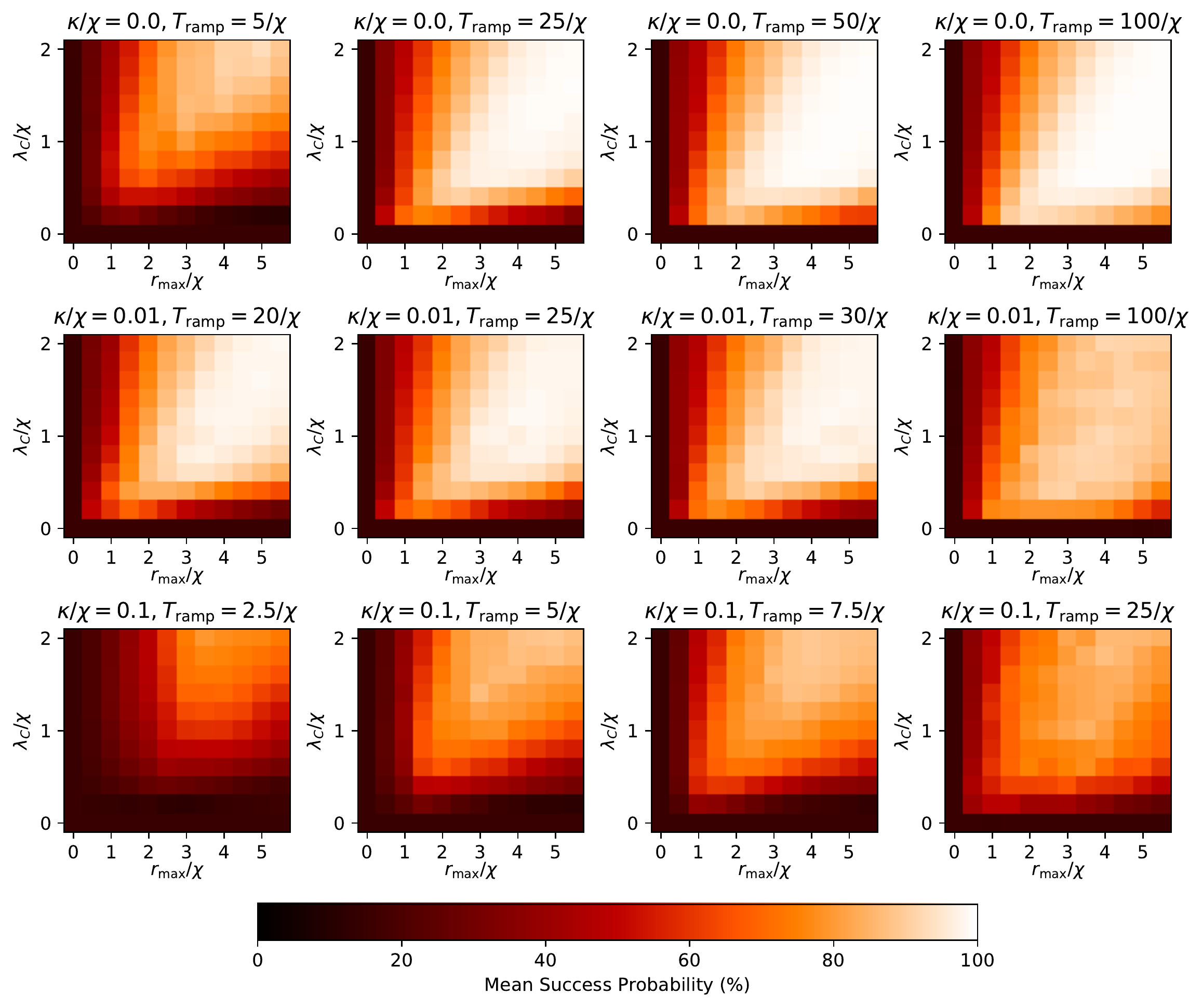}
    \caption{\textbf{Dependence of KPO network annealer on pump rate and ramp time.} Success probability of the statically-coupled KPO quantum annealer at solving SK7 problems with problem size $N = 4$, as functions of field decay rate $\kappa/\chi$ (rows) and the linear ramp time $T_\text{ramp} \chi$ (columns), as well as the maximum pump rate $r_\text{max}/\chi$ and problem-strength parameter $\lambda_\text{C}/\chi$.  The success probability is computed as a mean over ten problem instances.  For quantum simulation approach, see Section~\ref{sec:quantum}.}
    \label{fig:static-parameters}
\end{figure}

In this work, we also analyze the performance of the statically-coupled system with dissipation to ensure that the dynamical-coupling scheme is (at least comparably) robust to effects of linear loss. The presence of dissipation affects the choice for the parameters $T_\text{ramp}$, $r_\text{max}$ and $\lambda_\text{C}$ that  optimize success probability. To investigate these effects, we perform quantum simulations of the statically-coupled system and present them in Figure~\ref{fig:static-parameters}, which shows how the success probability of solving $N = 4$ SK7 problems varies for different settings of the parameters $\lambda_\text{C}$, $r_\text{max}$ and $T_\text{ramp}$, across the three different loss rates $\kappa/\chi \in \set{\num{0},\,\num{e-2},\,\num{e-1}}$, which are presented along each row respectively.

First, we see that the optimal choice of $T_\text{ramp}$ varies strongly with $\kappa$.  In the absence of loss, success probability monotonically improves as $T_\text{ramp}$ is increased, since the annealing procedure becomes more adiabatic.  However, with finite loss, larger $T_\text{ramp}$ causes a decrease in the success probability, as the number of photon emission events accumulate, causing decoherence. Based on the results of Figure~\ref{fig:static-parameters}, the ramp time should be set to $T_\text{ramp} = \min\paren{1/\kappa, 100/\chi}$ to maximize success probability, corresponding to order unity number of photon emission events per oscillator in the duration of the computation~\cite{Nigg2017b} (and where $100/\chi$ gives acceptable adiabaticity in the lossless case).

Next, $r_\text{max}$ determines the amplitude of the coherent states that constitute the multimode cat at the end of the computation.  Thus, it should be sufficiently large to ensure that the two constituent coherent states are distinguishable (i.e., nearly orthogonal). On the other hand, when there is loss, $r_\temax$ should also not be too large, as the number of photon emission events increases with the cat-state amplitude, reducing success probability. The numerical simulations suggest that $r_\temax = 5$ achieves reasonable success probabilities even in the presence of dissipation.
Finally, we note that there is not very strong dependence of $\lambda_\text{C}$ with dissipation, as we find that the value given by \eqref{eq:lambdaC} results in reasonably good success probabilities even in the presence of photon loss, at least for $N=4$.

Of course, in addition to these qualitative observations about the operating point of the statically-coupled system, one can perform additional fine-tuning.  For the quantum simulations in this work, we have chosen specific values of the parameters using Figure~\ref{fig:static-parameters} to maximize success probabilities; see Table~\ref{table:static-parameters}.


\newpage
\section{Bus-mediated JPO networks} \label{sec:bus-KPO}
\subsection{Josephson parametric oscillators}
A Josephson parametric oscillator (JPO) is a  Kerr resonator with an additional two-photon drive. JPOs can be implemented in different ways, such as with a SNAIL parametric amplifier \cite{Frattini2018} or by parametric modulation of a transmon oscillator's resonance frequency \cite{Nigg2017b, Puri2017b}. For this work, we focus on the latter implementation, where the magnetic flux $\Phi$ (henceforth normalized to the flux quantum $\Phi_0 \coloneqq h/2e$) applied to the DC SQUID is modulated instead of being held static. The system is described by the standard transmon Hamiltonian \cite{Koch2007}
\begin{equation} \label{eq:transmon}
   4\EC \hat q^2 - 2\EJ \cos(\Phi/2)\cos \hat\theta,
\end{equation}
where $\hat q, \hat \theta$ are canonically conjugate variables representing the charge on the capacitor and the phase difference of the two junctions, respectively, and $\EC$ and $\EJ$ are the charging and Josephson energies, respectively.

We posit that the flux $\Phi(t)$ has a time-dependence which can be separated into DC and AC components, as
\begin{equation} \label{eq:flux}
    \Phi(t)/2 \coloneqq \Phi_\dc + \Phi_\ac(t).
\end{equation}
Operationally, $\Phi_\ac$ is used for both parametric driving (at high frequencies) and for detuning modulation (at lower frequencies) to achieve dynamical coupling while $\Phi_\dc$ is used to set the operating frequency of each JPO in the absence of the modulation. Assuming the modulation amplitude is small (i.e., $\abs{\Phi_\ac(t)} \ll \Phi_\dc$), we can expand \cite{Nigg2017b}
\begin{align} \label{eq:cos-expand}
    \cos(\Phi(t)/2) &= \cos\sbrak{\Phi_\dc}\cos\sbrak{\Phi_\ac(t)} - \sin\sbrak{\Phi_\dc}\sin\sbrak{\Phi_\ac(t)} \nonumber \\
    & \approx \cos\sbrak{\Phi_\dc} - \sin\sbrak{\Phi_\dc}\Phi_\ac(t).
\end{align}

In the transmon limit of $\EJ \gg \EC$ \cite{Girvin2014a}, we can also expand \eqref{eq:transmon} in $\hat\theta$.  Dropping the zeroth-order (constant) term and inserting \eqref{eq:flux} after applying the approximation \eqref{eq:cos-expand}, we have, to fourth order in $\hat \theta$, the Hamiltonian
\begin{equation} \label{eq:jpo-expansion}
    4\EC \hat q^2 + \EJ \cos(\Phi_\dc) \, \hat \theta^2 - \EJ \sin(\Phi_\dc)\Phi_\ac(t) \, \hat\theta^2 - \frac{\EJ}{12} \cos\Phi_\dc \, \hat \theta^4.
\end{equation}

We next introduce bosonic operators $\hat a$ and $\hat a\dagg$ (satisfying $\sbrak{\hat a, \hat a\dagg} = 1$) via the canonical transformations \cite{Nigg2017b}
\begin{align} \label{eq:canonical-vars}
    \hat \theta &= \sbrak{\frac{\EC}{\EJ \cos(\Phi_\dc)}}^{1/4} \Paren{\hat a + \hat a\dagg}, &
   \hat q &= -\im \sbrak{\frac{\EJ \cos(\Phi_\dc)}{16\EC}}^{1/4} \Paren{\hat a - \hat a\dagg}.
\end{align}
Inserting \eqref{eq:canonical-vars} into \eqref{eq:jpo-expansion}, we arrive at the Hamiltonian
\begin{equation}
    4 \sqrt{\EC \EJ \cos(\Phi_\dc)} \, \hat a\dagg \hat a - \frac{\EC}{12} \Paren{\hat a + \hat a\dagg}^4 - \sqrt{\frac{\EC \EJ}{\cos(\Phi_\dc)}}\sin(\Phi_\dc) \Phi_\ac(t) \Paren{\hat a+\hat a\dagg}^2.
\end{equation}
Finally, after normal ordering and applying the rotating wave approximation \cite{Gardiner2004} to cut off frequencies much higher than twice the fundamental ($\omega_\dc$ in \eqref{eq:omega-spec}), we have the dynamically-controlled JPO Hamiltonian
\begin{equation} \label{eq:jpo-hamiltonian}
    \hat H_\jpo(t) \coloneqq \omega(t) \hat a\dagg\hat a - \frac{\chi}{2}\hat a^{\dagger2}\hat a^2 + \paren{\frac{\omega_\ac(t)}{2} \, \hat a^2 + \text{H.c.}}
\end{equation}
where we have defined the nonlinear Kerr coefficient $\chi \coloneqq \EC$ and the instantaneous frequency $\omega(t) \coloneqq \omega_\dc + \omega_\ac(t)$, which has DC and AC components given by
\begin{align}
    \label{eq:omega-spec}
    \omega_\dc &\coloneqq 4\sqrt{\EC \EJ \cos(\Phi_\dc)}-\EC, &
    \omega_\ac(t) &\coloneqq -2\sqrt{\frac{\EC \EJ}{\cos(\Phi_\dc)}}\sin(\Phi_\dc)\Phi_\ac(t).
\end{align}
These relations indicate that, for appropriate choices of $E_\text{C}$ and $E_\text{J}$, both the DC and AC components of the instantaneous frequency can be fully controlled in a JPO given appropriate control over $\Phi_\dc$ and $\Phi_\ac(t)$.

\subsection{Bus-mediated native couplings}
To couple together JPOs that are individually described by \eqref{eq:jpo-hamiltonian}, we utilize an architecture involving a common bus resonator.  We consider an excitation-exchange interaction between the JPOs and the bus, and the linear part of system Hamiltonian (i.e., omitting the individual JPO's parametric and Kerr terms) is \cite{Song2017}
\begin{equation}
    \hat H_\text{coupling} \coloneqq \sum_i \omega_i \hat a_i\dagg \hat a_i + \omega_\bus \hat b\dagg \hat b + \sum_i g_i \paren{\hat b\dagg \hat a_i + \hat b \hat a_i\dagg},
\end{equation}
where $\omega_\bus$ is the frequency of the bus resonator and $g_i$ is the coupling rate of the $i$th JPO to the bus.  As in \eqref{eq:jpo-hamiltonian}, the instantaneous frequency $\omega_i$ is in general a time-dependent parameter.

We start by introducing the bus-oscillator detunings
\begin{equation}
    \Delta_i \coloneqq \omega_\bus - \omega_i.
\end{equation}
Next, we make the assumption that, for $\varepsilon \ll 1$
\begin{equation}
    g_i / \Delta_i \sim \mathcal O(\varepsilon),
    \quad\text{and}\quad
    \chi_i / \Delta_i \sim {\omega_\ac}_i / \Delta_i \sim \mathcal O(\varepsilon^2),
\end{equation}
for all $i$ and at all times.  Provided these conditions are met, we can perform a Schrieffer-Wolff expansion \cite{Girvin2014a} in the rotating frame of the bus, to second order in $\varepsilon$.  The result of that expansion produces the coupling Hamiltonian
\begin{equation} \label{eq:bus-coupling-hamiltonian}
    \hat H_\text{coupling} \approx \sum_i \paren{\omega_i - \frac{g_i^2}{\Delta_i}} \hat a_i\dagg \hat a_i - \sum_{i\neq j} \frac{g_i g_j}{2} \paren{\frac 1{\Delta_i} + \frac 1{\Delta_j}} \hat a_i\dagg \hat a_j,
\end{equation}
where we have dropped the bus mode operators $\hat b$ in \eqref{eq:bus-coupling-hamiltonian} since the bus will no longer be populated in this limit.  Note that due to the Schrieffer-Wolff expansion, this Hamiltonian is only approximately in lab frame, as there is an $\mathcal O(\varepsilon)$ ``polariton angle'' between the JPO and bus modes: the $\hat a_i$ operators used in \eqref{eq:bus-coupling-hamiltonian} are hybridized with $\hat b$ to $\mathcal O(\varepsilon)$.  We make the standard assumption in circuit QED that this has negligible effect \cite{Girvin2014a}.

\subsection{JPO control problem}
\label{sec:JPO-control-problem}
Taking the results of the two above subsections together, the (approximately) lab-frame Hamiltonian of a system of JPOs with bus-mediated couplings is
\begin{equation}
    \sum_i \paren{\omega_i(t) - \frac{g_i^2}{\Delta_i(t)}} \hat a_i\dagg \hat a_i - \frac{\chi_i}{2} \hat a^{\dagger2}_i \hat a_i^2 + \paren{\frac{{\omega_\ac}_i(t)}{2} \hat a^2 + \text{H.c.}} - \sum_{j\neq i} \frac{g_ig_j}{2} \paren{\frac{1}{\Delta_i(t)} + \frac{1}{\Delta_j(t)}} \hat a_i\dagg \hat a_j.
\end{equation}

Our goal is to choose the (time-dependent) parameters such that this Hamiltonian approximates the generic KPO quantum annealing Hamiltonian \eqref{eq:kpo-network}. We first go into a rotating frame in which the $i$th JPO oscillates at a constant frequency of $-\delta_i$.  This produces the Hamiltonian
\[
-\sum_i \paren{\delta_i \hat a_i\dagg \hat a_i + \frac{\chi_i}{2} \hat a_i^{\dagger2}\hat a_i^2} + \hat H_\text{p} + \hat H_\text{c},
\]
where the pump and coupling terms are, respectively,
\begin{subequations} \label{eq:hamiltonian} \begin{align}
    \hat H_\text{p} &\coloneqq \sum_i \frac{{\omega_\ac}_i(t)}{2} \exp\sbrak{-2\im \int_0^t \paren{\omega_i(t') - \frac{g_i^2}{\Delta_i(t')} + \delta_i}\dif t'} \hat a_i^2 + \text{H.c.} \label{eq:hamiltonian-pumping} \\
    \hat H_\text{c} &\coloneqq -\sum_{i\neq j} \frac{g_i g_j}{2} \paren{\frac 1{\Delta_i(t)} + \frac 1{\Delta_j(t)}} \exp\sbrak{-\im\int_0^t \paren{ {\omega}_j(t') - \frac{g_j^2}{\Delta_j(t')} + \delta_j -{\omega}_i(t') + \frac{g_i^2}{\Delta_i(t')} - \delta_i}\dif t'} \hat a_i\dagg \hat a_j. \label{eq:hamiltonian-native}
\end{align} \end{subequations}
Comparing this Hamiltonian with \eqref{eq:kpo-network}, we see that the design problem is to obtain
\begin{subequations} \label{eq:problems} \begin{align}
    \hat H_\text{p} &\approx \sum_i \frac{r(t)}{2} \hat a_i^2 + \text{H.c.} \label{eq:jpo-pump-problem} \\
    \hat H_\text{c} &\approx -\lambda_\text{C} \sum_{i\neq j} C_{ij} \hat a_i\dagg \hat a_j. \label{eq:jpo-coupling-problem}
\end{align} \end{subequations}
We see that $\hat H_\text{c}$ is already written in the form of \eqref{eq:H-native}, so in principle, we can use our approach to dynamical coupling from Section~\ref{sec:dynamical-couplings} to parameterize $\omega_i(t)$ in terms of the Floquet design parameters $F^{(k)}_i$ and thus satisfy \eqref{eq:jpo-coupling-problem} by solving the Floquet design problem for $F^{(k)}_j$.

Unfortunately, for the specific case of JPOs, there is an additional complication because that same control signal $\omega_i$ is already used for performing the parametric pumping of the system: this is encapsulated by \eqref{eq:jpo-pump-problem} and \eqref{eq:hamiltonian-pumping}, which directly depends on the AC part of $\omega_i(t)$.  We emphasize that this additional complication is due to the hardware constraints of JPOs and is not inherent to the design of dynamical couplings.  (For example, in optical implementations of KPOs, a separate pump field is responsible for producing the parametric drive, opening a new degree of freedom for solving the design problem.)

To help get around this JPO-specific complication, we assume that the AC part of the instantaneous frequency has two ``bands'' of frequency content:
\begin{equation}
    \label{eq:omega-ac-split}
    \omega_\ac(t) = \omega_\slow(t) + \omega_\fast(t),
\end{equation}
where, roughly speaking, $\omega_\slow$ consists of frequency content near DC (the ``coupling band''), while $\omega_\fast$ consists of frequency content near $2\omega_\dc$ (the ``pumping band'').  We now proceed to solve the problem by assuming that the following two replacements can be made in \eqref{eq:hamiltonian}, at least in the context of trying to achieve the approximate equalities \eqref{eq:problems}:
\begin{subequations} \label{eq:decoupling} \begin{align}
    \frac{1}{\Delta_i(t)} &\longmapsto \frac{1}{\omega_\bus - \Paren{\omega_i^\dc + \omega_i^\slow(t)}} \\
    \exp\sbrak{\im\int_0^t \omega_i(t') \, \dif t'} &\longmapsto \exp\sbrak{\im \int_0^t \paren{\omega_i^\dc + \omega_i^\slow(t')} \dif t'}.
\end{align} \end{subequations}
This trick allows us to effectively decouple the pumping and coupling problems, at the cost of potential ``crosstalk'' between the two Hamiltonian terms once we insert our solution for $\omega_i(t)$.

We first use the approach of dynamical coupling to solve the design problem of \eqref{eq:jpo-coupling-problem}.  To match the abstract form of \eqref{eq:H-native}, we define the native coupling matrix and phase functions
\begin{subequations} \begin{align}
    J_{ij}(t) &\coloneqq \frac{g_i g_j}{2} \paren{\frac{1}{\omega_\bus - {\omega_\dc}_i - {\omega_\slow}_i(t)} + \frac{1}{\omega_\bus - {\omega_\dc}_i - {\omega_\slow}_i(t)}} \label{eq:J_def}\\
    \label{eq:phi_def}
    \phi_i(t) &\coloneqq \int_0^t \paren{{\omega_\dc}_i + {\omega_\slow}_i(t') - \frac{g_i^2}{\omega_\bus - {\omega_\dc}_i - {\omega_\slow}_i(t')} + \delta_i} \dif t' .
\end{align} \end{subequations}
These definitions identify for us, in the two-band approximation, the manifestation of the native-coupling matrix $J_{ij}(t)$ and oscillator phases $\phi_i(t)$ in the context of the bus-coupled JPO network.  Thus, following \eqref{eq:phi-ansatz}, we want to parameterize ${\omega_\dc}_i$ and ${\omega_\slow}_i(t)$ (on the left) in terms of the Floquet design parameters (on the right) according to
\begin{equation} \label{eq:control-problem-phi}
    \int_0^t \paren{{\omega_\dc}_i + {\omega_\slow}_i(t') - \frac{g_i^2}{\omega_\bus - {\omega_\dc}_i - {\omega_\slow}_i(t')} + \delta_i} \dif t' = \omega_0t - (i-1)\Lambda t - \sum_{k=1}^{N-1} F_i^{(k)}\sin(k\Lambda t),
\end{equation}
where the scalars $F^{(k)}_i$ are the solution to the Floquet design problem \eqref{eq:floquet-equation}.  Provided that we know the values of $F^{(k)}_j$ required, \eqref{eq:control-problem-phi} is a (hardware) design problem for the control variables ${\omega_\dc}_i$ and ${\omega_\slow}_i(t)$.  Differentiating both sides of \eqref{eq:control-problem-phi},
\begin{equation}
    {\omega_\dc}_i + {\omega_\slow}_i(t) - \frac{g_i^2}{\omega_\bus - {\omega_\dc}_i - {\omega_\slow}_i(t)} + \delta_i = \omega_0 -(i-1)\Lambda - \sum_{k=1}^{N-1} F_i^{(k)} \cos(k\Lambda t) k \Lambda.
\end{equation}
We then split this equation into time-independent (DC) and time-dependent (slow) parts:
\begin{subequations} \label{eq:coupling-equations} \begin{align}
    \label{eq:coupling-equation-dc-full}
    {\omega_\dc}_i - \frac{g_i^2}{\omega_\bus - {\omega_\dc}_i} + \delta_i &= \omega_0 -(i-1)\Lambda
    \\
    \label{eq:coupling-equation-slow-full}
    {\omega_\slow}_i(t) + \frac{g_i^2}{\omega_\bus - {\omega_\dc}_i} - \frac{g_i^2}{\omega_\bus - {\omega_\dc}_i - {\omega_\slow}_i(t)} &= - \sum_{k=1}^{N-1} F_i^{(k)} \cos(k\Lambda t) k \Lambda.
\end{align} \end{subequations}
The first equation can be converted into a quadratic equation for ${\omega_\dc}_i$ and solved, as can the second equation for ${\omega_\slow}_i(t)$ after given ${\omega_\dc}_i$.  (See subsection below for more details.)

As an aside, if $\delta_i \ll \Lambda$ and $g_i^2/\Delta_i(t) \ll \Lambda$ (which is necessary to satisfy the Schrieffer-Wolff conditions anyway), these equations \eqref{eq:coupling-equations} can be approximated by
\begin{subequations}
    \begin{align}
    {\omega_\dc}_i & \approx \omega_0 - (i-1)\Lambda \label{eq:coupling-eq-approx-dc} \\
    {\omega_\slow}_i(t) & \approx - \sum_{k=1}^{N-1} F_i^{(k)} \cos(k\Lambda t) k \Lambda, \label{eq:coupling-eq-approx-slow}
    \end{align}
\end{subequations}
which shows that this hardware manifestation supports our claim in Section~\ref{sec:dynamical-couplings} that the mean oscillator frequencies are approximately evenly spaced.  It also shows that the magnitudes of the slow component of the instantaneous frequencies directly correlate with the magnitude of $F_i^{(k)}$ and is intuitively given by the derivative of the instantaneous phase modulation $\dot{ \delta \phi}_i (t)$.

It is also worth noting that technically, the determination of the Floquet design parameters $F_i^{(k)}$ requires knowledge of the matrix $J_{ij}(t)$, and in this system, $J_{ij}(t)$ also depends on the control variables $\omega_i(t)$, as shown by \eqref{eq:J_def}.  Thus, the whole problem is, in principle, highly coupled, and we cannot assume we have the values of $F^{(k)}_j$ when trying to solve \eqref{eq:coupling-equations}!  A full numerical technique could iteratively solve the Floquet design problem \eqref{eq:floquet-equation} by proposing values for $F_i^{(k)}$, solving \eqref{eq:coupling-equations} for those proposed values, inserting the resulting value for $J_{ij}(t)$ into \eqref{eq:floquet-integral}, and repeating until converged.  Alternatively, it is possible to approximate $J_{ij}$ using an approximate solution to \eqref{eq:coupling-equations} that is \emph{independent} of $F_i^{(k)}$, and this latter approach will be discussed in Section~\ref{sec:native-coupling}.

Finally, turning to the pumping design problem in \eqref{eq:jpo-pump-problem}, the two-band approach says that we want
\begin{equation}
    \Paren{{\omega_\slow}_i(t) + {\omega_\fast}_i(t)} \exp\sbrak{-2\im \int_0^t \paren{{\omega_\dc}_i + {\omega_\slow}_i(t) - \frac{g_i^2}{\omega_\bus - {\omega_\dc}_i - {\omega_\slow}_i(t')} + \delta_i}\dif t'} \approx r(t).
\end{equation}
We note that the first term of this equation (proportional to ${\omega_\slow}_i(t)$) produces fast oscillations in the pumping band, whereas the right-hand side of the equation contains no high-frequency components.  Thus, we ignore the first term via the rotating-wave approximation.  Then all that remains is to choose ${\omega_\fast}_i(t)$ to cancel out the oscillations of the exponential term, producing a slow bias of the form $r(t)$.  One way to obtain this would be to take
\begin{equation} \label{eq:pumping-equation}
    {\omega_\fast}_i(t) = 2r(t)\cos\sbrak{2\paren{{\omega_\dc}_i + \delta_i} t  + 2\int_0^t \paren{{\omega_\slow}_i(t') - \frac{g_i^2}{\omega_\bus - {\omega_\dc}_i - {\omega_\slow}_i(t')}} \dif t'},
\end{equation}
using ${\omega_\dc}_i$ and ${\omega_\slow}_i(t)$ from \eqref{eq:coupling-equations}.  There is also a remaining oscillation of the phase at frequencies around $4\omega_\dc$, but those can also be neglected with a rotating wave approximation.

Equations \eqref{eq:coupling-equations} and \eqref{eq:pumping-equation}, together with the Floquet design parameters $F_i^{(k)}$, constitute a proposed solution to the bus-coupled JPO control problem for achieving dynamical coupling by designing the instantaneous frequency in the JPOs.

\subsubsection{Solving for the JPO control signals}
Here we summarize the procedure for obtaining the time-dependent instantaneous frequencies $\omega_i(t)$, as depicted, for example, in Figure~1 of the main manuscript.  The JPO control problem described by \eqref{eq:coupling-equations} and \eqref{eq:pumping-equation} can be explicitly solved as follows.

We assume that for a given problem size $N$, problem matrix $C$, bus frequency $\omega_\text{bus}$, maximum detuning $\Delta_\text{max}$, and maximum bus-oscillator coupling $g_\text{max}$, we have computed the values for $r_\text{max} = \widetilde r_\text{max}\chi$, $\delta_i = \widetilde \delta_i \chi$, $\Lambda = \widetilde\Lambda \chi$, $g_i = \widetilde g_i \chi$, ${\Delta_\text{nom}}_i = {{}{\widetilde\Delta}_\text{nom}}_i\chi$, and $F^{(k)}_j$ using the prescription described in Section~\ref{sec:parameter-summary}.  This makes use of various approximations and analytics that are described in later sections below.
\begin{enumerate}
\item The first step is to choose
\begin{equation}
\omega_0 = \omega_\text{bus} - {\Delta_\text{nom}}_1
\end{equation}
and solve \eqref{eq:coupling-equation-dc-full} for $\omega_{\dc,i}$ for each $i$ using the quadratic formula.
\item Then, with the solution to $\omega_{\dc,i}$, we can solve \eqref{eq:coupling-equation-slow-full} for $\omega_\text{\slow,i}(t)$, again using the quadratic formula, for each $i$ and every $t$.
\item Third, we can insert the above solutions for $\omega_{\dc,i}$ and $\omega_{\slow,i}$ into \eqref{eq:pumping-equation}, and numerically evaluate the integral to find $\omega_{\fast,i}(t)$ for each $i$ and every $t$.
\item Finally, the control signal for the $i$th JPO $\omega_i(t)$ is, using the two-band approximation \eqref{eq:omega-ac-split},
\begin{equation}
    \omega_i(t) = \omega_{\dc,i} + \omega_{\slow,i}(t) + \omega_{\fast,i}(t).
\end{equation}
\end{enumerate}

In Figure~1, we use $N = 5$, $\omega_\bus/2\pi = \SI{10}{GHz}$, $\Delta_\text{max}/2\pi = \SI{5}{GHz}$, and $g_\text{max}/2\pi = \SI{20}{MHz}$.  We choose an SK7 problem instance $C$ that results in prominent sidebands (for the purposes of illustration), which happens to be
\begin{equation}
    C = \frac{1}{7}\begin{pmatrix}
        0 & 2 & 1 & -4 & 4 \\
        2 & 0 & -5 & 7 & -6 \\
        1 & -5 & 0 & 5 & 3 \\
        -4 & 7 & 5 & 0 & -5 \\
        4 & -6 & 3 & -5 & 0
    \end{pmatrix}.
\end{equation}
To visualize the signals in Figure~1B (right), we compute the power spectral density of
\begin{equation}
    x_i(t) \coloneqq \cos\sbrak{\int_0^t \dif t' \omega_i(t')},
\end{equation}
which can be interpreted as the in-phase component of an oscillator with instantaneous frequency $\omega_i(t)$.

Finally, we note that for the value of $\eta$ prescribed by \eqref{eq:eta}, the modulation depth is strong enough to produce asymmetry in the sidebands, which often occurs for phase-modulated signals; for illustrative purposes, however, we reduced this asymmetry by arbitrarily using $\eta = 0.2/N$ instead; all other variables follow the usual procedure.

\newpage
\section{Native coupling matrix} \label{sec:native-coupling}
Here, we discuss the structure of the native coupling matrix $J_{ij}$.  As alluded to in the previous section, the exact form of $J_{ij}(t)$ is unknown until we have chosen the instantaneous frequencies $\omega_i(t)$; in our dynamical-coupling approach, these are parameterized by the Floquet design parameters $F^{(k)}_j$, which in turn require knowledge of $J_{ij}$ to obtain.  One way we can get around this set of coupled dependencies is to constrain ourselves to an approximate form for $J_{ij}(t)$ that is \emph{independent} of $F_i^{(k)}$.

We first address the time dependence of $J_{ij}(t)$. Using \eqref{eq:coupling-eq-approx-slow}, ${\omega_\slow}_i(t)$ can be heuristically bounded by
\begin{align}
    \label{eq:slow-freq-bound}
    \abs{{\omega_\slow}_i} &\lesssim \abs{{\omega_\slow}_1} \lesssim \max_{t \in (0, 2\pi/\Lambda)} \abs{\sum_{k=1}^{N-1} {f_1}^{(k)} \cos(k\Lambda t) k \Lambda} \\
    &\lesssim \delta{\omega_\text{bound}}_1 \coloneqq \Lambda \sum_{k=1}^{N-1} k|{f_i}^{(k)}|,
\end{align}
where ${f_i}^{(k)}$ is the first-order small-modulation solution \eqref{eq:F-taylor1-symmetric} to the design problem \emph{with $J_{ij}$ assumed uniform and time-independent}, as was done in Section~\ref{sec:small-modulation} for example.  This bound is a good heuristic provided that the native couplings are designed to be close to uniform, a condition that we will discuss shortly.  Since ${f_1}^{(k)}$ depends only on the elements $C_{ij}$ (as indicated by \eqref{eq:F-taylor1-symmetric}), $\delta{\omega_\text{bound}}_1$ can be readily computed.  In fact, if we assume \eqref{eq:F1k-analytic} as an approximate form for $f^{(k)}_1$ on average over random problem instances, then
\begin{equation}
    \label{eq:omega-bound-analytic}
    \delta{\omega_\text{bound}}_1 = \begin{cases}
        \eta \Lambda (N-1)^{3/2} & \text{for SK7} \\
        \frac12 \eta \Lambda (N-1)^2 & \text{for Dense MAX-CUT} \\
        3 \eta \Lambda (N-1)^2 N\inv & \text{for Cubic MAX-CUT}
    \end{cases} .
\end{equation}

In any case, regardless of how we compute $\delta{\omega_\text{bound}}_1$, we see that if we impose that $\delta{\omega_\text{bound}}_1 \ll \omega_\bus - {\omega_\dc}_1$, then we can neglect the effects of \eqref{eq:coupling-eq-approx-slow}, upon which \eqref{eq:coupling-eq-approx-dc} implies that the oscillator detunings are nominally (i.e., approximately) equally spaced and given by
\begin{align}
    \Delta_i(t) \approx {\Delta_\nom}_i \coloneqq {\Delta_\nom}_1 + (i-1)\Lambda,
\end{align}
where ${\Delta_\nom}_1 \coloneqq \omega_\bus - {\omega_\dc}_1 = \omega_\bus - \omega_0$ is the nominal detuning of the first oscillator.  Under this approximation, the native coupling matrix can be approximated using the nominal detunings via
\begin{equation} \label{eq:approx-J}
    J_{ij} \approx \frac{g_i g_j}{2}\paren{\frac{1}{{\Delta_\nom}_i} + \frac{1}{{\Delta_\nom}_j}},
\end{equation}
which is a time-independent expression.  Thus, so long as $J_{ij}$ is approximately uniform, we can ensure time-independence by choosing $\delta{\omega_\text{bound}}_1/{\Delta_\nom}_1 \ll 1$.

\begin{figure}
    \centering
    \includegraphics[width=0.9\textwidth]{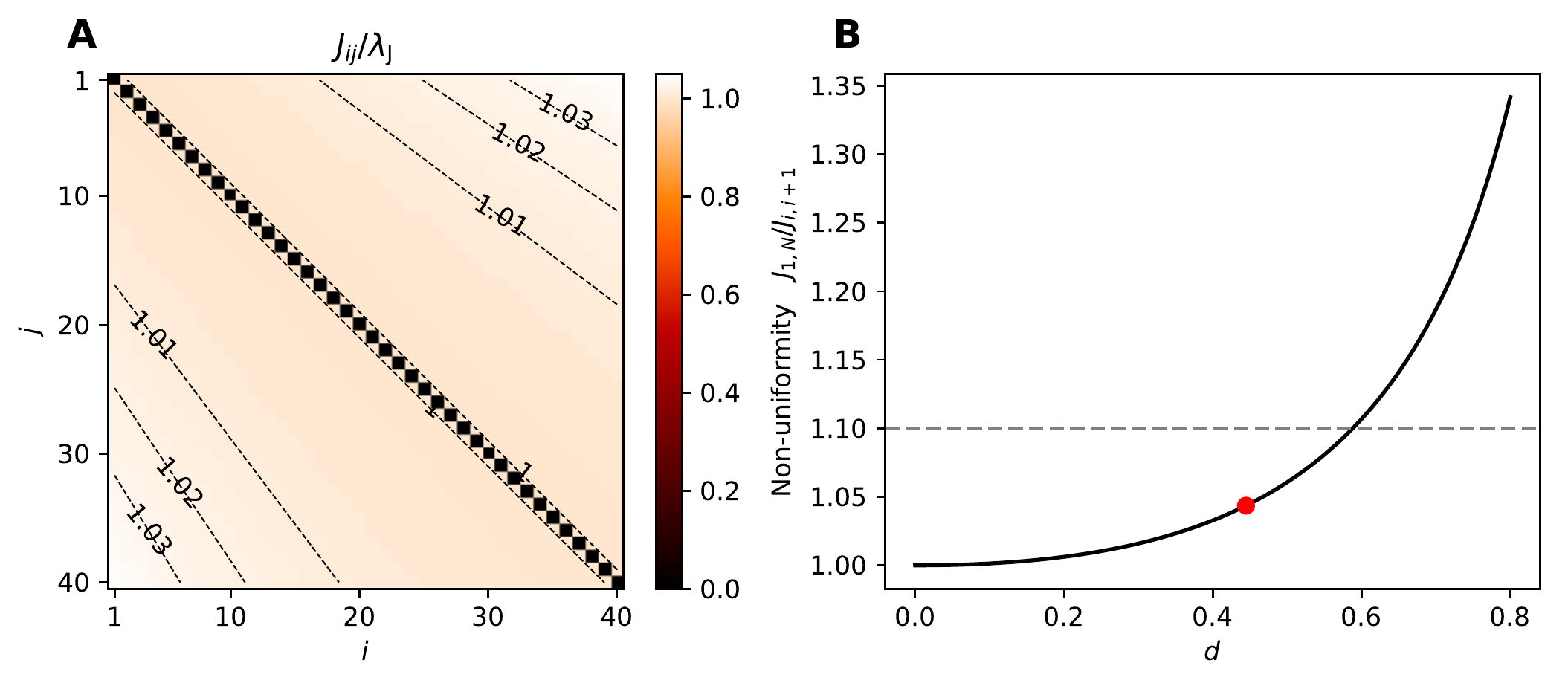}
    \caption{\textbf{Structure of native coupling matrix.} (\textbf{A}) The approximation to the native coupling matrix $J_{ij}$ given by \eqref{eq:approx-J} for $N=40$ and with nominal detunings set via \eqref{eq:Delta_nom} for a filling ratio of $d = \num{0.45}$. (\textbf{B}) Measure of non-uniformity in the native coupling matrix \eqref{eq:non-uni} as a function of the filling ratio $d$. The red dot represents the point $d=0.45$ used in (A), while the dashed line indicates a nominal non-uniformity of \SI{10}{\percent}, corresponding to $d \approx \num{0.6}$.}
    \label{fig:J}
\end{figure}

We now turn to the question of how to ensure the matrix $J_{ij}$ is uniform, in order to justify \eqref{eq:slow-freq-bound}.  The uniformity is primarily dictated by the distribution of the bus-resonator couplings $g_i$.  Looking at the form of \eqref{eq:approx-J}, one reasonable choice of the resonator-bus couplings is
\begin{equation}
    \label{eq:g}
    g_i = \sqrt{{\Delta_\nom}_i \lambda_\J}.
\end{equation}
where $\lambda_\text{J}$ is a scalar chosen to represent the approximate value of the elements in the first diagonal of $J_{ij}$. (If we succeed in obtaining a small non-uniformity, $\lambda_\text{J}$ is also the approximate value of every nondiagonal entry of $J$.)  We show in Figure~\ref{fig:J}A the approximation to the native coupling matrix \eqref{eq:approx-J} using the choice of resonator-bus coupling \eqref{eq:g}.  We observe that the overall matrix is close in magnitude to $\lambda_\text{J}$, but there is some degree of non-uniformity.

A good measure for the non-uniformity under this choice of couplings is to look at the ratio between the most off-diagonal element $J_{1N}$ and $\lambda_\J$, which can be expressed in terms of the nominal detunings ${\Delta_\nom}_i$ by
\begin{align}
    \label{eq:non-uni}
    \frac{J_{1N}}{\lambda_\J} \approx \frac{({\Delta_\nom}_1 + {\Delta_\nom}_N)/2}{\sqrt{{\Delta_\nom}_1{\Delta_\nom}_N}} = \frac{2-d}{2 \sqrt{1-d}}
\end{align}
where
\begin{equation}
    d \coloneqq \frac{{\Delta_\nom}_N-{\Delta_\nom}_1}{{\Delta_\nom}_N}
\end{equation}
is the filling ratio, a dimensionless quantity that characterizes how much of the bandwidth between $\omega_\bus$ and ${\Delta_\nom}_N$ is utilized by the oscillators.  In Figure~\ref{fig:J}B, we plot \eqref{eq:non-uni} and see that the non-uniformity scales with $d$.  More concretely, if we allow for a maximum non-uniformity of, say, \SI{10}{\percent}, the filling ratio should be at most \num{0.6}.  Thus, so long as $J_{ij}$ is approximately time-independent, we can ensure uniformity by choosing $d$ sufficiently small.

Having considered the above, we want to work in a regime in which $J_{ij}$ is both approximately time-independent and uniform.  Together with the Schrieffer-Wolff requirements, these considerations can be summarized by a set of constraints involving $d$.  We introduce parameters
\begin{enumerate}
    \item $\epsilon_\text{sw} \coloneqq g_1/{\Delta_\nom}_1$, which specifies the tolerance for the Schrieffer-Wolff condition $g_1/(\omega_\bus-\omega_i) \ll 1$ in the uniform, time-independent approximation;
    \item $\epsilon_\text{ti} \coloneqq \delta{\omega_\text{bound}}_1/{\Delta_\nom}_1$, which specifies the tolerance for time-independence in the uniform approximation; and
    \item $\epsilon_\text{uni} \coloneqq (2-d)/(2\sqrt{1-d}) - 1$, which specifies the tolerance for uniformity in the time-independent approximation.
\end{enumerate}
If we recast all these parameters in terms of $d$, we see that in order to satisfy all three constraints, $d$ should satisfy
\begin{equation} \label{eq:d}
   d = \text{min}\sbrak{
       1-\paren{1+\epsilon_{\text{sw}}^2\frac{2(N-1)\Lambda}{\lambda_\text{J}}}\inv,\;
       1-\paren{1+\epsilon_{\text{ti}}\frac{(N-1)\Lambda}{{\delta\omega_\text{bound}}_1}}\inv,\;
        1-\paren{1 + \sqrt{(1+\epsilon_{\text{uni}})^2-1}+\epsilon_{\text{uni}}}^{-2}
       }.
\end{equation}
In the above, we only technically need $d$ to be bounded by the right-hand side (i.e., a smaller $d$ still satisfies all three constraints).  However, that would result in a smaller nonlinear rate $\chi$ (as will be discussed in Section~\ref{sec:nonlinear-rate}), which in turn implies more stringent requirements on the field decay rate.  Thus, we choose $d$ to saturate its bound above.

Having chosen $d$, the nominal detunings are then given by
\begin{align}
    \label{eq:Delta_nom}
    {\Delta_\nom}_i = \frac{d}{1-d}(N-1)\Lambda + (i - 1)\Lambda.
\end{align}

\newpage
\section{Scaling Considerations for Dynamical Coupling} \label{sec:scaling-considerations}
Having discussed now our principles for achieving dynamical coupling as well as the key features of the hardware model we would like to demonstrate our principles on, we address in this section how the various parameters, approximations, and constraints come together to specify a concrete system and dynamical-control scheme in a bus-coupled JPO network.  First, we summarize our approach in a prescription for choosing dimensionless parameters given any problem instance.  Then we analyze how the nonlinear timescale $\chi$ should be chosen, which sets the dimensionful realization of a JPO network for consideration.  Finally, we analyze how our approach scales with the problem size, particularly in terms of requirements on the field decay rate.

\subsection{Summary and prescription of system parameters}
\label{sec:parameter-summary}
Up to the nonlinear timescale $\chi$, we are able to supply values for the dimensionless system parameters in order to satisfy all the conditions needed to apply our dynamical-coupling scheme to a bus-coupled system of JPOs.  Given the desired target problem couplings $C_{ij}$ with problem size $N$, we use the following prescription throughout this work:

\begin{enumerate}
    \item Compute the problem-strength parameter $\widetilde\lambda_\text{C} \coloneqq \lambda_\text{C}/\chi$ via \eqref{eq:lambdaC} in Section~\ref{sec:static-parameters}:
    \begin{equation} \label{eq:lambdaC-dimless}
        \widetilde\lambda_\text{C} = \begin{cases}
        4/N & \text{for SK7} \\
        8/N & \text{for Dense MAX-CUT} \\
        2 & \text{for Cubic MAX-CUT}
    \end{cases}.
    \end{equation}
    This choice ensures the coupling term in the target Hamiltonian scales appropriately with problem size.
    \item Set the maximum pump rate $\widetilde r_\temax \coloneqq r_\temax/\chi$ to its optimal value $\widetilde r_\temax = \num{5}$, as discussed in Section~\ref{sec:static-parameters}.  This choice is empirically found to work well for the target quantum annealer.
    \item Compute the oscillator frame-detunings $\widetilde\delta_i \coloneqq \delta_i/\chi$ via \eqref{eq:goto-detunings} in Section~\ref{sec:static-parameters}:
    \begin{equation}
        \label{eq:dimless-goto-detunings}
        \widetilde\delta_i = \widetilde\lambda_\text{C} \sum_j |C_{ij}|.
    \end{equation}
    This choice ensures that the target Hamiltonian operates correctly as a continous-variable quantum annealer.
    \item Compute the dynamical-coupling parameter $\eta$ via \eqref{eq:eta} in Section~\ref{sec:eta}:
    \begin{equation} \label{eq:eta-dimless}
        \eta = \begin{cases}
        0.8 \cdot N\inv & \text{for SK7} \\
        1.2 \cdot N\inv(1+\log N)\inv & \text{for Dense MAX-CUT} \\
        0.12 \cdot (1+\log N)\inv & \text{for Cubic MAX-CUT}
    \end{cases}.
    \end{equation}
    This choice ensures that the Floquet design equation can be numerically solved with acceptable error $\mathcal E_\text{C} \lesssim \SI{3}{\percent}$.
    \item Compute the dimensionless native-coupling strength parameter via $\widetilde\lambda_\text{J} = \widetilde\lambda_\text{C}/\eta$.  This parameter controls the nominal strength of the native couplings $J_{ij}$ and follows from the definition of $\eta$.
    \item Compute the dimensionless Floquet frequency $\widetilde\Lambda \coloneqq \Lambda / \chi$ via \eqref{eq:floquet-requirement} in Section~\ref{sec:Lambda} (specialized here for the JPO system):
    \begin{equation} \label{eq:floquet-requirement-dimless}
        \widetilde\Lambda = A \times \max(1, \widetilde \delta_i, \widetilde r_\temax, \widetilde \lambda_\text{J}),
    \end{equation}
    where $A \gtrsim \num{100}$; this value of $A$ is empirically found to ensure that the Floquet approximation holds.
    \item Compute the dimensionless approximate bound on the slow component of the instantaneous frequency ${{}\widetilde{\delta\omega}_\text{bound}}_1 \coloneqq {\delta\omega_\text{bound}}_1 / \chi$ via \eqref{eq:omega-bound-analytic} in Section~\ref{sec:native-coupling}:
    \begin{equation} \label{eq:deltaomega-dimless}
    {{}\widetilde{\delta\omega}_\text{bound}}_1 =
        \begin{cases}
            \eta \cdot \widetilde \Lambda \cdot (N-1)^{3/2} & \text{for SK7} \\
            \frac{\eta}{2} \cdot \widetilde \Lambda \cdot (N-1)^2 & \text{for Dense MAX-CUT} \\
            3 \eta \cdot \widetilde\Lambda \cdot (N-1)^2 N\inv & \text{for Cubic MAX-CUT}
        \end{cases}.
    \end{equation}
    This quantity estimates the time dependence of the native coupling using a first-order, uniform approximation for the modulations.
    \item Compute the filling ratio $d$ via \eqref{eq:d} in Section~\ref{sec:native-coupling}:
    \begin{small} \begin{equation}
    \label{eq:d-dimensionless}
    d = \text{min}\sbrak{
        1-\paren{1+\epsilon_{\text{sw}}^2\frac{2(N-1)\widetilde\Lambda}{\widetilde\lambda_\text{J}}}\inv,\;
        1-\paren{1+\epsilon_{\text{ti}}\frac{(N-1)\widetilde\Lambda}{{{}\widetilde{\delta\omega}_\text{bound}}_1}}\inv,\;
            1-\paren{1 + \sqrt{(1+\epsilon_{\text{uni}})^2-1}+\epsilon_{\text{uni}}}^{-2}
        }.
    \end{equation}\end{small}
    \!\!where $\epsilon_{\text{sw}}$, $\epsilon_{\text{ti}}$, and $\epsilon_{\text{uni}}$ must be small. (In this work, we use $\epsilon_{\text{sw}} = \epsilon_{\text{ti}} = \epsilon_{\text{uni}} = 1/10$.)
    This parameter controls the relative magnitudes of the bus-oscillator detunings in order to ensure that the resulting native couplings can be approximated as both uniform and time-independent (in addition to obeying the Schrieffer-Wolff condition).
    \item Compute the dimensionless nominal detunings ${{}\widetilde\Delta_{\nom}}_i \coloneqq {\Delta_{\nom}}_i/\chi $ via \eqref{eq:Delta_nom} in Section~\ref{sec:native-coupling}:
    \begin{align} \label{eq:Delta_nom-dimensionless}
    {{}{\widetilde\Delta}_\nom}_i = \widetilde\Lambda \sbrak{\frac{d}{1-d}(N-1)+i-1}.
    \end{align}
    These nominal detunings approximate the bus-oscillator detunings as evenly spaced after ignoring small shifts due to coupling non-uniformity and time-dependence.
    \item Compute the dimensionless bus-resonator coupling strengths $\widetilde g_{i} = g_i/\chi$ via \eqref{eq:g} in Section~\ref{sec:native-coupling}:
    \begin{align}
        \label{eq:g-dimensionless}
        \widetilde g_i = \sqrt{{{}\widetilde\Delta_\nom}_i \widetilde \lambda_\J}.
    \end{align}
    This form for the coupling produces a set of native couplings which approximately have the uniform strength $\lambda_\J$, provided that the ${{}\widetilde\Delta_\nom}_i$ and $d$ have been chosen accordingly.
    \item Approximate the dimensionless native couplings $\widetilde J_{ij}(t) \coloneqq J_{ij}(t) / \chi$ by \eqref{eq:approx-J} in Section~\ref{sec:native-coupling}:
    \begin{align}
        \label{eq:dimless-native-J}
    \widetilde J_{ij} \approx \frac{\widetilde\lambda_\J}{2} \frac{{{}{\widetilde \Delta}_\nom}_i+{{}\widetilde\Delta_\nom}_j}{\sqrt{{{}\widetilde\Delta_\nom}_i{{}\widetilde\Delta_\nom}_j}}.
    \end{align}
    Although the true native couplings $\widetilde J_{ij}(t)$ are functions of $\Delta_i(t)$ and are thus functions of time, we approximate them by this time-independent expression, which is valid provided that ${{}\widetilde\Delta_\nom}_i$ are chosen according to the steps above; in this latter case, this approximation also has the property that $\widetilde J_{ij} \approx \lambda_\J$.
    \item Solve for the Floquet design parameters $F_i^{(k)}$ with the procedure(s) outlined in Section~\ref{sec:dynamical-couplings}.  To do so, make the formal replacement $\lambda_\text{C}/J_{ij} \mapsto \widetilde\lambda_\text{C}/\widetilde J_{ij}$ in the Floquet integral \eqref{eq:floquet-integral}, where \eqref{eq:dimless-native-J} is used to approximate $\widetilde J_{ij}$.  This latter step is a good approximation provided that $d$ has been chosen accordingly, and it crucially allows us to sidestep the problem of $F^{(k)}_j$ being dependent on $\Delta_i(t)$, which is in turn dependent on $F^{(k)}_j$ and so on.  Note that for scalability and sufficiently small error $\mathcal E_\text{C} \lesssim \SI{3}{\percent}$ as discussed in Section~\ref{sec:eta}, the full-order objective function \eqref{eq:optimizeF-column} should be optimized when $N < 100$, while for $N>100$, second-order approach \eqref{eq:optimizeF-column-taylor2} should be used instead.
\end{enumerate}

By following this above prescription for any given problem instance, we arrive at a concrete set of dimensionless parameters (namely, $\widetilde\delta_i$, $\widetilde r_\temax$, $\widetilde g_i$, $\widetilde J_{ij}$, and $F^{(k)}_j$, the others being secondary to these) which allow us to construct a bus-mediated, dynamically-coupled JPO quantum annealing Hamiltonian (2), at least up to an overall timescale $\chi$.  This prescription is thus sufficient for performing simulations (e.g., as discussed in Section~\ref{sec:quantum}) to study the performance of the scheme.  Furthermore, after selecting the timescale $\chi$ as described below, we also arrive at a concrete hardware realization, complete with control signals (as parameterized by $F^{(k)}_j$)---additionally see Section~\ref{sec:JPO-control-problem} for more details on this latter point.

It is worth emphasizing again that this prescription is characterized by our particular approach to obtaining the Floquet design parameters $F^{(k)}_j$: we solve the Floquet design problem using a nominal, time-independent approximation for the native couplings $J_{ij}$ rather than their ``true'' values $J_{ij}(t)$.  To enable this, the other system parameters (e.g., $g_i$ and so on) have been chosen in very particular ways (e.g., through $d$ and its constrained values) such that this approximation holds well.  As such, for the purposes of running simulations, we take the approximation \eqref{eq:dimless-native-J} to hold \emph{exactly}, and while the design parameters $F^{(k)}_j$ resulting from this procedure could, in principle, be further refined by iterated time-dependent optimization, we forgo doing so for both our simulations and our analysis of hardware realizations/scaling.

Finally, we note that the above prescription is technically constructed for a \emph{lossless} system, in that the field decay rate $\kappa$ is not considered to affect any of the dimensionless parameters mentioned.  This goes hand-in-hand with the fact that the parameter $T_\text{ramp}$---which governs how adiabatically the pump is ramped up---also has not been discussed, as it should ideally be set to an arbitrarily high value to ensure adiabaticity.  However, when the timescale $\widetilde\kappa \coloneqq \kappa / \chi$ is also brought into consideration, we prescribe that one should pick
\begin{equation}
\widetilde T_\text{ramp} \coloneqq T_\text{ramp}/\chi = 1/\widetilde\kappa
\end{equation}
to obtain order unity number of photon emission events per oscillator, in accordance with the observations made in Section~\ref{sec:static-parameters}.  In addition, when photon loss is encountered, one can also empirically tweak the above-prescribed values $\widetilde r_\temax$ and $\widetilde \lambda_\J$ (among others) to further enhance success probability.  Although we do not include such hyper-optimization here as part of any general procedure, we note that we do incorporate such an approach in some of our quantum simulations (see Section~\ref{sec:quantum} for more details).

\subsection{Nonlinear (Kerr) rate}
Up to this point, all our discussions have been implicitly in units of the nonlinear rate $\chi$.  Thus, selecting a value for $\chi$ produces an instantiation of all the other parameters in a dimensionful hardware system.  Provided a fixed field decay rate, it is clear that we want to maximize the value of $\chi$ in order to obtain fast system rates relative to the field decay rate.  However, we cannot pick the rate arbitrarily high, as it is limited by two experimental constraints:
\begin{enumerate}
  \item $g_i \leq g_\text{max}$, the maximal bus-resonator coupling achievable, and
  \item $\Delta_N \leq \Delta_\text{max}$, the maximal detuning possible due to bandwidth constraints.
\end{enumerate}

To treat the first constraint, we use our form for the bus-oscillator couplings \eqref{eq:g} to write it as
\begin{equation}
    \label{eq:g-constraint}
    \sqrt{{\Delta_\nom}_i \lambda_\J} \leq g_\temax .
\end{equation}
Since ${\Delta_\nom}_i \leq {\Delta_\nom}_N$ by the construction in \eqref{eq:Delta_nom}, the oscillator with the largest detuning (i.e., $i = N$) has the largest bus-resonator coupling, and it is thus the most relevant oscillator for this constraint.  We substitute the expression for ${\Delta_\nom}_N$ from \eqref{eq:Delta_nom-dimensionless} into the constraint above and express the nonlinear rate in terms of dimensionless variables, namely $\chi = \Lambda/\widetilde \Lambda = \lambda_\text{J}/\widetilde \lambda_\text{J}$.  Doing this, we arrive at the constraint
\begin{align}
    \label{eq:chi_g}
    \chi \leq g_\temax\sqrt{\frac{2d}{\widetilde \Lambda \widetilde \lambda_\text{J}(N-1)}}.
\end{align}

To treat the second constraint, we approximate the detunings ${\Delta}_i$ by the nominal (i.e., evenly spaced) detunings ${\Delta_\nom}_i$.  Then using $\Delta_N \approx {\Delta_\nom}_N$ in the second constraint and using again \eqref{eq:Delta_nom-dimensionless}, we arrive at
\begin{equation} \label{eq:chi_detuning}
    \chi \lesssim \Delta_\temax\frac{d}{(N-1)\widetilde \Lambda}.
\end{equation}

Thus, to satisfy both constraints \eqref{eq:chi_g} and \eqref{eq:chi_detuning} simultaneously, the nonlinear rate should be given by
\begin{align} \label{eq:chi}
    \chi \leq \chi_\temax \equiv \min \paren{g_\temax\sqrt{\frac{2d}{\widetilde \Lambda \widetilde \lambda_\text{J}(N-1)}}, \, \Delta_\temax\frac{d}{(N-1)\widetilde \Lambda}},
\end{align}
where $\chi_\temax$ is the maximum allowable nonlinear rate (absolute-scale) before the physical constraints on the bus-resonator couplings and the maximal detuning are violated. As previously mentioned, given a fixed field decay rate, it is preferable to set the nonlinear rate to the maximum, i.e., $\chi = \chi_\text{max}$. However, this is not technically necessary; for example, if, in a particular hardware implementation, $\kappa$ is linked to $\chi$ in such a way that decreasing $\chi$ yields an improvement in $\chi/\kappa$, such an approach would be compatible with all the constraints we have considered.

As we will see in the following subsection, $\chi_\text{max}$ generally decreases with the problem size $N$ at sufficiently large $N$. This can be contrasted with the typical scenario for gate-based transmon qubit architectures, where it is desirable to use the maximum nonlinear rate allowed by the hardware (while maintaining acceptable field decay rates), more or less independently of $N$. Instead, in our dynamical-coupling scheme at sufficiently large $N$, the nonlinear rate $\chi$ is forced to be generally much smaller than the hardware-limited maximum, essentially because of the need to fit $N$ oscillators, all distinct in frequency, into a limited bandwidth while still obeying the Floquet approximation, etc. Thus, at least at large $N$, the key experimental challenge is to obtain field decay rates low enough to support such a small nonlinear rate (i.e., at most $\chi_\text{max}$).

\subsection{Scaling of dynamical coupling scheme with problem size} \label{sec:scaling}
In this subsection, we explore quantitatively the implications of the above results for the scaling of the dynamical coupling scheme with problem size (at least for the JPO networks we consider in this work). To do so, we analyze the scaling of the maximum allowable nonlinear rate $\chi_\text{max}$ of the system since this parameter ultimately sets the requirements on the field decay rate.  In consideration of \eqref{eq:chi}, we first discuss the scaling of the three dimensionless parameters on which $\chi_\text{max}$ depends: $\widetilde \lambda_\text{J}$, $\widetilde \Lambda$, and $d$.

\subsubsection{Scaling of native coupling strength and Floquet frequency with problem size}
First, we consider the nominal native coupling strength $\widetilde\lambda_\text{J}$.  Using the forms we prescribe for $\eta = \widetilde\lambda_\text{C}/\widetilde\lambda_\text{J}$ in \eqref{eq:eta-dimless} and $\widetilde\lambda_\text{C}$ in \eqref{eq:lambdaC-dimless}, we get
\begin{equation} \label{eq:tilde_lambdaJ}
    \widetilde \lambda_\text{J} = \begin{cases}
        5 & \text{for SK7} \\
        (20/3)\paren{1+\log N} &  \text{for Dense MAX-CUT} \\
        (50/3)\paren{1+\log N} & \text{for Cubic MAX-CUT}
    \end{cases}.
\end{equation}
In particular, because $\widetilde \lambda_\text{J}$ shows up in the denominator of \eqref{eq:chi}, this result implies that SK7 problems have the smallest native coupling strength requirement, while Cubic MAX-CUT problems have the largest.

Next, due to the Floquet timescale requirements, $\widetilde \Lambda = A \max(1, \widetilde \delta_i, \widetilde r_\temax, \widetilde \lambda_\text{J})$ by \eqref{eq:floquet-requirement-dimless}.  In the case where $\widetilde \lambda_\text{J}$ dominates, $\widetilde\lambda_\text{J}$ and $\widetilde\Lambda$ are linearly related, in which case $\widetilde\lambda_\text{J}$ and $\widetilde\Lambda$ will have the same functional form.  For our prescription, this is the most relevant situation, since $\widetilde r_\temax$ is constant with $N$ in our prescription, so unless $\widetilde\Lambda$ is a constant, it will be limited by $\widetilde\lambda_\text{J}$ at sufficiently large $N$.

\begin{figure}
    \includegraphics[width=1.0\linewidth]{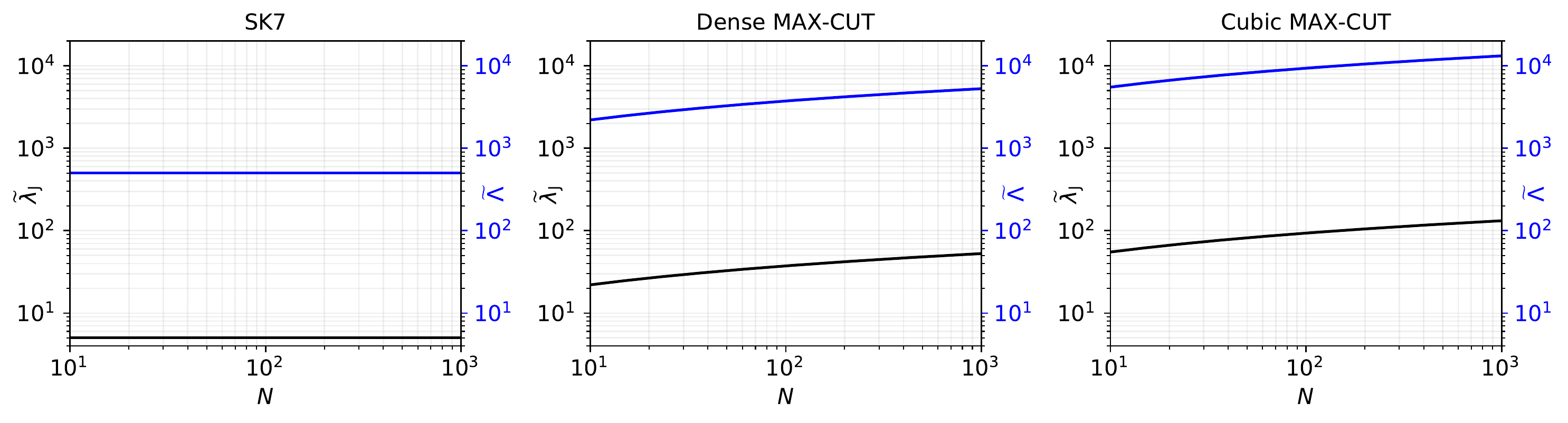}
    \caption{\textbf{Scaling of native coupling strength and Floquet frequency.} In black, the dimensionless native coupling strength $\widetilde \lambda_\text{J}$ following \eqref{eq:tilde_lambdaJ}, and in blue, the dimensionless Floquet frequency $\widetilde \Lambda$ following \eqref{eq:floquet-requirement-dimless}, both as functions of problem size $N$.  These results follow the prescription outlined in Section~\ref{sec:parameter-summary}, wherein $r_\temax = 5$ and $A = 100$.}
    \label{fig:lambdaJ_Lambda}
\end{figure}

In Figure~\ref{fig:lambdaJ_Lambda}, we show the values of $\widetilde\lambda_\text{J}$ following \eqref{eq:tilde_lambdaJ} and $\widetilde\Lambda$ following \eqref{eq:floquet-requirement-dimless} as a function of problem size.  From the figure, it turns out $\widetilde \lambda_\text{J} \geq 5$ in all cases, and since $\widetilde r_\temax = 5$ in our prescription (see also Section~\ref{sec:static-quantum-annealing}), $\widetilde\lambda_\text{J}$ limits the value of $\widetilde \Lambda$ in all cases. (Note also that $\widetilde \delta_i$ does not limit $\widetilde \Lambda$ provided we follow \eqref{eq:lambdaC-dimless} and \eqref{eq:dimless-goto-detunings}.)

\subsubsection{Scaling of filling ratio with problem size} \label{sec:d-scaling}
In \eqref{eq:chi}, the value of $\chi$ increases for larger filling ratio $d$, which implies that it is important for $d$ to be large to maximize the performance of the dynamical-coupling scheme.  In Section~\ref{sec:native-coupling}, we showed that the filling ratio is determined by three different constraints, which is summarized by \eqref{eq:d-dimensionless} in our prescription. We can rewrite this as
\begin{gather}
    \label{eq:d-expanded}
    d = \min\sbrak{\frac{D_{\text{sw}}}{1+D_{\text{sw}}},\; \frac{D_{\text{ti}}}{1+D_{\text{ti}}},\; d_{\text{uni}}},\quad\text{where} \\
    D_{\text{sw}} \coloneqq 2\epsilon_{\text{sw}}^2(N-1)\widetilde\Lambda/\widetilde\lambda_\text{J}, \qquad
    D_{\text{ti}} \coloneqq \epsilon_{\text{ti}}(N-1)\widetilde\Lambda/{\widetilde\delta\omega_\text{bound}}_1, \qquad
    d_\text{uni} \coloneqq 1-\paren{1 + \sqrt{(1+\epsilon_{\text{uni}})^2-1}+\epsilon_{\text{uni}}}^{-2}. \nonumber
\end{gather}
In this form, we see that the uniformity constraint (dictated by $d_\text{uni}$) does not depend on the problem size $N$.  As discussed in Section~\ref{sec:native-coupling}, this uniformity constraint fundamentally limits the bandwidth utilization to about 60\% (i.e., $d \leq d_{\text{uni}}=0.6$) for a non-uniformity in the native coupling matrix of 10\% (i.e., $\epsilon_{\text{uni}} = 0.1$).  On the other hand, both the Schrieffer-Wolff and time-independence conditions vary with $N$ and are both described by a similar functional form, where $d$ increases for larger $D_{\text{sw}}$ or $D_\text{ti}$.  Thus, for sufficiently large $D_{\text{sw}}$ and $D_{\text{ti}}$, we have the simple picture that the filling ratio $d$ is limited by the uniformity constraint to give $d=d_\text{uni}$.  To understand when this is the case, we study $D_{\text{sw}}$ and $D_{\text{ti}}$ below.

The simpler of the two is the Schrieffer-Wolff condition.  Assuming, as in the previous subsection, that $\widetilde \Lambda = A \widetilde \lambda_\text{J}$, we can first rewrite
\begin{align}
    \label{eq:D-sw}
    D_{\text{sw}} = 2\epsilon_{\text{sw}}^2 A (N-1)\; .
\end{align}
Thus, $D_\text{sw}$ is linearly proportional to the problem size and does not depend on the problem class.  For sufficiently large $N$, this condition ceases to limit $d$.

Turning to the time-independence constraint, we can substitute the form for ${\widetilde{\delta\omega}}_\text{bound}$ from \eqref{eq:deltaomega-dimless} in our prescription into $D_\text{ti}$, upon which we see that $D_\text{ti}$ depends on $N$ and $\eta$.  Then using the form for $\eta$ from \eqref{eq:eta-dimless} in our prescription, we arrive at
\begin{equation}
    \label{eq:D-ti}
    D_{\text{ti}} = \begin{cases}
        (5/4)\epsilon_{\text{ti}} \cdot N^{1/2} \cdot \sqrt{N/(N-1)} & \text{for SK7} \\
        (5/3)\epsilon_{\text{ti}} \cdot (1+\log N) \cdot N/(N-1) & \text{for Dense MAX-CUT} \\
        (25/9)\epsilon_{\text{ti}} \cdot (1+\log N) \cdot N/(N-1) & \text{for Cubic MAX-CUT}
    \end{cases}.
\end{equation}
We can see that $D_{\text{ti}}$ increases with the problem size in all cases, so the time-independence constraint also ceases to limit $d$ at sufficiently large $N$.  Moreover, for the SK7 problems, $D_{\text{ti}}$ increases as $N^{1/2}$ while for the Dense and Cubic MAX-CUT problems, $D_{\text{ti}}$ increases more slowly.  This implies that for the SK7 problems, $d$ varies more drastically with respect to $N$ than for Dense and Cubic MAX-CUT problems.

\begin{figure}
    \includegraphics[width=1.0\linewidth]{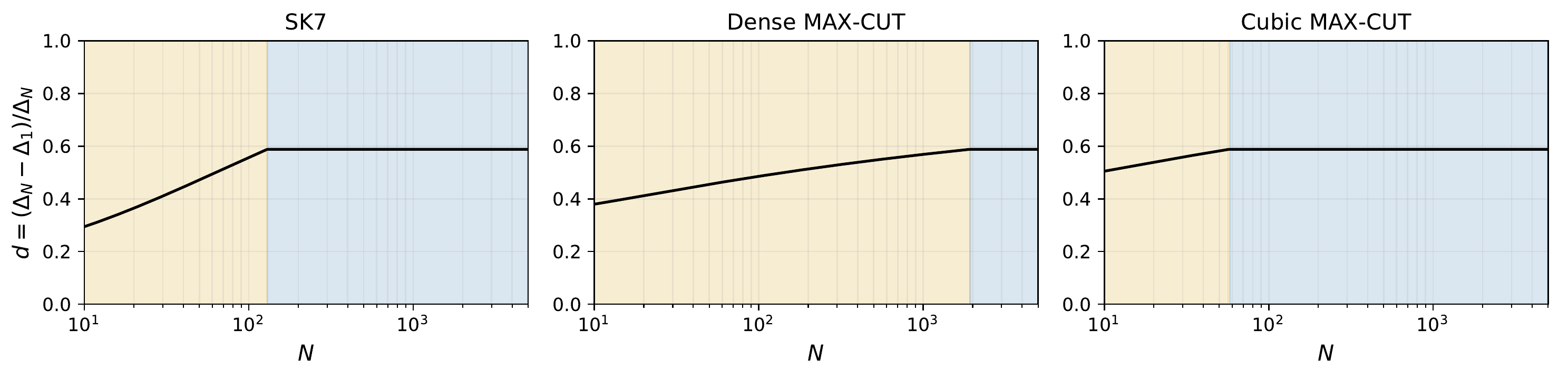}
\caption{\textbf{Scaling of filling ratio.} The filling ratio $d$ as given by \eqref{eq:d-expanded} as a function of $N$, with $D_\text{sw}$ given by \eqref{eq:D-sw} and $D_\text{ti}$ given by \eqref{eq:D-ti}. The gold and blue regions denote where $d$ is constrained by the time-independence and uniformity requirements, respectively.  These results follow the prescription outlined in Section~\ref{sec:parameter-summary}, wherein $\epsilon_{\text{sw}} = \epsilon_{\text{ti}} = \epsilon_{\text{uni}} = 0.1$. }
    \label{fig:d}
\end{figure}

In Figure~\ref{fig:d}, we show the filling ratio $d$ as a function of problem size following the above analysis.  For the parameters we consider, it turns out that $d$ is never limited by the Schrieffer-Wolff condition.  Rather, it is dominated by the time-independence condition at small $N$ before being clamped by the uniformity condition at approximately 60\%, as argued above.  Finally, these results also verify that $d$ depends more weakly on $N$ for the MAX-CUT problems.

\subsubsection{Scaling of nonlinear rate with problem size} \label{sec:nonlinear-rate}
Putting the above results together via \eqref{eq:chi} produces the maximum allowable nonlinear rate $\chi_\text{max}$ for each desired problem size $N$. Since the effects of $\widetilde\lambda_\text{J}$ and $\widetilde\Lambda$ are to decrease $\chi_\text{max}$ with $N$, while the effects of $d$ eventually saturates with $N$, we ultimately expect that $\chi_\text{max}$ decreases with $N$.

\begin{figure}[hb]
    \includegraphics[width=0.66\linewidth]{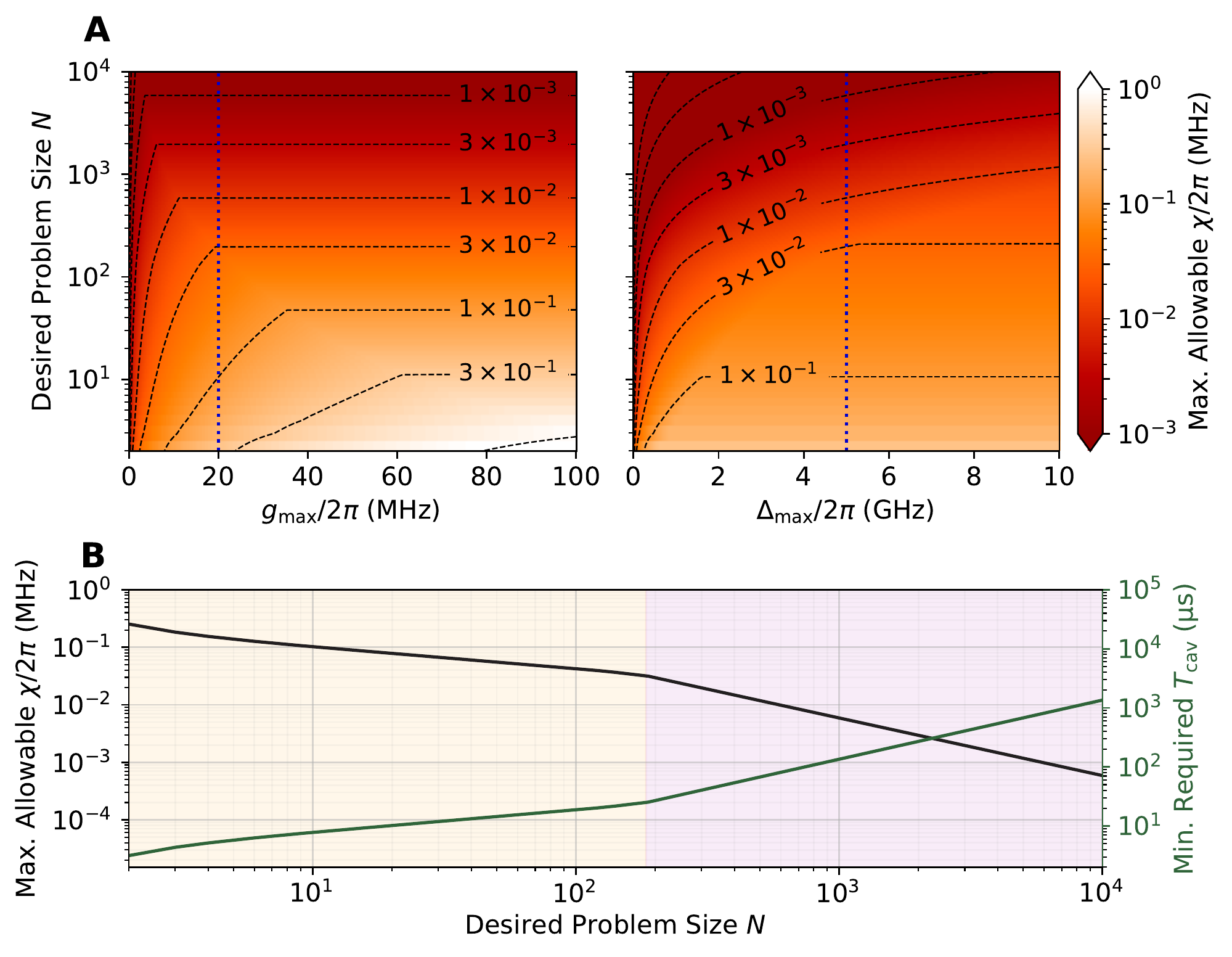}
    \caption{\textbf{Scaling requirements for SK7 problems.} (Reproduced from Figure~5 of main manuscript) (\textbf{A}) The maximum allowable nonlinear rate $\chi_\text{max}$ according to \eqref{eq:chi}, as a function of the desired problem size $N$ and the maximum achievable bus-oscillator coupling rate $g_\temax$ (left, with $\Delta_\temax/2\pi = \SI{5}{GHz}$) and the maximum achievable bus-oscillator detuning $\Delta_\temax$ (right, with $g_\temax/2\pi = \SI{20}{MHz}$). The black dotted lines are level curves of constant $\chi_\temax$, while the dashed blue lines indicate the slice of these plots shown in (B) below. (\textbf{B}) On the left axis (black curve), $\chi_\temax$ as a function of problem size for $\Delta_\temax/2\pi = \SI{5}{GHz}$ and $g_\temax/2\pi = \SI{20}{MHz}$. On the right axis (green curve), the minimum required cavity lifetime $T_\text{cav}$, assuming we use $\chi = \chi_\temax$ and that we need $T_\text{cav} = 5/\chi$ for the quantum annealer to perform effectively. The beige shaded region indicates the regime in which $\chi_\temax$ is limited by $g_\temax$ while the pink shaded region indicates where it is limited by $\Delta_\temax$. All parameters of the system (e.g., $\widetilde \Lambda$, $\widetilde \lambda_\J$ and $d$) follow the prescription specified in Section~\ref{sec:parameter-summary}.}
    \label{fig:scaling-sk7}
\end{figure}

\begin{figure}[ht!]
    \includegraphics[width=0.68\linewidth]{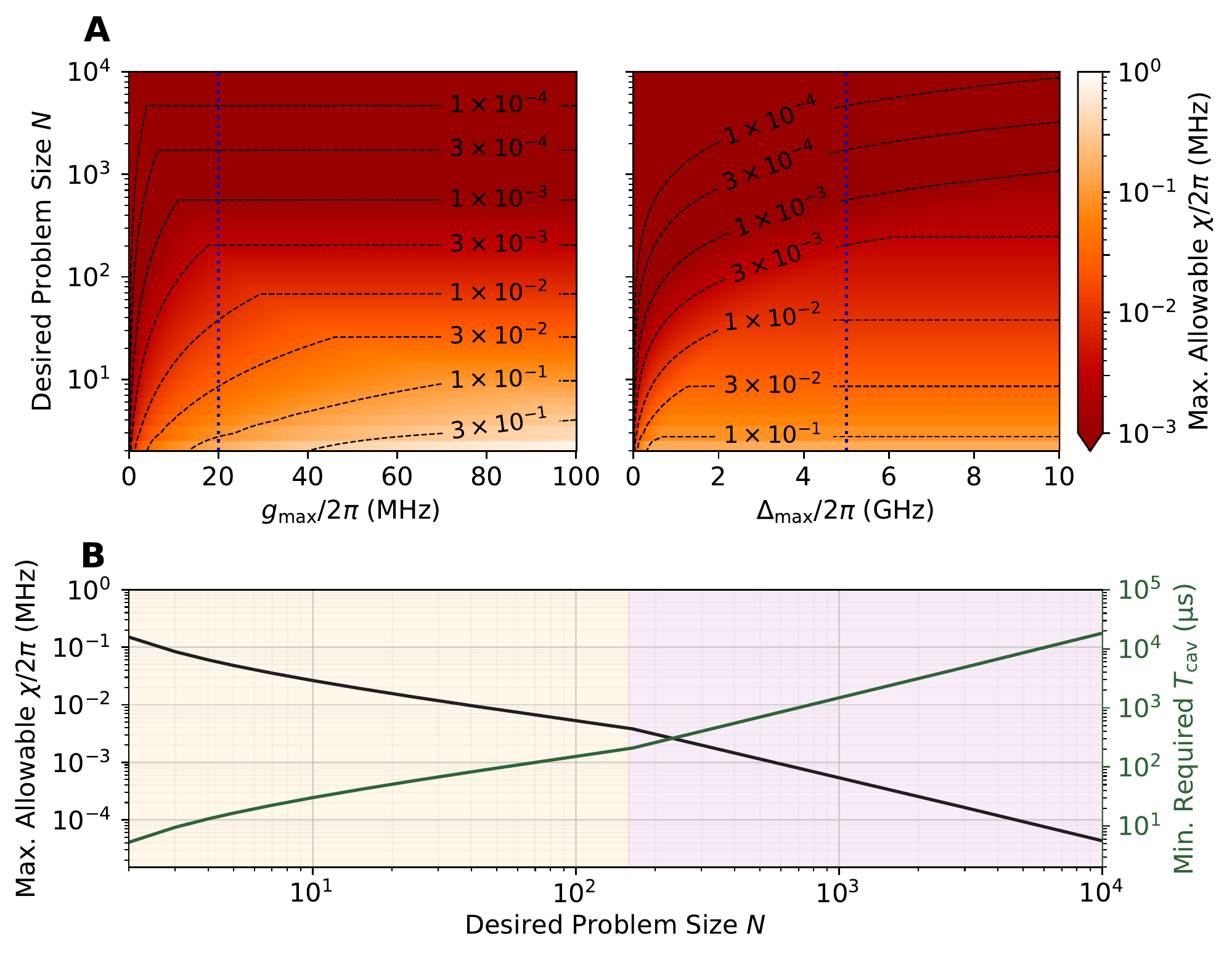}
    \caption{\textbf{Scaling requirements for Dense MAX-CUT problems.} See caption of Figure~\ref{fig:scaling-sk7}.}
    \label{fig:scaling-densemaxcut}
\end{figure}

\begin{figure}[hb!]
    \includegraphics[width=0.68\linewidth]{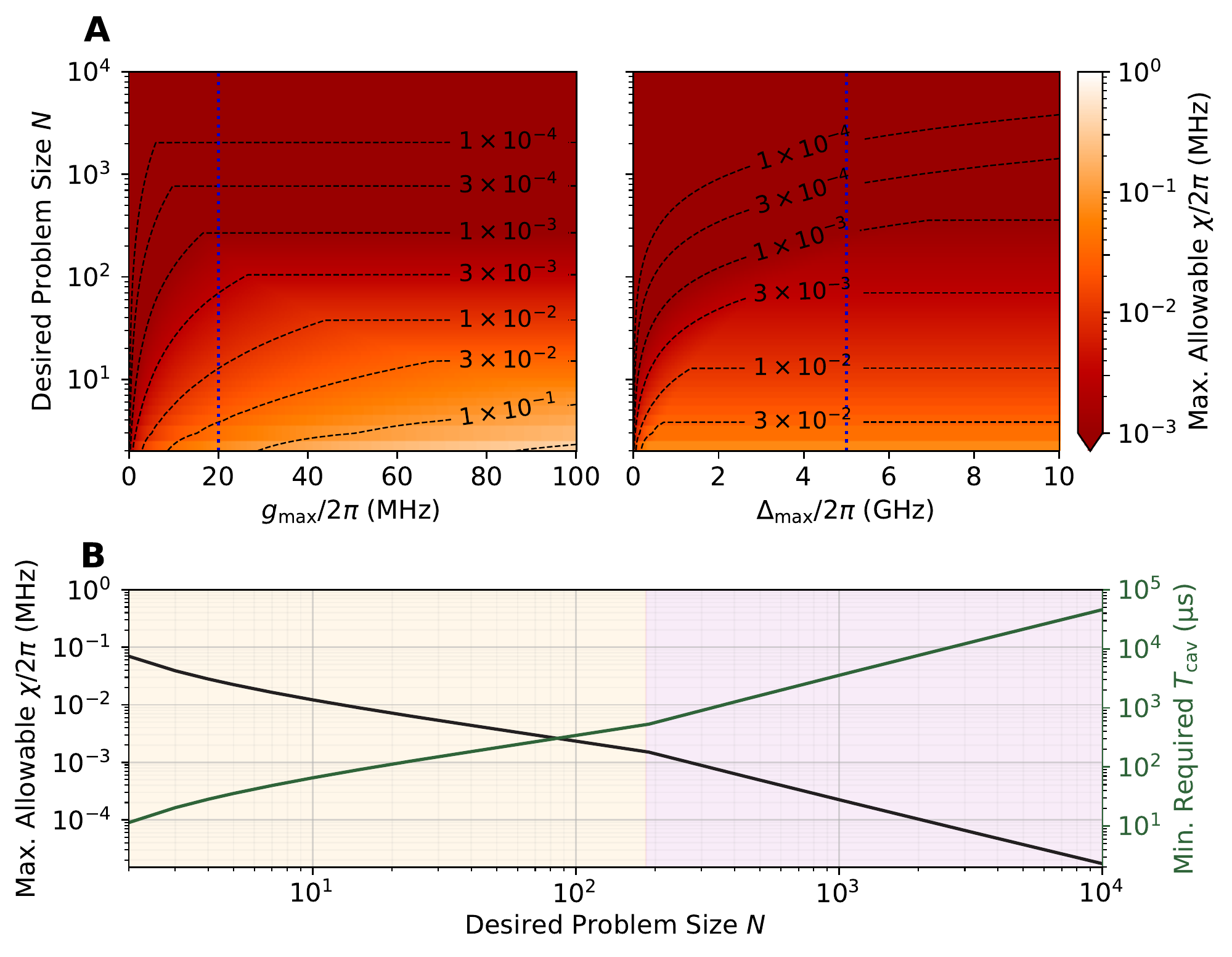}
    \caption{\textbf{Scaling requirements for Cubic MAX-CUT problems.} See caption of Figure~\ref{fig:scaling-sk7}.}
     \label{fig:scaling-cubicmaxcut}
\end{figure}

This trend can be quantitatively seen in Figures~\ref{fig:scaling-sk7}, \ref{fig:scaling-densemaxcut} and \ref{fig:scaling-cubicmaxcut}. In addition to the general trend of decreasing $\chi_\text{max}$ with $N$, we also see that the particular scaling behavior of $\chi_\text{max}$ with $N$ changes dramatically when the system switches from being limited by the maximum coupling strength $g_\temax$ (where it is more gentle) to being limited by the maximum available bandwidth $\Delta_\temax$ (where it is sharper).

While Figures~\ref{fig:scaling-sk7}, \ref{fig:scaling-densemaxcut} and \ref{fig:scaling-cubicmaxcut} use the exact formula for $\chi_\temax$ given by \eqref{eq:chi}, we can also qualitatively understand much of the scaling behavior using the arguments already presented. First, based on what we found for the scaling of $d$, we can make the simplifying assumption that $d$ is constant (this holds at large $N$, and even when $d$ increases at small $N$, it is a slow function). Then we can insert the expressions for $\widetilde\lambda_\text{J}$ from \eqref{eq:tilde_lambdaJ} along with $\widetilde\Lambda \propto \widetilde\lambda_\text{J}$ as argued above. In doing so, we find that when $\chi_\temax$ is constrained by $g_\temax$, we have (with $N-1 \approx N$)
\begin{equation} \label{eq:chi_gmax}
    \chi_\temax \propto \begin{cases}
        N^{-1/2} & \text{for SK7} \\
        N^{-1/2} \cdot \paren{1 + \log N}\inv & \text{for Dense MAX-CUT} \\
        N^{-1/2} \cdot \paren{1 + \log N}\inv & \text{for Cubic MAX-CUT}
    \end{cases}.
\end{equation}
On the other hand, when the system is constrained by $\Delta_\temax$, we can apply the same simplifying assumptions to get
\begin{equation} \label{eq:chi_Deltamax}
    \chi_\temax \propto \begin{cases}
        N\inv & \text{for SK7} \\
        N\inv \cdot \paren{1 + \log N}\inv & \text{for Dense MAX-CUT} \\
        N\inv \cdot \paren{1 + \log N}\inv & \text{for Cubic MAX-CUT}
    \end{cases}.
\end{equation}
Thus, at large $N$, the bandwidth constraint due to $\Delta_\temax$ is the one that limits the scaling of $\chi_\temax$, while the milder coupling constraint due to $g_\temax$ is potentially relevant for intermediate values of $N$. The latter result is especially interesting for near-term demonstrations of dynamical coupling as \eqref{eq:chi_gmax} implies that the scaling requirements for $\chi_\temax$ (and thus ultimately of the field decay rate) are less stringent than the $1/N$ scaling that one might naively expect, and which indeed dominates at large $N$.

Finally, we note that both \eqref{eq:chi_gmax} and \eqref{eq:chi_Deltamax} show that the scaling requirements are more favorable for the SK7 problem class. This observation is supported quantitatively by Figures~\ref{fig:scaling-sk7}, \ref{fig:scaling-densemaxcut} and \ref{fig:scaling-cubicmaxcut}. This advantage stems from the much smaller Floquet timescale requirement for $\widetilde \Lambda$ for SK7 problems, as discussed in Section~\ref{sec:Lambda}.

\newpage
\section{Models and Methodologies for Numerical Simulations} \label{sec:methodology}
In this section, we specify and review the mathematical framework we use to model the quantum states and dynamics of the KPO system, under both static coupling and dynamical coupling. We describe the quantum simulations we perform and the procedure we use to interpret the resulting state trajectories as success probabilities. We also describe the classical EOMs we use to obtain intuition about how the observations made in the quantum simulations scale for larger problem sizes.

\subsection{Quantum Simulations}
\label{sec:quantum}
We perform quantum simulations in this work to analyze the performance of the dynamical-coupling scheme relative to the statically-coupled system, especially in the presence of dissipation. Our dissipation model is described by the standard Lindblad master equation
\begin{align}
    \label{eq:master}
    \dot{\hat\rho} = -\im[\hat H, \hat\rho] + \sum_m \paren{ \hat L_m \hat \rho \hat L_m\dagg - \frac{1}{2}\hat L_m\dagg \hat L_m \hat \rho -  \frac{1}{2}\hat\rho \hat L_m\dagg \hat L_m }
\end{align}
where $\hat H$ is the Hamiltonian of the system, and $\hat L_m$ are the Lindblad operators that describe the effect of dissipation due to coupling to the environment; the time dependence of all operators have been omitted for brevity. For the statically-coupled scheme, the Hamiltonian is given by \eqref{eq:kpo-network} while the Lindblad operators are given by \eqref{eq:lindblad}. For the dynamical-coupling scheme, the Hamiltonian is given by (3) in the main manuscript, while the Lindblad operators, due to being in the rotating frame, are given by $\hat L_m = \sqrt{2\kappa} \hat a_m \exp\Paren{-\im \int_0^t \paren{\omega_m(t') - g_m^2/\Delta_m(t') + \delta_m} \dif t'}$. However, by direct substitution of this Lindblad operator into \eqref{eq:master}, we see that the overall phase is always canceled with its complex conjugate; this implies that the overall phase in the Lindblad operator does not have any observable consequences for the internal state of the open quantum system. Thus, without loss of generality, we take the Lindblad operator of the dynamical-coupling scheme to be $\hat L_m = \sqrt{2\kappa}\hat a_m$, as for the statically-coupled system.

Directly simulating \eqref{eq:master} is difficult due to the large Hilbert-space dimension of the quantum state. To circumvent the memory requirement of storing a full density matrix, we utilize the Monte-Carlo wavefunction method (i.e., quantum jump method) to approximately solve \eqref{eq:master} via stochastic sampling \cite{Wiseman2010}. Given a finite number of trajectories $N_{\text{traj}}$, this method approximates the density matrix $\hat \rho(t)$ using the ensemble mean of the trajectories $\ket{\psi^{(\ell)}(t)}$ via
\begin{align}
    \label{eq:rho_psi}
    \hat \rho(t) \approx \frac{1}{N_{\text{traj}}}\sum_{\ell=1}^{N_\text{traj}} \ket{\psi^{(\ell)}(t)} \bra{\psi^{(\ell)}(t)}.
\end{align}
It should be noted that for the idealized lossless case of $\kappa=0$, the number of trajectories should be taken to be $N_\text{traj} = 1$ as there is no stochasticity due to the lack of dissipation to the environment; this is equivalent to solving the Schr\"odinger equation.

Given a spin configuration $\sigma = \paren{\sigma_1, \ldots, \sigma_N}$ where $\sigma_i = \pm 1$, we can define, for the $\ell$th trajectory, the probability of the quantum state $\ket{\psi^{(\ell)}(t)}$ being in the specific spin configuration $\sigma$ as \cite{Goto2016b}
\begin{align}
    \label{eq:prob-conf}
    \text{P}_{\sigma}^{(\ell)}(t) \coloneqq \prod_{i=1}^N \paren{\int_{\sgn x_i = \sigma_i} \hspace{-15pt}\dif x_i }  \abs{\psi^{(\ell)}(x_1, \ldots, x_N, t)}^2
\end{align}
where $\psi^{(\ell)}(x_1, x_2, \ldots, x_N, t) \coloneqq \braket{x_1, x_2, \ldots, x_N | \psi^{(\ell)}(t)}$ is the wavefunction in the position basis.  (Note: We use the convention $\hat x_i \coloneqq (\hat a_i + \hat a_i\dagg)/2$.)  For our numerical approach to computing \eqref{eq:prob-conf}, see Section~\ref{sec:wavefunction} below.

This definition of the success probability is motivated by the fact that the final state of the KPO system is approximately a superposition of coherent states $\ket{\pm\alpha_i}$ for each KPO, and we interpret these coherent states to encode the Ising spins $\sigma_i = \pm 1$.  In practice, we obtain the result of the computation by an in-phase homodyne measurement of $\hat x$, and the sign of the result on the $i$th KPO is interpreted as $\sigma_i$.  Under this definition of the spin configuration probability, the vacuum state at the beginning of the computation has equal probability on all possible spin configurations, and as the computation progresses, the displacements of the KPOs evolve---in a correlated way---to remove the ambiguity, pushing the probability distribution into ideally the ground-state quadrants of phase space.

Using \eqref{eq:prob-conf}, we define the success probability of the $\ell$th trajectory as
\begin{align}
    \label{eq:prob-success-traj}
    \text{P}_\text{success}^{(\ell)}(t) \coloneqq \sum_{\sigma \in \mathcal{S}_{\text{gs}}} \text{P}_\sigma^{(\ell)}(t)
\end{align}
where $\mathcal{S}_{\text{gs}}$ is set of spin configurations that are ground states of the specified Ising problem.

These trajectory-wise definitions for the configuration and success probabilities are distinct from the ensemble-averaged configuration and success probabilities which would be computed from \eqref{eq:master}.  We estimate these ensemble-averaged quantities, $\text{P}_\sigma$ and $\text{P}_\text{success}$, respectively, by taking the sample mean of the trajectory-wise quantities:
\begin{align}
    \label{eq:prob-configuration-ensemble}
    \text{P}_{\sigma}(t) &\approx \frac{1}{N_{\text{traj}}}\sum_{\ell=1}^{N_\text{traj}} \text{P}_{\sigma}^{(\ell)}(t), \\
    \label{eq:prob-success-ensemble}
    \text{P}_{\text{success}}(t) &\approx \frac{1}{N_{\text{traj}}}\sum_{\ell=1}^{N_\text{traj}} \text{P}_{\text{success}}^{(\ell)}(t).
\end{align}
Of course, since $N_\text{traj}$ is finite, there is statistical sampling uncertainty on this estimation, which we henceforth take to be the standard error of the mean. We show the standard error of the mean as a colored fill around the mean on Figures~\ref{fig:gs_lowloss_1}--\ref{fig:gs_highloss_2},\ref{fig:staconf_lowloss_1}--\ref{fig:staconf_highloss_2}, and \ref{fig:dynconf_lowloss_1}--\ref{fig:dynconf_highloss_2}.

Our quantum simulation procedure is as follows. Given a problem matrix $C_{ij}$, we chose the parameters $\widetilde r_\temax$, $\widetilde T_\text{ramp}, \widetilde \lambda_\text{C}$ to maximize the success probabilities of the static annealer, summarized in Table~\ref{table:static-parameters} for the simulations in this work. Then we use Steps 3--12 of the prescription outlined in Section~\ref{sec:parameter-summary} to calculate the system parameters $\widetilde\delta_i$, $\widetilde J$, and $\widetilde \Lambda$ and $F_j^{(k)}$.  Since we take the unit of time to be $1/\chi$, the simulations are independent of dimensionful constraints associated with particular physical realizations (i.e., due to $g_\temax$ and $\Delta_\temax$).  Having obtained these dimensionless system parameters, we simulate both the statically- and dynamically-coupled systems using the Monte-Carlo wavefunction method with the QuTiP library (version 4.3.1) in Python \cite{Johansson2013}.  To reliably compare the stochastic results from the statically- and dynamically-coupled systems, we use the same noise process (i.e., random number generator) to ensure that both systems physically experience the same instantiation of quantum noise.

As is the standard approach, we run our simulations with various truncation settings for the Fock basis of each mode, and we choose a setting which does not result in significant changes in the results upon increase. (Specifically, for all the quantum simulations in this work, the truncation was set to \num{12} basis states per mode, corresponding to $12^4$-dimensional states for $N=4$ simulations.) As is also standard, we decrease the time step taken in our solver until no significant changes in the result occur upon further decreasing the step size.

This approach was taken to generate Figure~2 in the main manuscript. For the simulations with no cavity photon loss ($\kappa=0$), $N_{\text{traj}}=1$; while for the low-loss ($\kappa = 0.01\chi$) and high-loss ($\kappa = 0.1\chi$) cases, $N_{\text{traj}} = 10$. For the \num{2}-mode wavefunctions shown in Figure~2A, we use the approach described below in Section~\ref{sec:wavefunction}. The success probability used in the histogram is defined as the success probability at the end of the computation $\text{P}_{\text{success}} \coloneqq \text{P}_{\text{success}}(T)$.  Note that even though the histogram in Figure~2D does not show the uncertainty of the success probabilities, these uncertainties can be obtained by examining the trajectory plots of each problem instances in Figures~\ref{fig:gs_lowloss_1}--\ref{fig:gs_highloss_2}.

\setlength\extrarowheight{3pt}

\begin{table}
    \begin{center}
        \begin{tabular}{ l | r r r }
    Simulation &  $\widetilde \lambda_\text{C}$ & $\widetilde T_\text{ramp}$ & $\widetilde r_\temax$ \\ \hline\hline
            $N = 2$, $\kappa/\chi =0$ & 2 & 100 & 5.0 \\
            $N = 4$, $\kappa/\chi =0$ & 1 & 100 & 5.0 \\
            $N = 4$, $\kappa/\chi =\num{e-1}$ & 1 & 25 & 4.5 \\
            $N = 4$, $\kappa/\chi =\num{e-2}$ & 1 & 5 & 4.0 \\
        \end{tabular}
        \label{table:static-parameters}
    \end{center}
    \caption{System parameters associated with the static annealer (see Section\ref{sec:static-parameters}) used for the quantum simulations in this work, chosen to optimize the success probability. These optimized parameters also carry over to the quantum simulation of the corresponding dynamically-coupled system.}
\end{table}

\subsubsection{Continuous-variable representations of state}
\label{sec:wavefunction}
One simple way to visualize the correlations among the oscillators is to look at the quantum state in a continuous-variable basis.  For example, in position space, the population (i.e., diagonal) elements of the density operator $\hat\rho$ can be represented as a function over $x \in \mathbb{R}^N$ as $p_\text{x}(x) \coloneqq \bra{x}\hat \rho\ket{x}$.  (For pure states $\hat \rho = \ket{\psi}\bra{\psi}$, $p_\text{x}(x) = \abs{\braket{x|\psi}}^2$.)  Of course, to obtain non-classical information about the state, we also need to look at a conjugate variable; for example, we can also consider the momentum-basis population $p_\text{p}(p)$, computed in the same way but with $\ket{x} \mapsto \ket{p}$.

In practice, since $\hat \rho$ is represented (e.g., in QuTiP) in the Fock basis, a basis transformation is needed to compute such matrix elements. The state $\ket{x}$ can be obtained in the Fock basis via a standard basis transformation:
\begin{equation} \label{eq:x-eigenstate}
  \ket{x} = \sum_{n\in\mathbb{N}^N} \braket{n|x}\ket{n}
  = \bigotimes_{i=1}^N \sum_{n_i=0}^\infty \braket{n_i|x_i}\ket{n_i},
\end{equation}
which can be readily computed since $\braket{x_i|n_i} = \phi^{(\text{x})}_{n_i}(x_i)$, where for $x \in \mathbb R$,
\begin{equation}\label{eq:qho-wavefunction}
\phi^{(\text{x})}_n(x) \coloneqq \frac{(-1)^n}{\paren{\sqrt{\pi}2^n n!}^{1/2}} \exp\paren{x^2/2} \frac{\dif^n}{\dif x^n} \exp\paren{-x^2}
\end{equation}
is the position-basis wavefunction of the $n$th eigenstate of the quantum harmonic oscillator (i.e., the $n$th Fock state).  Similarly, if we are interested in $\braket{p|n}$, we can perform the same calculation using the Fock states in the momentum basis instead, using $\phi^{(\text{p})}_n(p) \coloneqq (-\im)^n \phi^{(\text{x})}_n(p)$.

Explicitly, one can in practice (e.g., in QuTiP) compute $p_\text{x}(x_1,\ldots,x_N)$ by evaluating the ``matrix element'' of the ``density matrix'' $P_{n'n} \coloneqq \bra{n'_1,\ldots,n'_N}\hat\rho\ket{n_1,\ldots,n_N}$ against the ``state vector'' $c_n(x) \coloneqq \braket{n_1,\ldots,n_N|x_1,\ldots,x_N}$.  Furthermore, we note that, in practice, the summations (e.g., in \eqref{eq:x-eigenstate}) are limited by truncation in the Fock basis, with upper limits $M_n$ such that the support of $\bra{n_1,\ldots,n_N}\hat\rho\ket{n_1,\ldots,n_N}$ is negligible for $n_i > M_i$.

\subsubsection{Evaluation of success probabilities}
\label{sec:quantum-success-evaluation}
A similar consideration is needed to evaluate the success probabilities defined in \eqref{eq:prob-conf} when the quantum state is represented in the Fock basis.  To do so, we note that (focusing on the case of $N = 1$ so that $x \in \mathbb R$)
\begin{equation} \label{eq:projector-1}
    \int_{x_\text{min}}^{x_\text{max}} \abs{\psi(x)}^2 \, \dif x = \int_{x_\text{min}}^{x_\text{max}} \dif x \braket{\psi|x}\braket{x|\psi} = \Vbrak{\hat \Pi_{x_\text{min}}^{x_\text{max}}}_{\ket{\psi}},
\end{equation}
where $\hat \Pi_{x_\text{min}}^{x_\text{max}} \coloneqq \int_{x_\text{min}}^{x_\text{max}} \ket{x}\bra{x} \, \dif x$ is the projector onto the set of $x$-eigenstates between $x_\text{min}$ and $x_\text{max}$.  We need to obtain the Fock representation of this projector.  For any two Fock states $\ket{n'}$ and $\ket{n}$, its matrix elements are
\begin{equation} \label{eq:projector-matrix-elements}
    \braket{n'|\hat \Pi_{x_\text{min}}^{x_\text{max}}|n} = \int_{x_\text{min}}^{x_\text{max}} \braket{n'|x}\braket{x|n} \, \dif x,
\end{equation}
which can be readily evaluated using \eqref{eq:qho-wavefunction}.  From this, we can continue \eqref{eq:projector-1} by writing
\begin{equation}
    \Vbrak{\hat \Pi_{x_\text{min}}^{x_\text{max}}}_{\ket{\psi}} = \sum_{n'=0}^{\infty} \sum_{n=0}^\infty \braket{\psi|n'} \braket{n'|\hat \Pi_{x_\text{min}}^{x_\text{max}}|n} \braket{n|\psi}.
\end{equation}
Explicitly, one can compute this in practice (e.g. in QuTiP) as an ``expectation value'' of the ``operator matrix'' $\Paren{P_{x_\text{min}}^{x_\text{max}}}_{n'n} \coloneqq \braket{n'|\hat \Pi_{x_\text{min}}^{x_\text{max}}|n}$ on the ``state vector'' $c_n \coloneqq \braket{n|\psi}$.

For $N > 1$, the system composition rule of quantum mechanics implies that
\begin{equation}
    \prod_{i=1}^N\int_{{x_\text{min}}_i}^{{x_\text{max}}_i} \hspace{-6pt}\dif x_i \, \abs{\psi(x_1, \ldots, x_N)}^2 = \prod_{i=1}^N \sum_{n_i,n_i'=0}^\infty \overbrace{\bra{n'_1}\hat\Pi_{{x_\text{max}}_1}\ket{n_1}\cdots\bra{n'_N}\hat\Pi_{{x_\text{max}}_N}\ket{n_N}}^{\Sbrak{P_{{x_\text{min}}_1}^{{x_\text{max}}_1} \otimes \cdots \otimes P_{{x_\text{min}}_N}^{{x_\text{max}}_N}}_{n'_1,\ldots,n'_N,n_1,\ldots, n_N}}
    \overbrace{\braket{\psi|n'_1,\ldots,n'_N}}^{c\conj_{n'_1,\ldots,n'_N}} \overbrace{\braket{n_1,\ldots,n_N|\psi}}^{c_{n_1,\ldots,n_N}},
\end{equation}
where $\otimes$ indicates the Kronecker product of matrices.  Again, as indicated by the braced quantities above the expression, this quantity can be evaluated in practice (e.g., in QuTiP) as an expectation value of an operator matrix on a state vector (both over an $N$-mode Fock space).

In practice, there are two more issues we need to address: discretization and truncation in position basis. (Truncation in the Fock basis has already been discussed above and simply consists of restricting the upper limit of all the summations.) In order to evaluate the integral in \eqref{eq:projector-matrix-elements}, we can discretize the interval $x_\text{min} < x < x_\text{max}$ into $L$ sections, and we approximate the matrix element by
\begin{subequations}
\begin{equation} \label{eq:projector-truncated}
    \Paren{P_{x_\text{min}}^{x_\text{max}}}_{n'n} \approx \sum_{j=0}^{L-1} \braket{n'|x^{(j)}}\braket{x^{(j)} | n} \Delta x,
\end{equation}
where
\begin{equation}
    \Delta x \coloneqq \frac{x_\text{max}-x_\text{min}}{L}
    \quad\text{and}\quad
    x^{(j)} \coloneqq x_\text{min} + \paren{j+\frac 12}\Delta x.
\end{equation}
\end{subequations}
Note that if our Fock basis is truncated to $M$ basis elements, \eqref{eq:projector-truncated} is simply a multiplication of an $M \times L$ matrix with an $L \times M$ matrix (the latter representing the change of basis from truncated-Fock to discretized-position, and the former vice-versa).  Finally, the position basis is also truncated in practice, so to represent the improper integral $\int_{\mathbb R^+} \abs{\psi(x)}^2\, \dif x$, we need to pick $x_\text{min} = 0$ and $x_\text{max}$ sufficiently large as to cover the non-negligible support of $\psi(x>0)$, while to represent $\int_{\mathbb R^-} \abs{\psi(x)}^2 \, \dif x$, we pick $x_\text{min}$ sufficiently negative as to cover the non-negligible support of $\psi(x<0)$ and $x_\text{max} = 0$.

\subsection{Classical Simulations} \label{sec:classical}
We also perform classical simulations to analyze how the performance of the dynamical-coupling scheme scales with respect to problem size $N$. The idea here is to derive equations of motion (EOMs) with system dimension linear in $N$, in order to make numerical analysis tractable.  We start with the quantum Heisenberg-Langevin EOMs
\begin{align}
    \label{eq:qsde}
    \frac{\dif\hat a_i}{\dif t} = -\im [\hat a_i, \hat H] + \frac{1}{2}\sum_m\paren{\hat L_m\dagg [\hat a_i, \hat L_m] + [\hat L_m\dagg, \hat a_i]\hat L_m} + \text{Noise terms}
\end{align}
describing the quantum system and then perform the formal substitution $\hat a_i \rightarrow \alpha_i$, where $\alpha_i \in \mathbb{C}$; the noise terms are then neglected.  Although these classicalized EOMs are not a physically correct model for the KPO network in the quantum regime, such an approach has been used in, e.g., Ref.~\cite{Nigg2017b} to understand the operational mechanism of such quantum devices, and models of this kind have been shown to be effective heuristic Ising solvers in and of themselves \cite{Goto2019}.  We utilize the same approach to help inform the scaling of the Floquet requirements at larger problem sizes (of the quantum model).

For the statically-coupled system, this classicalization procedure produces the EOMs
\begin{align}
    \label{eq:eom-static}
    \frac 1 \chi \frac{\dif\alpha_i}{\dif t} = - \widetilde\kappa \alpha_i + \im\paren{\widetilde\delta_i + \abs{\alpha_i}^2}\alpha_i - \im \widetilde r \alpha_i\conj + \im \widetilde \lambda_\text{C} \sum_{i \neq j} C_{ij} \alpha_j.
\end{align}
Using the same procedure, the dynamically-coupled system has the following EOMs:
\begin{align}
    \label{eq:eom-dynamical}
    \frac 1 \chi \frac{\dif\alpha_i}{\dif t} = - \widetilde\kappa \alpha_i + \im\paren{ \widetilde\delta_i + \abs{\alpha_i}^2}\alpha_i - \im \widetilde r \alpha_i\conj + \im \widetilde \lambda_\text{C} \sum_{j \neq i} J_{ij}  \exp\Sbrak{-\im\Lambda(j-i)t +\im\delta\phi_j(t)-\im\delta\phi_i(t)}\alpha_j .
\end{align}
For all the classical simulations, we use $\widetilde\kappa = 0$.  From empirical investigations, the value of $\widetilde\kappa$ does not significantly affect the qualitative distribution of the success probabilities in this classicalized model.

Since $\alpha_i = 0$ is an unstable equilibrium, we follow the approach of Reference~\cite{Goto2016b} and sample random initial-field amplitudes $\alpha_i(t=0)$, according to
\begin{align}
    \label{eq:initial-field}
    \alpha_i(t = 0) = \sqrt{\frac{n_0}{2}}z_i \e{2\im\pi u_i},
\end{align}
where $z_i$ are iid random variables with the standard normal distribution while $u_i$ are iid random variables with the uniform distribution on the interval $[0,1]$.  For all classical simulations, we choose $n_0 = \num{e-2}$.

As with the quantum simulations, we simulate the classical EOMs on a finite number of trajectories $N_{\text{traj}}$. The $\ell$th trajectory $\alpha^{(\ell)}(t)$ is an $N$-dimensional vector at each time point $t$, and each $\ell$ corresponds with a random sample from \eqref{eq:initial-field} for the initial condition of the trajectory.  Note that in contrast to the quantum case, we always need to take $N_\text{traj} > 1$ in all cases, even though $\widetilde\kappa = 0$.

In this approach, we identify the spin configuration of a trajectory as a function of time using the relation
\begin{align}
    \label{eq:classical-sigma}
    \sigma^{(\ell)}_i(t) = \sgn\Paren{\alpha^{(\ell)}_i(t)}.
\end{align}
The success of a given trajectory $\ell$ at time $t$ is then given by the indicator variable
\begin{equation}
    \label{eq:classical-success}
    s^{(\ell)}(t) \coloneqq
    \begin{cases}
        1 & \text{if} \;\; \sigma^{(\ell)}(t) \in \mathcal{S}_{\text{gs}}\\
        0 & \text{if} \;\; \sigma^{(\ell)}(t) \notin \mathcal{S}_{\text{gs}}
    \end{cases},
\end{equation}
where $\mathcal{S}_{\text{gs}}$ is the set of ground-state spin configurations.  Then, as in the quantum case, we define the success probability of the ensemble of trajectories to be the sample mean
\begin{align}
    \label{eq:success-classical}
    \text{P}_{\text{success}}(t) = \frac{1}{N_{\text{traj}}} \sum_\ell s^{(\ell)}(t).
\end{align}
Again, there is a statistical sampling uncertainty associated with the sample mean, which we take to be the usual standard error on the mean.  We show the standard error of the mean as a colored fill around the mean on Figure~3A.

Our classical simulation procedure is as follows. Given a problem matrix $C$, we use step 1--12 in the prescription outlined in \ref{sec:parameter-summary} to calculate the system parameters $\widetilde r_\temax$, $\widetilde\delta_i$, $\widetilde \lambda_\text{C}$, $\widetilde J$, $\widetilde T$ and $\widetilde \Lambda$ and $F_i^{(k)}$.  Since we take the unit of time to be $1/\chi$, the simulations are independent of dimensionful constraints associated with particular physical realizations (i.e., due to $g_\temax$ and $\Delta_\temax$). Finally, for simplicity, we use $\tilde \kappa=0$ for these simulations.

For each trajectory index, both the statically- and dynamically-coupled systems are simulated with the same (random) initial field amplitude. We then simulate both the statically- and dynamically-coupled systems using the ODE solver library DifferentialEquations.jl (version 4.5.0) in Julia \cite{Rackauckas2017a}.   As is the standard approach, we decrease the time step taken in our solver until no significant changes in the results occur upon further decrease of the step size. Note that since the timescale of the dynamics (i.e., $1/\tilde\Lambda$) become smaller with increasing $N$, the time step required for convergence decreases as $\sim 1/N$.

In Figure~3A, we simulate the classical EOMs \eqref{eq:eom-static} and \eqref{eq:eom-dynamical} with $N_{\text{traj}} = 5000$.  This large number of trajectory was chosen to ensure that the error on the sample mean is negligible small.  In Figure~3B, we show one of the simulations in the ensemble used to generate Figure~3A (i.e., showing a specific value of $k$).  In order to show more fine-grained information, we plotted the in-phase quadrature $\RE\Paren{\alpha^{(k)}_i(t)}$ instead of $s^{(k)}_i(t)$ (which is just the sign of the former).  To generate Figure~3C, the classical EOMs are simulated for problem sizes $N$ up to $N=50$. Here, we use for each problem instance $N_{\text{traj}} = 100$ trajectories; this was done as the larger problem sizes required intensive computation time (on the order of one day of computation per trajectory for $N=50$).  As in the quantum simulations, the success probability used in the histogram is defined as the success probability at the end of the computation $\text{P}_{\text{success}} = \text{P}_{\text{success}}(T)$.

\section{Ising problem classes and problem instance generation}
\label{sec:Ising-problem-classes}
In this work, we consider instances drawn from three specific classes of Ising problems, characterized by the statistical distribution of the $C_{ij}$ couplings specifying the Ising problem. Note that we choose the convention that the $C_{ij}$ couplings are normalized such that $\max_{ij} \abs{C_{ij}} = 1$.

The instances are generated according to the following prescriptions:

\begin{itemize}
    \item Integer Sherrington-Kirkpatrick graphs with range 7 (SK7) \cite{Rønnow2014}: Every upper-triangular element of $C$ is independently chosen with equal probability from \num{14} discrete values $\set{-7, \ldots, -1, 1, \ldots, 7}$. Then $C$ is normalized by its maximum amplitude: $C \mapsto C/\max_{ij}(|C_{ij}|)$.
    \item Dense MAX-CUT graphs: Every upper-triangular element of $C$ is independently chosen with \SI{50}{\percent} probability to be either 0 or 1.
    \item Cubic MAX-CUT graphs: We sample uniformly from the set of 3-regular graphs, where every vertex has degree 3. These graphs were generated with the LightGraphs.jl package \cite{Bromberger17} in Julia).
\end{itemize}

\newpage
\section{Additional Data Figures} \label{sec:data-dump}
In this section, we provide additional figures which depict in greater detail the data generated from the quantum and classical simulations performed in this work. The more comprehensive statistics provided by these figures support the conclusion made in the main manuscript that the dynamical-coupling scheme results in time-evolution of success probabilities in a way which closely matches that of the corresponding statically-coupled system. They also provide a check that the figures presented in the main manuscript are statistically representative of the entire data set generated from our simulations.

\begin{figure}
    \includegraphics[width=0.9\linewidth]{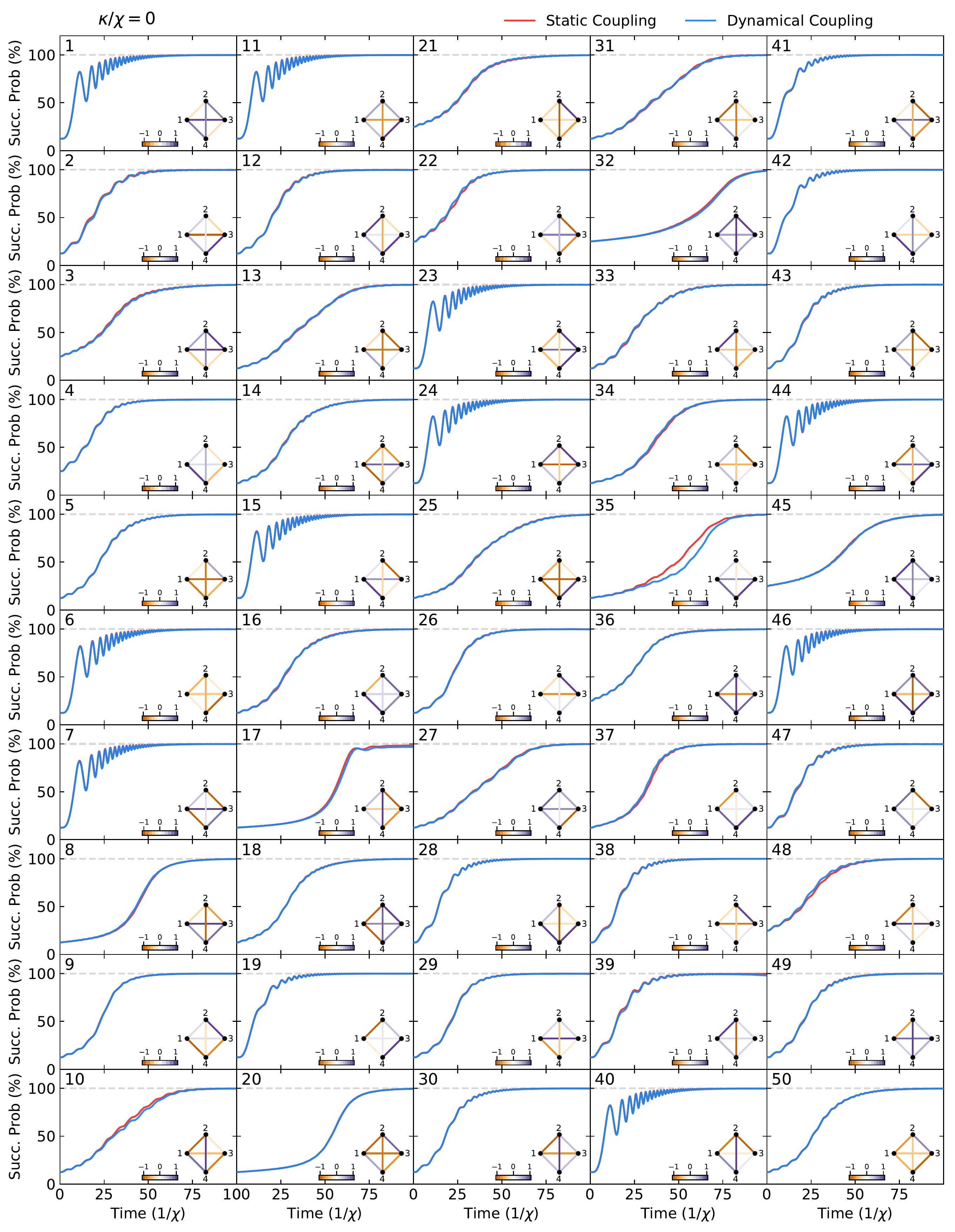}
    \caption{ \textbf{Quantum simulation of success probability, lossless case (part 1/2)}. Evolution of the success probability $\text{P}_\text{success}(t)$ of finding a ground state of the inset Ising problem instance (one for each sub-axis, numbered), assuming the cavity field decay rate $\kappa = 0$. In red is shown the success probability of a statically-coupled system implementing the Ising couplings exactly, while in blue is shown the success probability of the corresponding dynamically-coupled system. The system parameters are chosen according to the prescription described in Section~\ref{sec:parameter-summary} for each Ising problem instance. See Section~\ref{sec:quantum} for additional details of our quantum simulation methodology. }
    \label{fig:gs_noloss_1}
\end{figure}

\begin{figure}
    \includegraphics[width=0.9\linewidth]{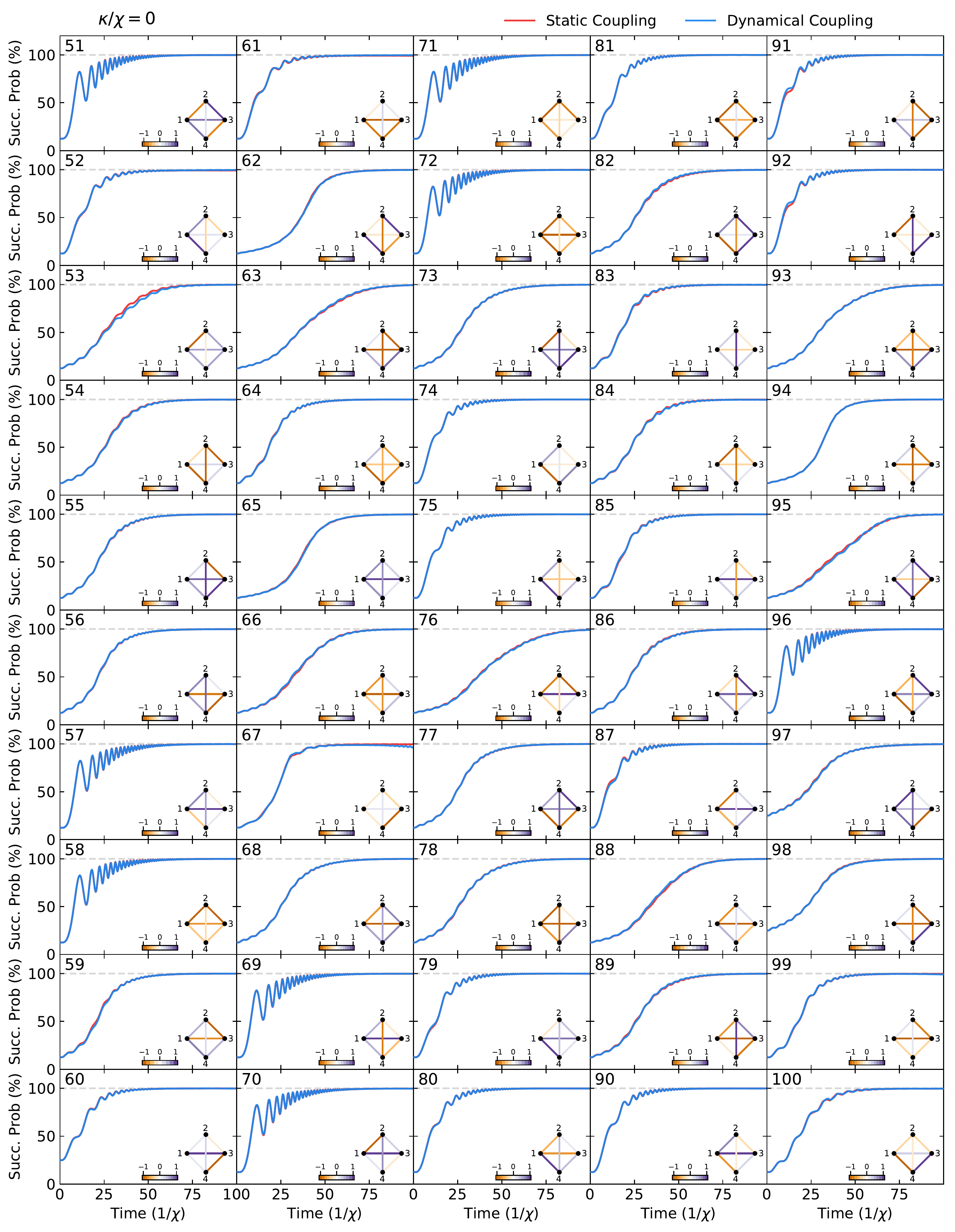}
    \caption{ \textbf{Quantum simulation of success probability, lossless case (part 2/2)}. Evolution of the success probability $\text{P}_\text{success}(t)$ of finding a ground state of the inset Ising problem instance (one for each sub-axis, numbered), assuming the cavity field decay rate $\kappa = 0$. In red is shown the success probability of a statically-coupled system implementing the Ising couplings exactly, while in blue is shown the success probability of the corresponding dynamically-coupled system. The system parameters are chosen according to the prescription described in Section~\ref{sec:parameter-summary} for each Ising problem instance. See Section~\ref{sec:quantum} for additional details of our quantum simulation methodology. }
    \label{fig:gs_noloss_2}
\end{figure}

\begin{figure}
    \includegraphics[width=0.9\linewidth]{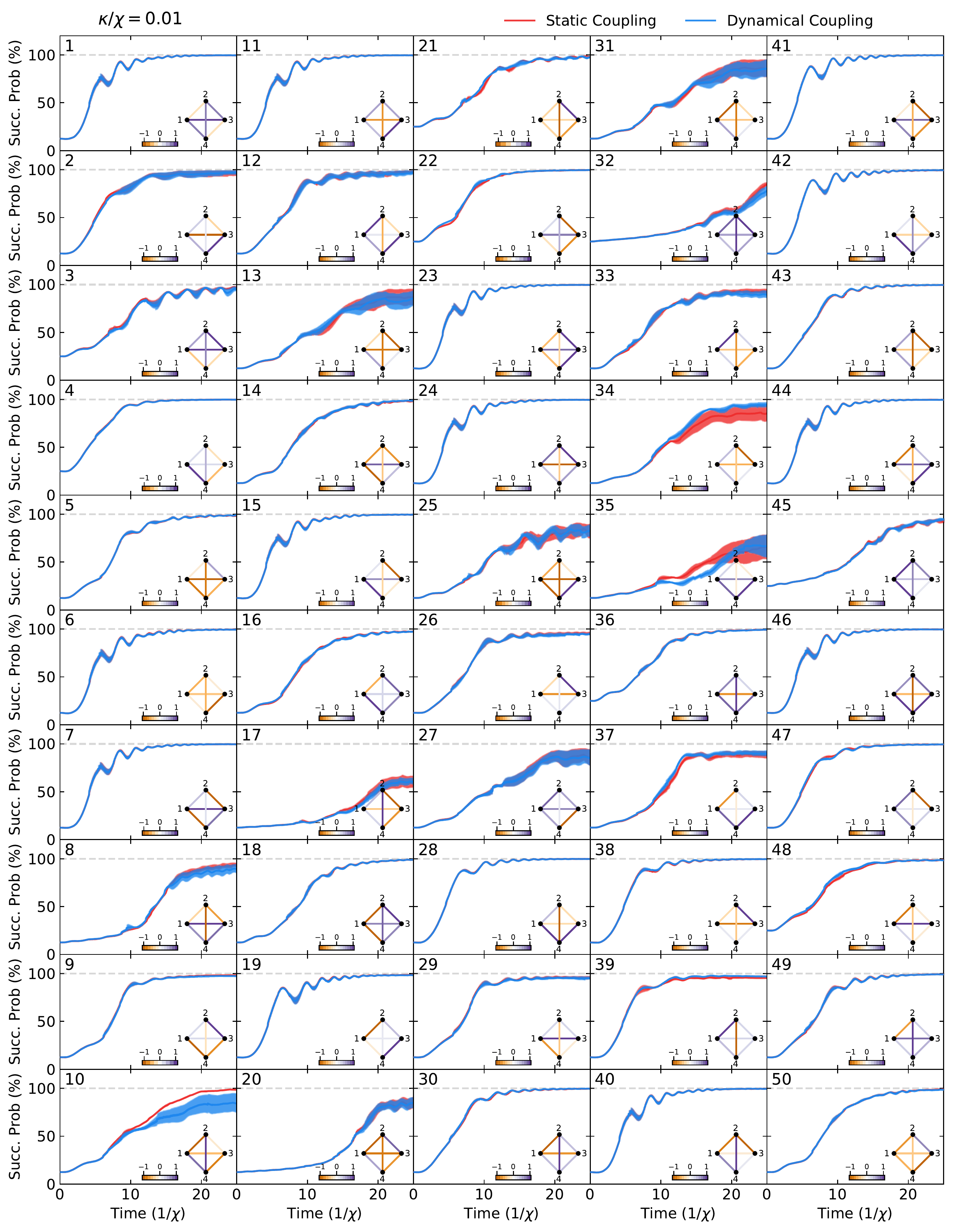}
    \caption{ \textbf{Quantum simulation of success probability, low-loss case (part 1/2)}. Evolution of the success probability $\text{P}_\text{success}(t)$ of finding a ground state of the inset Ising problem instance (one for each sub-axis), assuming the cavity field decay rate $\kappa = 0.01\chi$. In red is shown the success probability of a statically-coupled system implementing the Ising couplings exactly, while in blue is shown the success probability of the corresponding dynamically-coupled system. The system parameters are chosen according to the prescription described in Section~\ref{sec:parameter-summary} for each Ising problem instance. See Section~\ref{sec:quantum} for additional details of our quantum simulation methodology. Due to the stochastic nature of the simulation, $N_\text{traj} = 10$ trajectories are run for each problem instance; the solid line indicates the mean over the trajectories, while the fill indicates the standard error of the mean, as described in Section~\ref{sec:quantum}. }
    \label{fig:gs_lowloss_1}
\end{figure}

\begin{figure}
    \includegraphics[width=0.9\linewidth]{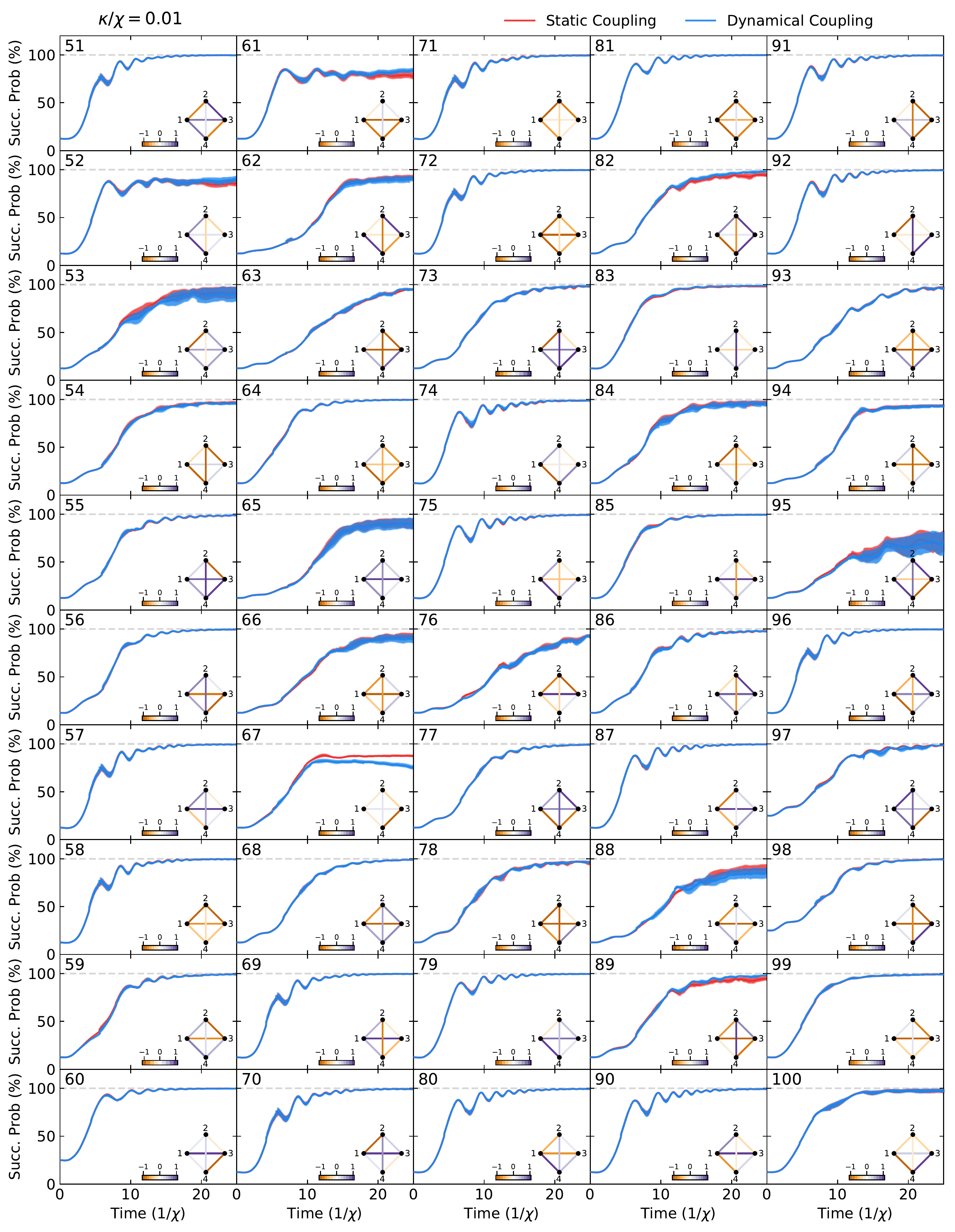}
    \label{fig:gs_lowloss_2}
    \caption{ \textbf{Quantum simulation of success probability, low-loss case (part 2/2)}. Evolution of the success probability $\text{P}_\text{success}(t)$ of finding a ground state of the inset Ising problem instance (one for each sub-axis), assuming the cavity field decay rate $\kappa = 0.01\chi$. In red is shown the success probability of a statically-coupled system implementing the Ising couplings exactly, while in blue is shown the success probability of the corresponding dynamically-coupled system. The system parameters are chosen according to the prescription described in Section~\ref{sec:parameter-summary} for each Ising problem instance. See Section~\ref{sec:quantum} for additional details of our quantum simulation methodology. Due to the stochastic nature of the simulation, $N_\text{traj} = 10$ trajectories are run for each problem instance; the solid line indicates the mean over the trajectories, while the fill indicates the standard error of the mean, as described in Section~\ref{sec:quantum}. }
\end{figure}

\begin{figure}
    \includegraphics[width=0.9\linewidth]{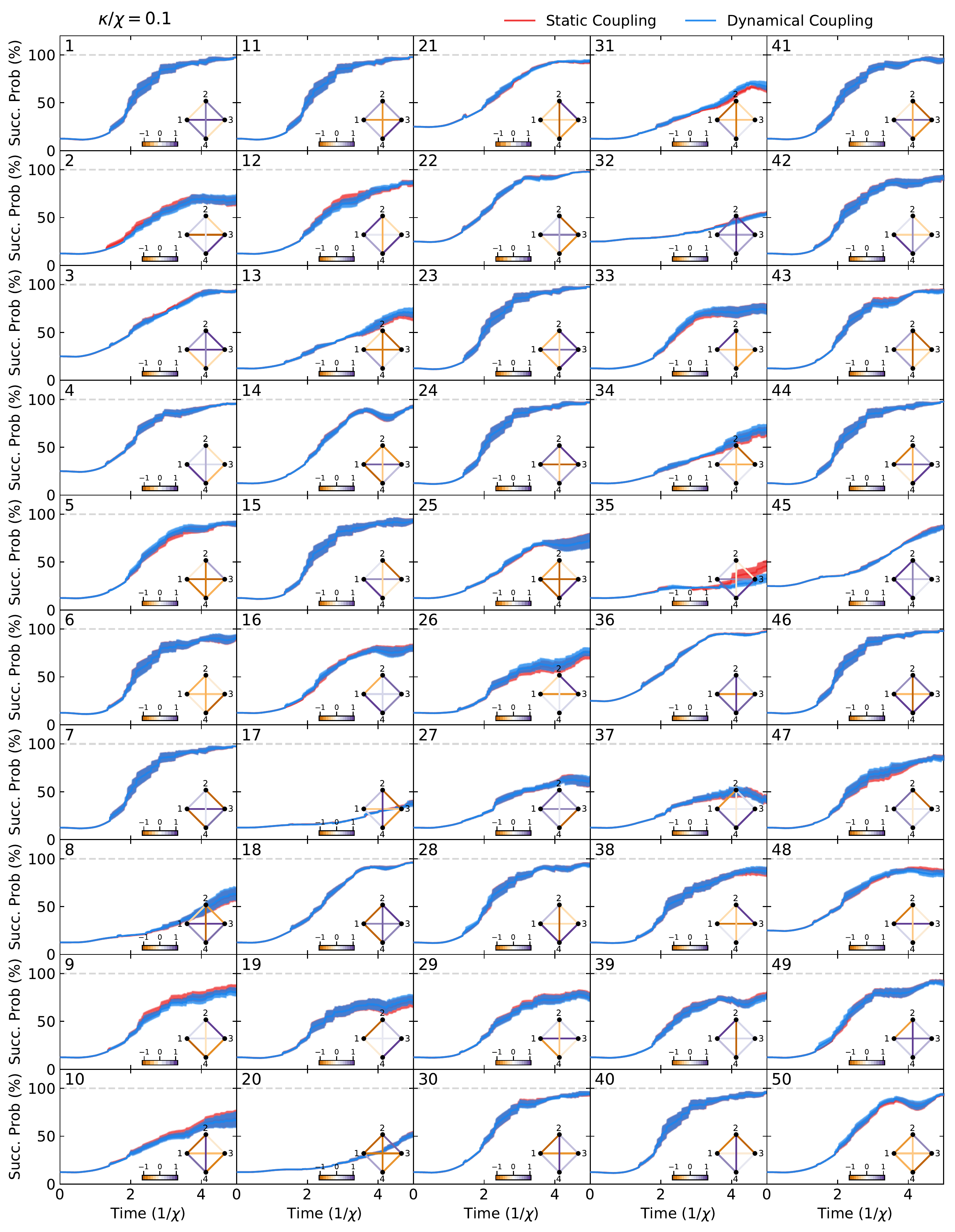}
    \caption{ \textbf{Quantum simulation of success probability, high-loss case (part 1/2)}. Evolution of the success probability $\text{P}_\text{success}(t)$ of finding a ground state of the inset Ising problem instance (one for each sub-axis), assuming the cavity field decay rate $\kappa = 0.1\chi$. In red is shown the success probability of a statically-coupled system implementing the Ising couplings exactly, while in blue is shown the success probability of the corresponding dynamically-coupled system. The system parameters are chosen according to the prescription described in Section~\ref{sec:parameter-summary} for each Ising problem instance. See Section~\ref{sec:quantum} for additional details of our quantum simulation methodology. Due to the stochastic nature of the simulation, $N_\text{traj} = 10$ trajectories are run for each problem instance; the solid line indicates the mean over the trajectories, while the fill indicates the standard error of the mean, as described in Section~\ref{sec:quantum}. }
    \label{fig:gs_highloss_1}
\end{figure}

\begin{figure}
    \includegraphics[width=0.9\linewidth]{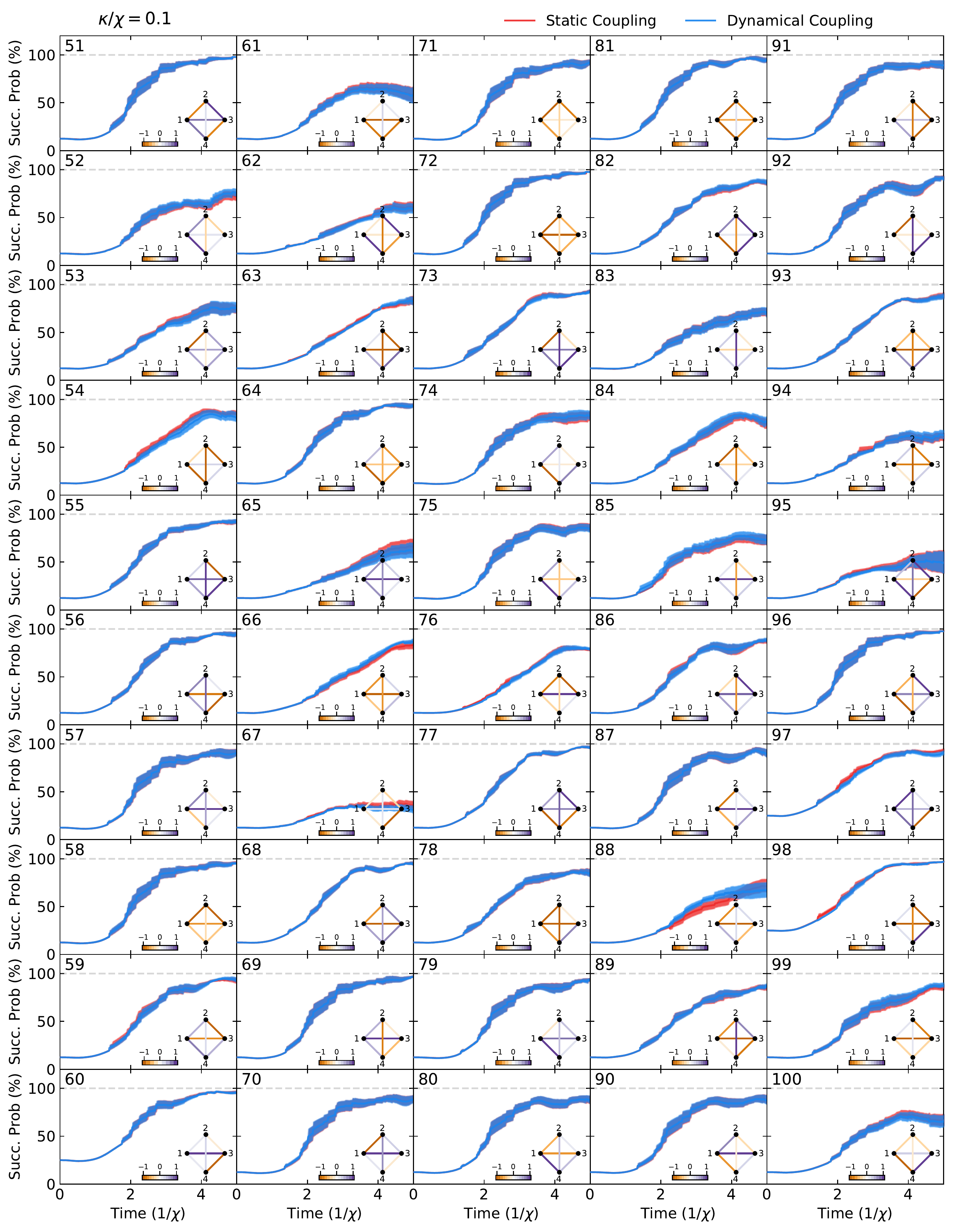}
    \caption{ \textbf{Quantum simulation of success probability, high-loss case (part 2/2)}. Evolution of the success probability $\text{P}_\text{success}(t)$ of finding a ground state of the inset Ising problem instance (one for each sub-axis), assuming the cavity field decay rate $\kappa = 0.1\chi$. In red is shown the success probability of a statically-coupled system implementing the Ising couplings exactly, while in blue is shown the success probability of the corresponding dynamically-coupled system. The system parameters are chosen according to the prescription described in Section~\ref{sec:parameter-summary} for each Ising problem instance. See Section~\ref{sec:quantum} for additional details of our quantum simulation methodology. Due to the stochastic nature of the simulation, $N_\text{traj} = 10$ trajectories are run for each problem instance; the solid line indicates the mean over the trajectories, while the fill indicates the standard error of the mean, as described in Section~\ref{sec:quantum}. }
    \label{fig:gs_highloss_2}
\end{figure}

\begin{figure}
    \includegraphics[width=0.9\linewidth]{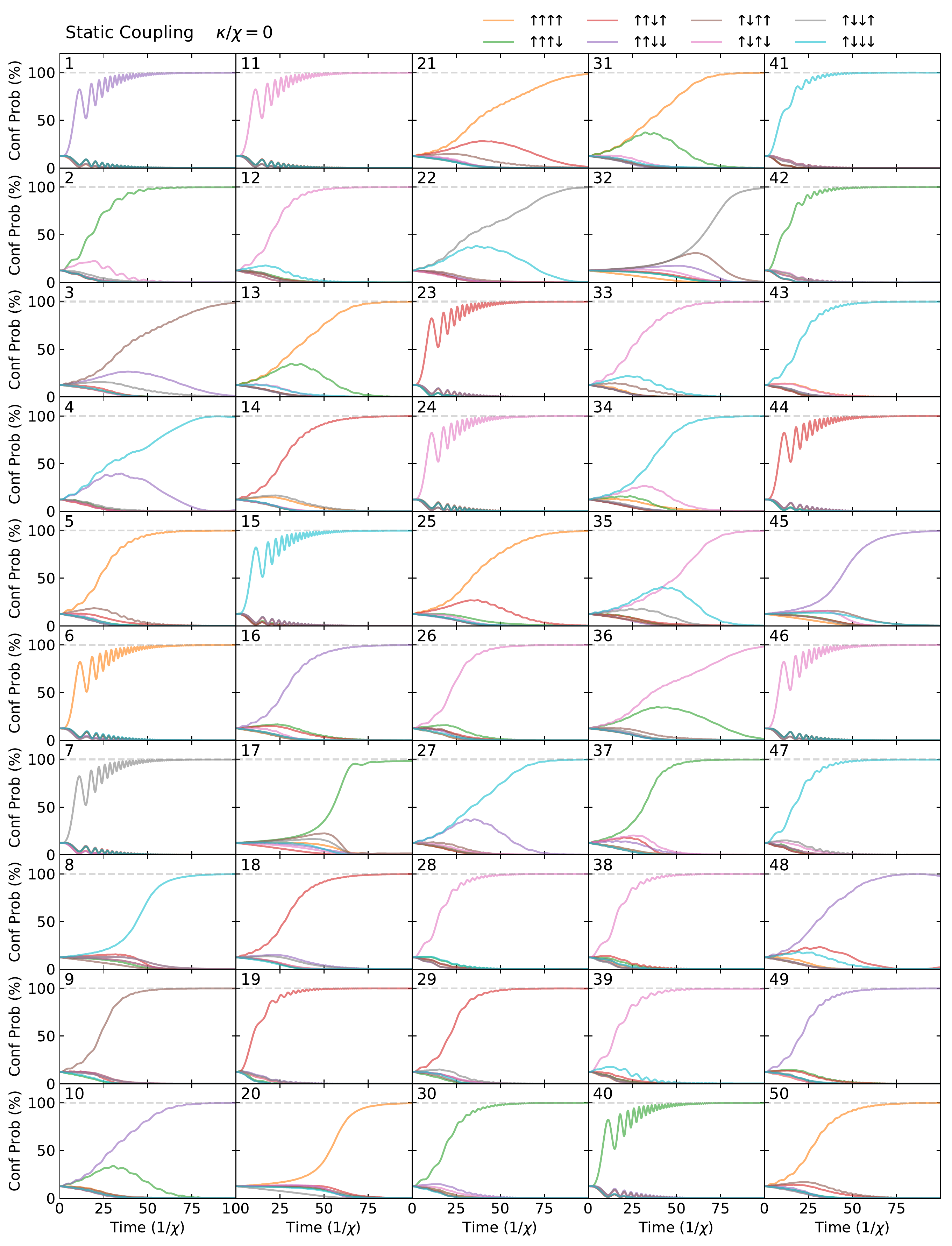}
    \caption{ \textbf{Configuration probabilities of statically-coupled system, lossless case (part 1/2)}. Evolution of the probabilities $\text{P}_{\sigma}(t)$ of obtaining each of the possible spin configurations in the statically-coupled system as a function of the evolution time, for the problem instances depicted inset in the corresponding sub-axis (see upper-left corner) of Figure~\ref{fig:gs_noloss_1}. The configuration probability is defined according to \ref{eq:prob-conf} in Section~\ref{sec:quantum}. Note that due to the symmetries of the system, the configuration probability is exactly the same upon flipping every spin, so we take the convention that each curve represents the sum of the probabilities for the two degenerate configurations. The system parameters are chosen according to the prescription described in Section~\ref{sec:parameter-summary} for each Ising problem instance. See Section~\ref{sec:quantum} for additional details of our quantum simulation methodology. }
    \label{fig:staconf_noloss_1}
\end{figure}

In Figures~\ref{fig:gs_noloss_1} through~\ref{fig:gs_highloss_2}, we compare the success probability $\text{P}_\text{success}(t)$ of the statically- and dynamically-coupled systems for each of the 100 problem instances we used in our quantum simulations. These figures investigate, respectively, the zero-loss, low-loss, and high-loss cases, indicating that, in most cases, the dynamical-coupling scheme produces the desired evolution, at least in distribution. Corresponding to each of the plots, Figures~\ref{fig:staconf_noloss_1} through~\ref{fig:staconf_highloss_2} show, for the statically-coupled system, the configuration probabilities $\text{P}_\sigma(t)$ for each of the spin configurations $\sigma$, while Figures~\ref{fig:dynconf_noloss_1} through~\ref{fig:dynconf_highloss_2} show the corresponding configuration probabilities for the dynamically-coupled system. Again, we see that the trajectories match well, at least in distribution.

Finally, in Figure~\ref{fig:classical-hist}, we show the same correlation/histogram plots used in the bottom of Figure~3C of the main manuscript, for all the problem sizes that we have numerically simulated. Note that the projected histograms correspond to the histogram shown in Figure~3B of the main manuscript. The data shows good correspondence in the success probability between the dynamically- and statically-coupled systems for all problem sizes considered.

\begin{figure}
    \includegraphics[width=0.9\linewidth]{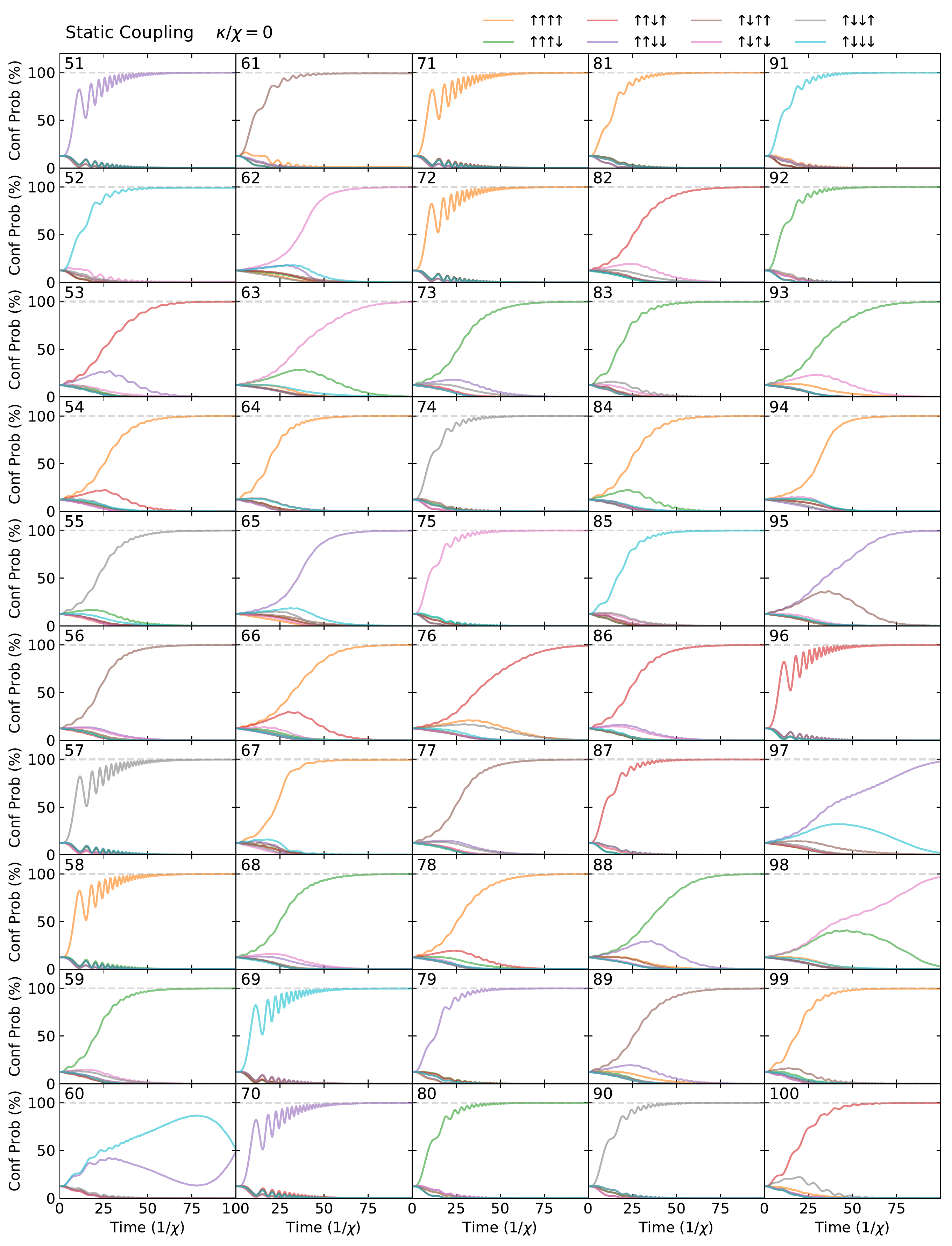}
    \caption{ \textbf{Configuration probabilities of statically-coupled system, lossless case (part 2/2)}. Evolution of the probabilities $\text{P}_{\sigma}(t)$ of obtaining each of the possible spin configurations in the statically-coupled system as a function of the evolution time, for the problem instances depicted inset in the corresponding sub-axis (see upper-left corner) of Figure~\ref{fig:gs_noloss_2}. The configuration probability is defined according to \ref{eq:prob-conf} in Section~\ref{sec:quantum}. Note that due to the symmetries of the system, the configuration probability is exactly the same upon flipping every spin, so we take the convention that each curve represents the sum of the probabilities for the two degenerate configurations. The system parameters are chosen according to the prescription described in Section~\ref{sec:parameter-summary} for each Ising problem instance. See Section~\ref{sec:quantum} for additional details of our quantum simulation methodology. }
\end{figure}

\begin{figure}
    \includegraphics[width=0.9\linewidth]{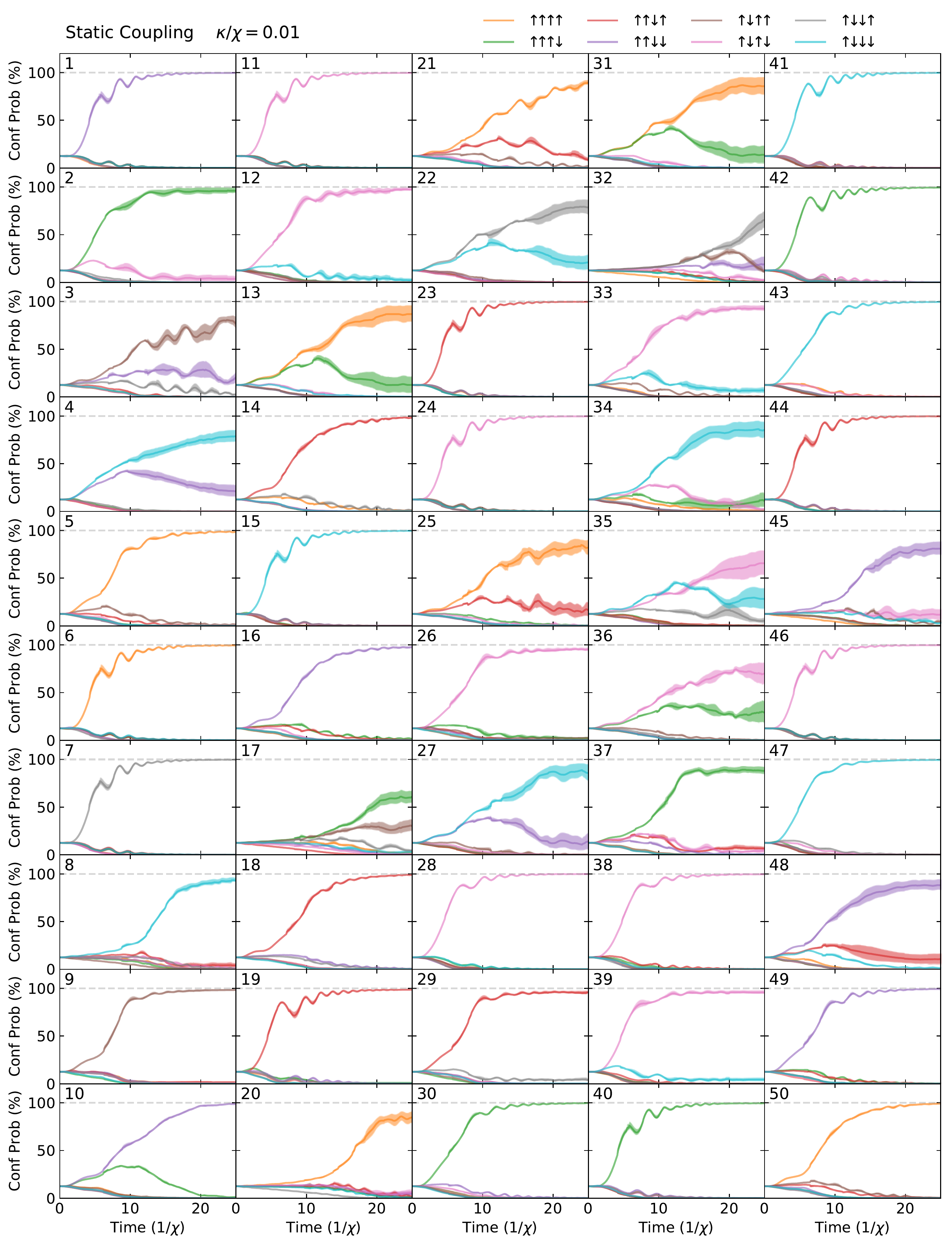}
    \caption{ \textbf{Configuration probabilities of statically-coupled system, low-loss case (part 1/2)}. Evolution of the probabilities $\text{P}_{\sigma}(t)$ of obtaining each of the possible spin configurations in the statically-coupled system as a function of the evolution time, for the problem instances depicted inset in the corresponding sub-axis (see upper-left corner) of Figure~\ref{fig:gs_noloss_1}. The configuration probability is defined according to \ref{eq:prob-conf} in Section~\ref{sec:quantum}. Note that due to the symmetries of the system, the configuration probability is exactly the same upon flipping every spin, so we take the convention that each curve represents the sum of the probabilities for the two degenerate configurations. The system parameters are chosen according to the prescription described in Section~\ref{sec:parameter-summary} for each Ising problem instance. See Section~\ref{sec:quantum} for additional details of our quantum simulation methodology. Due to the stochastic nature of the simulation, $N_\text{traj} = 10$ trajectories are run for each problem instance; the solid line indicates the mean over the trajectories, while the fill indicates the standard error of the mean, as described in Section~\ref{sec:quantum}.
    \label{fig:staconf_lowloss_1}
    }
\end{figure}

\begin{figure}
    \includegraphics[width=0.9\linewidth]{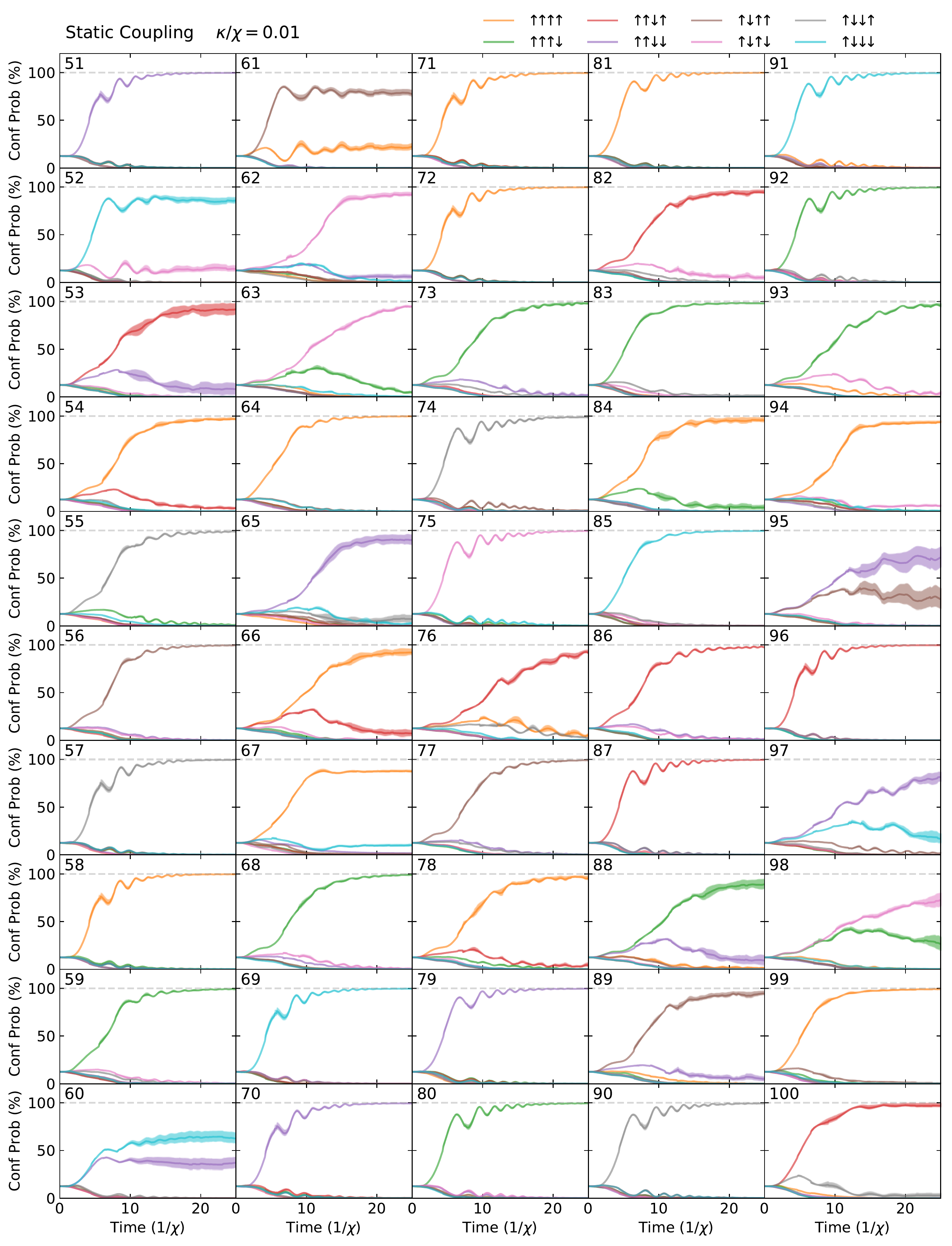}
    \caption{ \textbf{Configuration probabilities of statically-coupled system, low-loss case (part 2/2)}. Evolution of the probabilities $\text{P}_{\sigma}(t)$ of obtaining each of the possible spin configurations in the statically-coupled system as a function of the evolution time, for the problem instances depicted inset in the corresponding sub-axis (see upper-left corner) of Figure~\ref{fig:gs_noloss_2}. The configuration probability is defined according to \ref{eq:prob-conf} in Section~\ref{sec:quantum}. Note that due to the symmetries of the system, the configuration probability is exactly the same upon flipping every spin, so we take the convention that each curve represents the sum of the probabilities for the two degenerate configurations. The system parameters are chosen according to the prescription described in Section~\ref{sec:parameter-summary} for each Ising problem instance. See Section~\ref{sec:quantum} for additional details of our quantum simulation methodology. Due to the stochastic nature of the simulation, $N_\text{traj} = 10$ trajectories are run for each problem instance; the solid line indicates the mean over the trajectories, while the fill indicates the standard error of the mean, as described in Section~\ref{sec:quantum}. }
\end{figure}

\begin{figure}
    \includegraphics[width=0.9\linewidth]{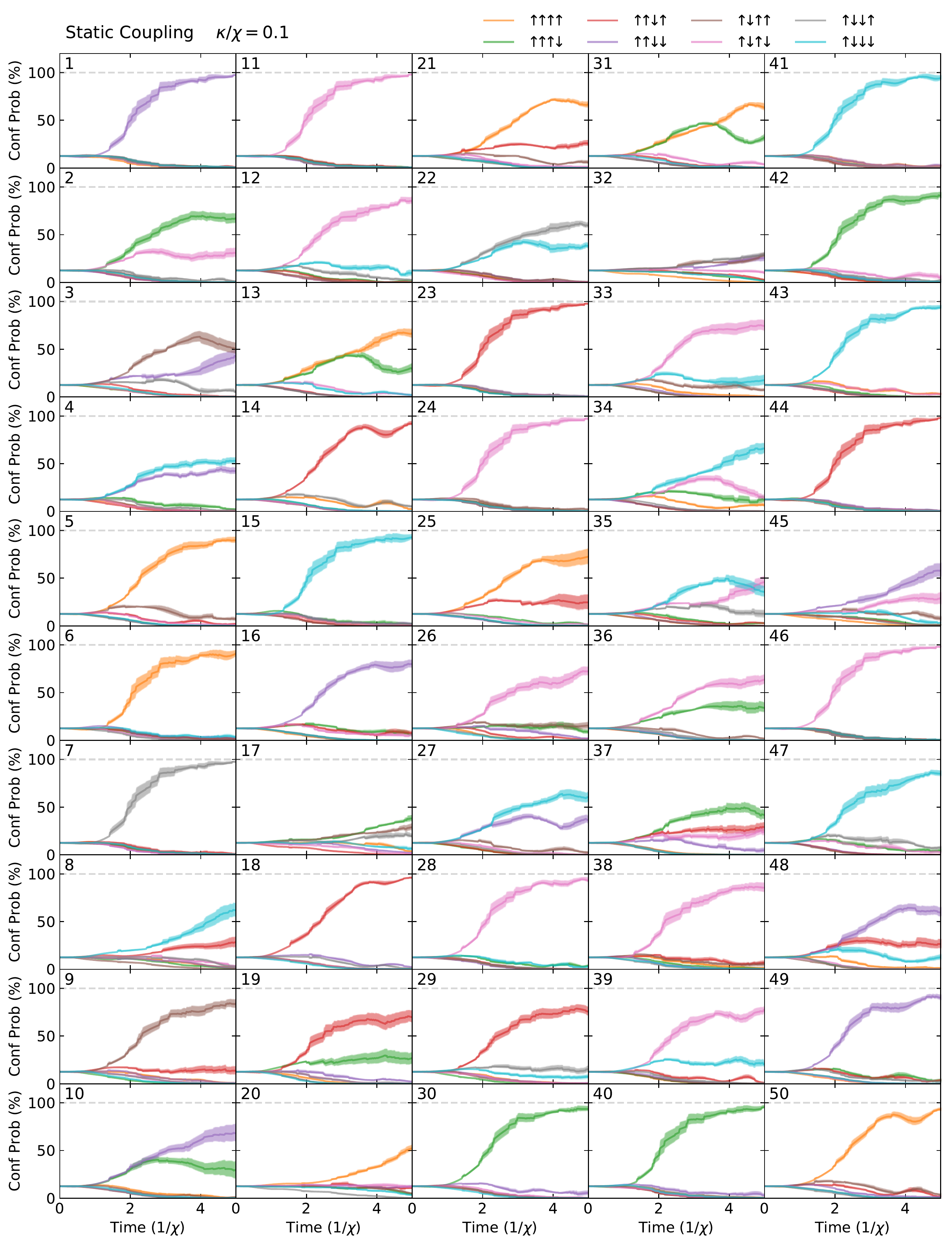}
    \caption{ \textbf{Configuration probabilities of statically-coupled system, high-loss case (part 1/2)}. Evolution of the probabilities $(\text{P}_{\sigma}(t))$ of obtaining each of the possible spin configurations in the statically-coupled system as a function of the evolution time, for the problem instances depicted inset in the corresponding sub-axis (see upper-left corner) of Figure~\ref{fig:gs_noloss_1}. The configuration probability is defined according to \eqref{eq:prob-conf} in Section~\ref{sec:quantum}. Note that due to the symmetries of the system, the configuration probability is exactly the same upon flipping every spin, so we take the convention that each curve represents the sum of the probabilities for the two degenerate configurations. The system parameters are chosen according to the prescription described in Section~\ref{sec:parameter-summary} for each Ising problem instance. See Section~\ref{sec:quantum} for additional details of our quantum simulation methodology. Due to the stochastic nature of the simulation, $N_\text{traj} = 10$ trajectories are run for each problem instance; the solid line indicates the mean over the trajectories, while the fill indicates the standard error of the mean, as described in Section~\ref{sec:quantum}. }
\end{figure}

\begin{figure}
    \includegraphics[width=0.9\linewidth]{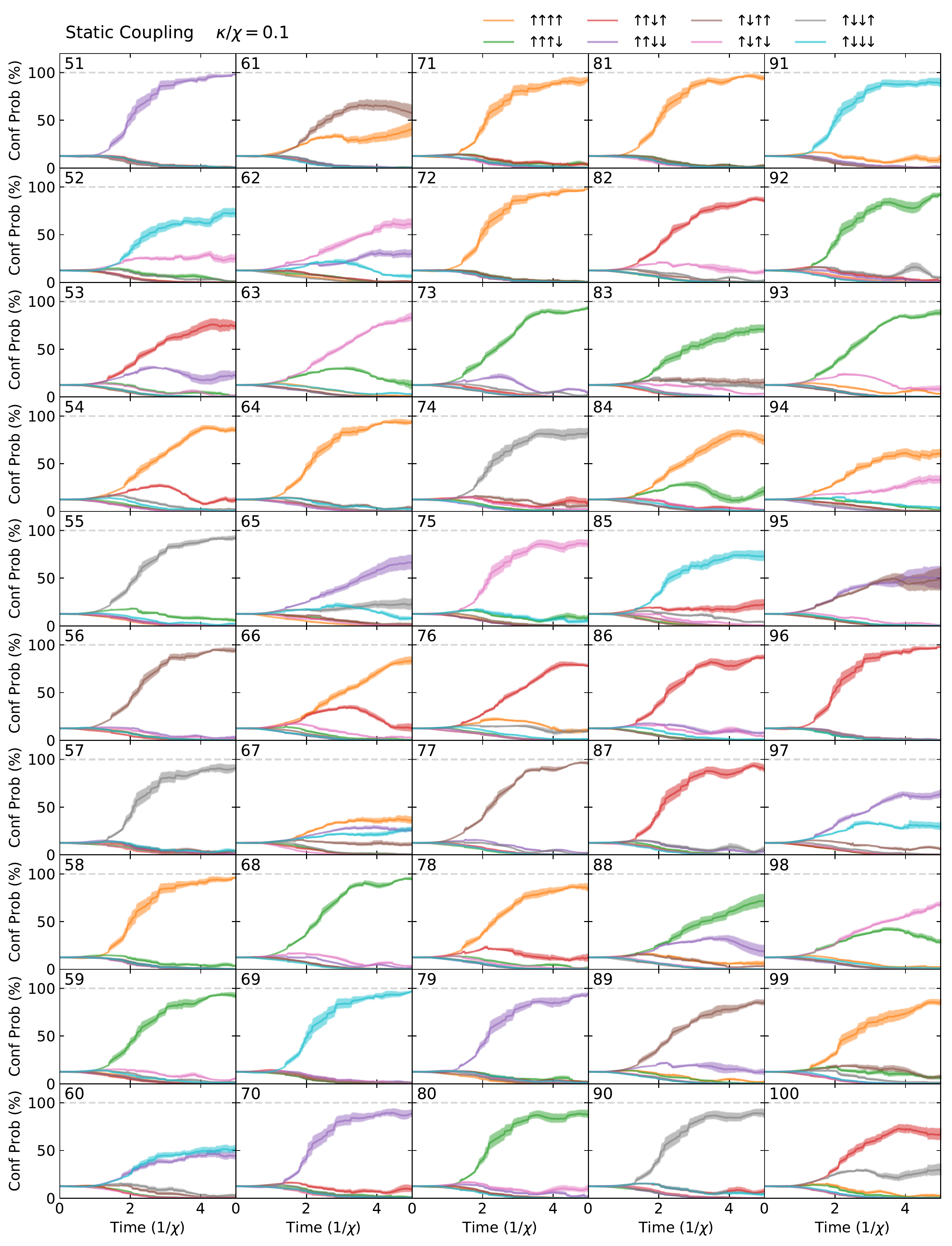}
    \caption{ \textbf{Configuration probabilities of statically-coupled system, high-loss case (part 2/2)}. Evolution of the probabilities $(\text{P}_{\sigma}(t))$ of obtaining each of the possible spin configurations in the statically-coupled system as a function of the evolution time, for the problem instances depicted inset in the corresponding sub-axis (see upper-left corner) of Figure~\ref{fig:gs_noloss_2}. The configuration probability is defined according to \eqref{eq:prob-conf} in Section~\ref{sec:quantum}. Note that due to the symmetries of the system, the configuration probability is exactly the same upon flipping every spin, so we take the convention that each curve represents the sum of the probabilities for the two degenerate configurations. The system parameters are chosen according to the prescription described in Section~\ref{sec:parameter-summary} for each Ising problem instance. See Section~\ref{sec:quantum} for additional details of our quantum simulation methodology. Due to the stochastic nature of the simulation, $N_\text{traj} = 10$ trajectories are run for each problem instance; the solid line indicates the mean over the trajectories, while the fill indicates the standard error of the mean, as described in Section~\ref{sec:quantum}. }
    \label{fig:staconf_highloss_2}
\end{figure}

\begin{figure}
    \includegraphics[width=0.9\linewidth]{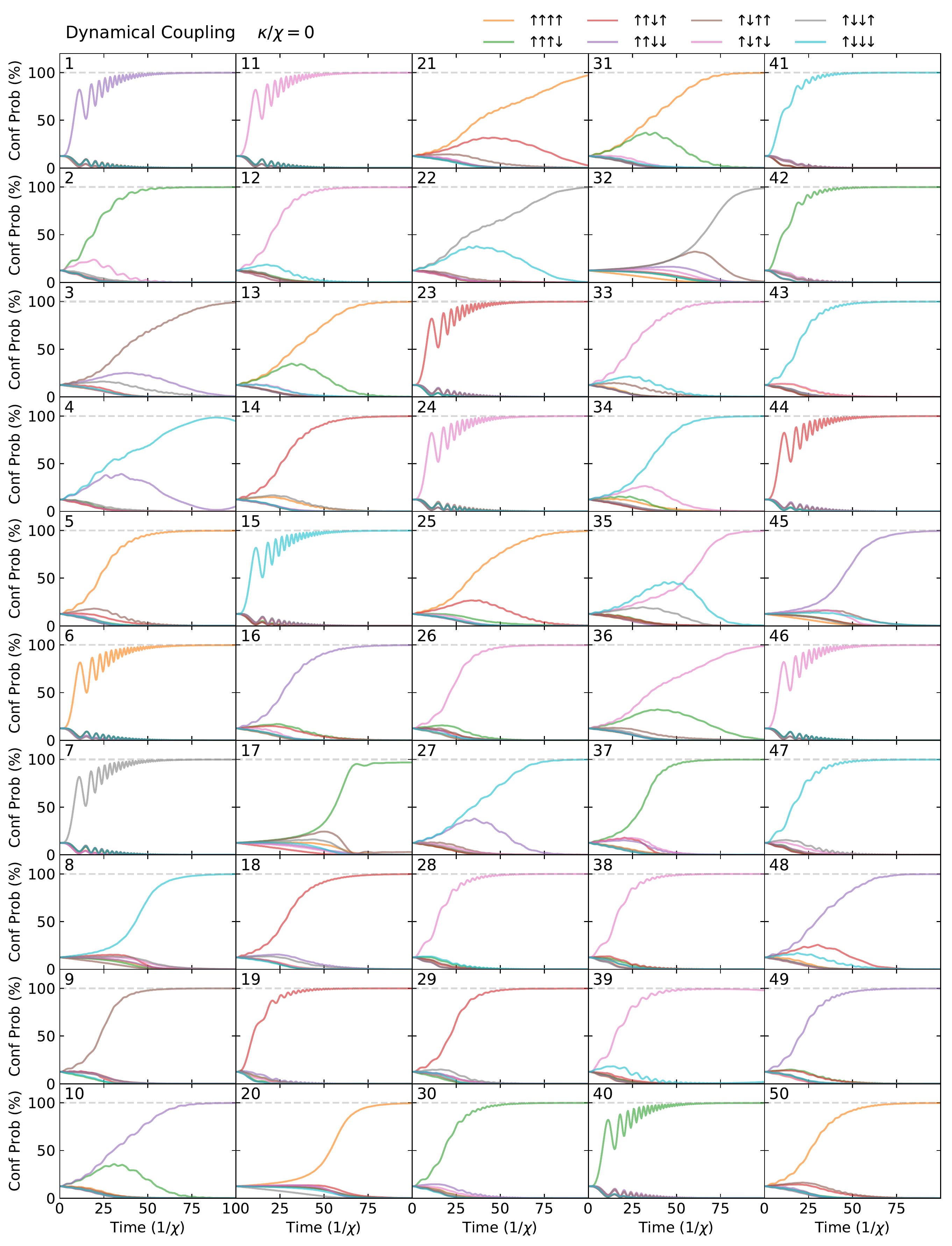}
    \caption{ \textbf{Configuration probabilities of dynamically-coupled system, lossless case (part 1/2)}. Evolution of the probabilities $\text{P}_{\sigma}(t)$ of obtaining each of the possible spin configurations in the dynamically-coupled system as a function of the evolution time, for the problem instances depicted inset in the corresponding sub-axis (see upper-left corner) of Figure~\ref{fig:gs_noloss_1}. The configuration probability is defined according to \eqref{eq:prob-conf} in Section~\ref{sec:quantum}. Note that due to the symmetries of the system, the configuration probability is exactly the same upon flipping every spin, so we take the convention that each curve represents the sum of the probabilities for the two degenerate configurations. The system parameters are chosen according to the prescription described in Section~\ref{sec:parameter-summary} for each Ising problem instance. See Section~\ref{sec:quantum} for additional details of our quantum simulation methodology. }
    \label{fig:dynconf_noloss_1}
\end{figure}

\begin{figure}
    \includegraphics[width=0.9\linewidth]{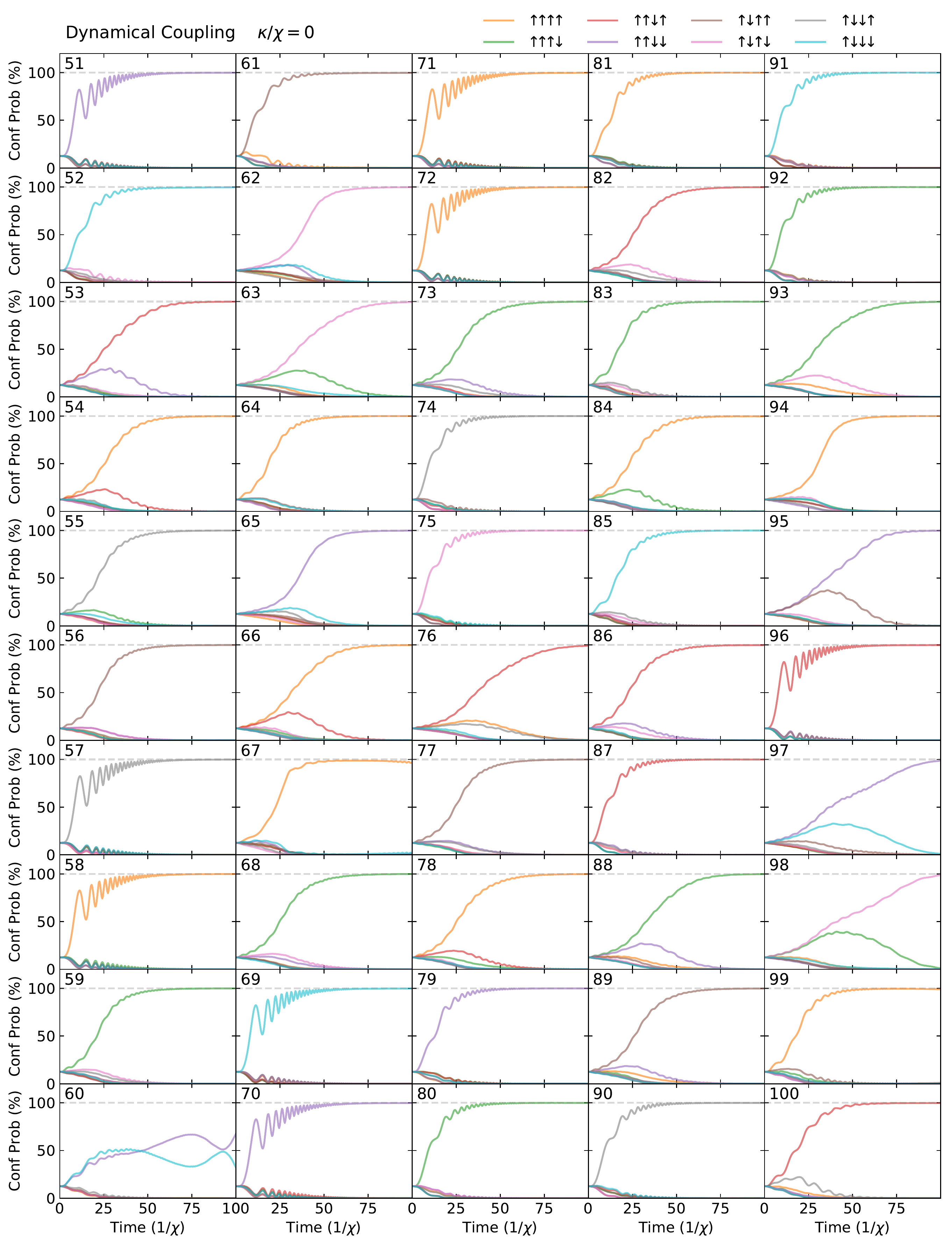}
    \caption{ \textbf{Configuration probabilities of dynamically-coupled system, lossless case (part 2/2)}. Evolution of the probabilities $\text{P}_{\sigma}(t)$ of obtaining each of the possible spin configurations in the dynamically-coupled system as a function of the evolution time, for the problem instances depicted inset in the corresponding sub-axis (see upper-left corner) of Figure~\ref{fig:gs_noloss_2}. The configuration probability is defined according to \eqref{eq:prob-conf} in Section~\ref{sec:quantum}. Note that due to the symmetries of the system, the configuration probability is exactly the same upon flipping every spin, so we take the convention that each curve represents the sum of the probabilities for the two degenerate configurations. The system parameters are chosen according to the prescription described in Section~\ref{sec:parameter-summary} for each Ising problem instance. See Section~\ref{sec:quantum} for additional details of our quantum simulation methodology. }
\end{figure}

\begin{figure}
    \includegraphics[width=0.9\linewidth]{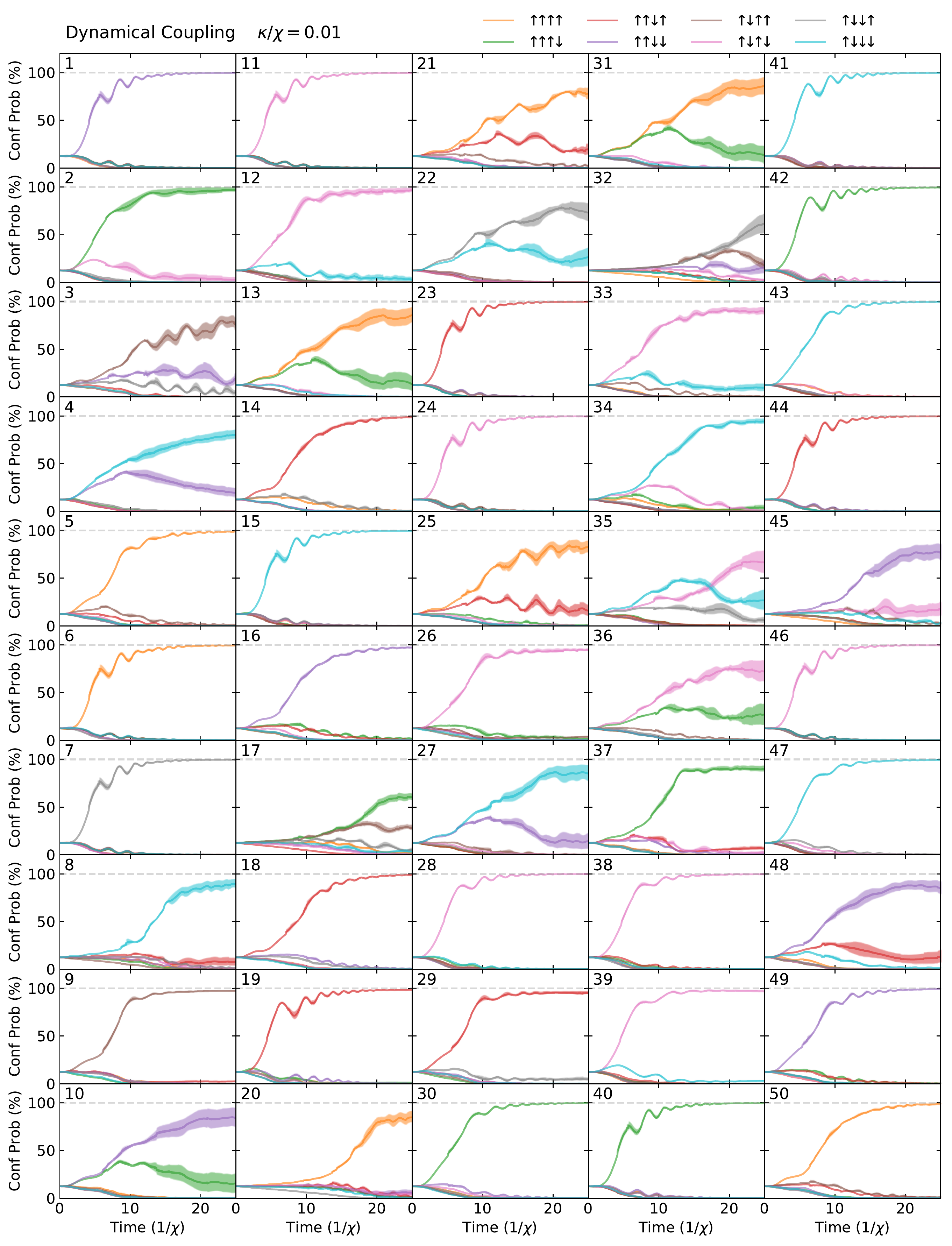}
    \caption{ \textbf{Configuration probabilities of dynamically-coupled system, low-loss case (part 1/2)}. Evolution of the probabilities $\text{P}_{\sigma}(t)$ of obtaining each of the possible spin configurations in the dynamically-coupled system as a function of the evolution time, for the problem instances depicted inset in the corresponding sub-axis (see upper-left corner) of Figure~\ref{fig:gs_noloss_1}. The configuration probability is defined according to \eqref{eq:prob-conf} in Section~\ref{sec:quantum}. Note that due to the symmetries of the system, the configuration probability is exactly the same upon flipping every spin, so we take the convention that each curve represents the sum of the probabilities for the two degenerate configurations. The system parameters are chosen according to the prescription described in Section~\ref{sec:parameter-summary} for each Ising problem instance. See Section~\ref{sec:quantum} for additional details of our quantum simulation methodology. Due to the stochastic nature of the simulation, $N_\text{traj} = 10$ trajectories are run for each problem instance; the solid line indicates the mean over the trajectories, while the fill indicates the standard error of the mean, as described in Section~\ref{sec:quantum}. }
    \label{fig:dynconf_lowloss_1}
\end{figure}

\begin{figure}
    \includegraphics[width=0.9\linewidth]{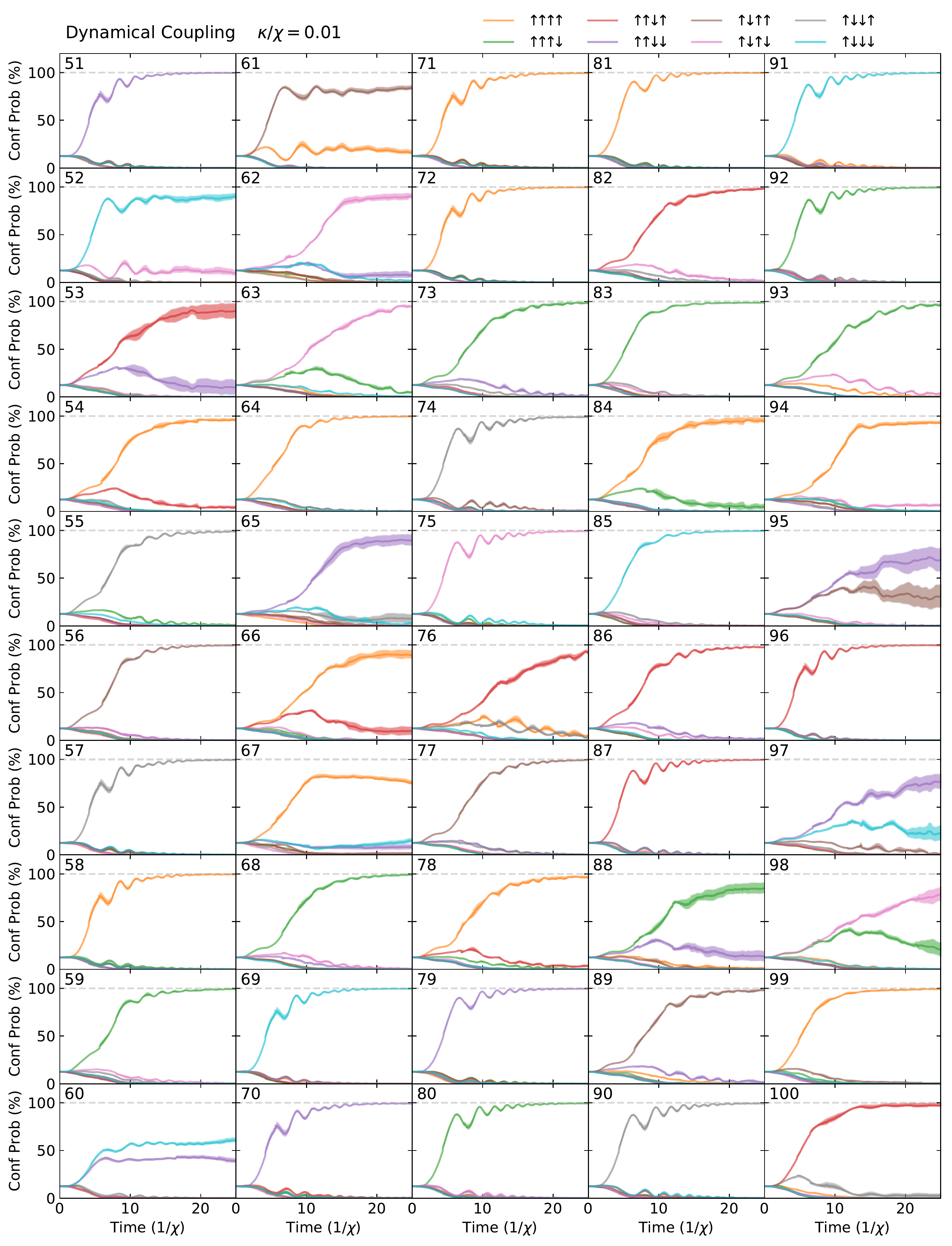}
    \caption{ \textbf{Configuration probabilities of dynamically-coupled system, low-loss case (part 2/2)}. Evolution of the probabilities $\text{P}_{\sigma}(t)$ of obtaining each of the possible spin configurations in the dynamically-coupled system as a function of the evolution time, for the problem instances depicted inset in the corresponding sub-axis (see upper-left corner) of Figure~\ref{fig:gs_noloss_2}. The configuration probability is defined according to \eqref{eq:prob-conf} in Section~\ref{sec:quantum}. Note that due to the symmetries of the system, the configuration probability is exactly the same upon flipping every spin, so we take the convention that each curve represents the sum of the probabilities for the two degenerate configurations. The system parameters are chosen according to the prescription described in Section~\ref{sec:parameter-summary} for each Ising problem instance. See Section~\ref{sec:quantum} for additional details of our quantum simulation methodology. Due to the stochastic nature of the simulation, $N_\text{traj} = 10$ trajectories are run for each problem instance; the solid line indicates the mean over the trajectories, while the fill indicates the standard error of the mean, as described in Section~\ref{sec:quantum}. }
\end{figure}

\begin{figure}
    \includegraphics[width=0.9\linewidth]{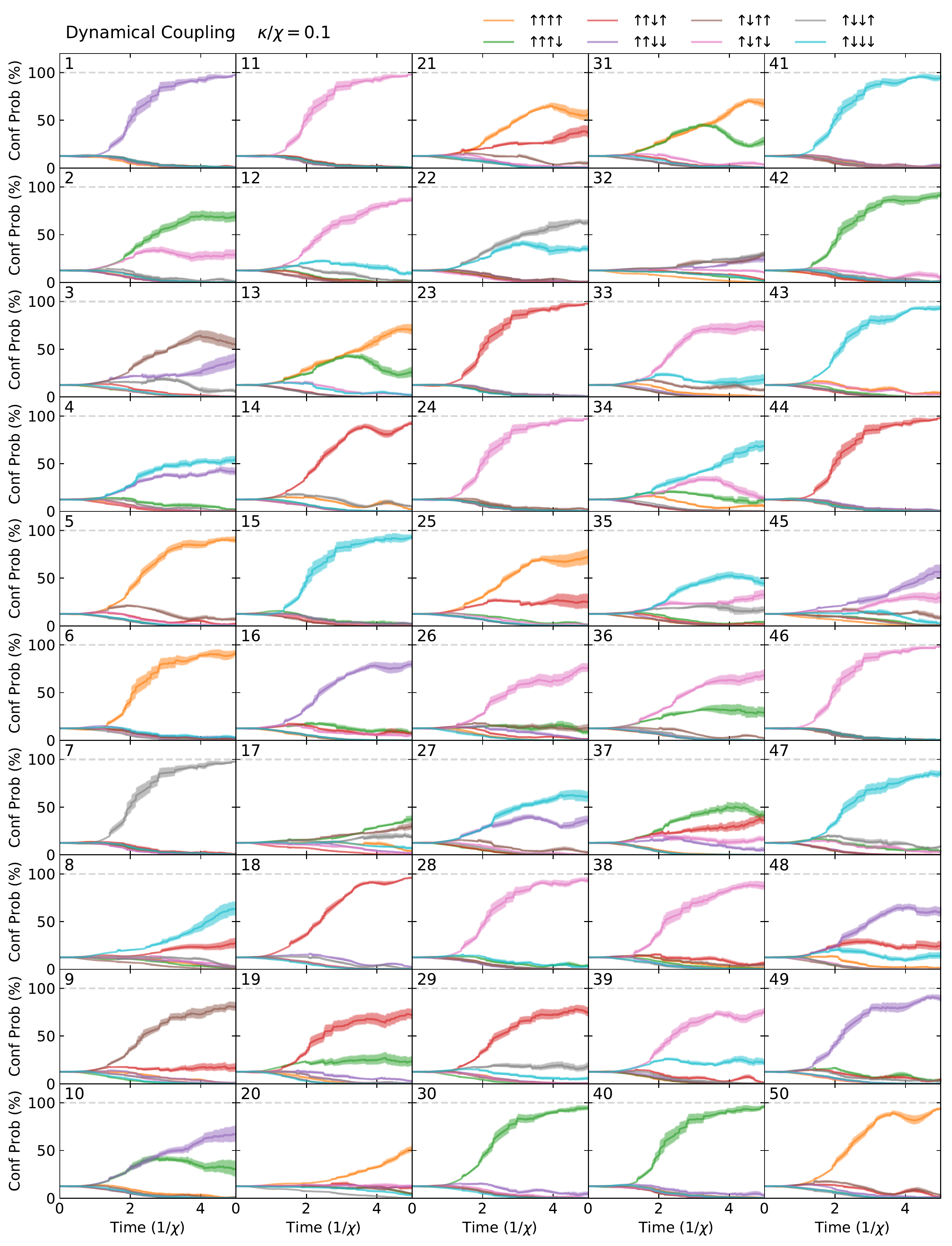}
    \caption{ \textbf{Configuration probabilities of dynamically-coupled system, high-loss case (part 1/2)}. Evolution of the probabilities $\text{P}_{\sigma}(t)$ of obtaining each of the possible spin configurations in the dynamically-coupled system as a function of the evolution time, for the problem instances depicted inset in the corresponding sub-axis (see upper-left corner) of Figure~\ref{fig:gs_noloss_1}. The configuration probability is defined according to \eqref{eq:prob-conf} in Section~\ref{sec:quantum}. Note that due to the symmetries of the system, the configuration probability is exactly the same upon flipping every spin, so we take the convention that each curve represents the sum of the probabilities for the two degenerate configurations. The system parameters are chosen according to the prescription described in Section~\ref{sec:parameter-summary} for each Ising problem instance. See Section~\ref{sec:quantum} for additional details of our quantum simulation methodology. Due to the stochastic nature of the simulation, $N_\text{traj} = 10$ trajectories are run for each problem instance; the solid line indicates the mean over the trajectories, while the fill indicates the standard error of the mean, as described in Section~\ref{sec:quantum}. }
\end{figure}

\begin{figure}
    \includegraphics[width=0.9\linewidth]{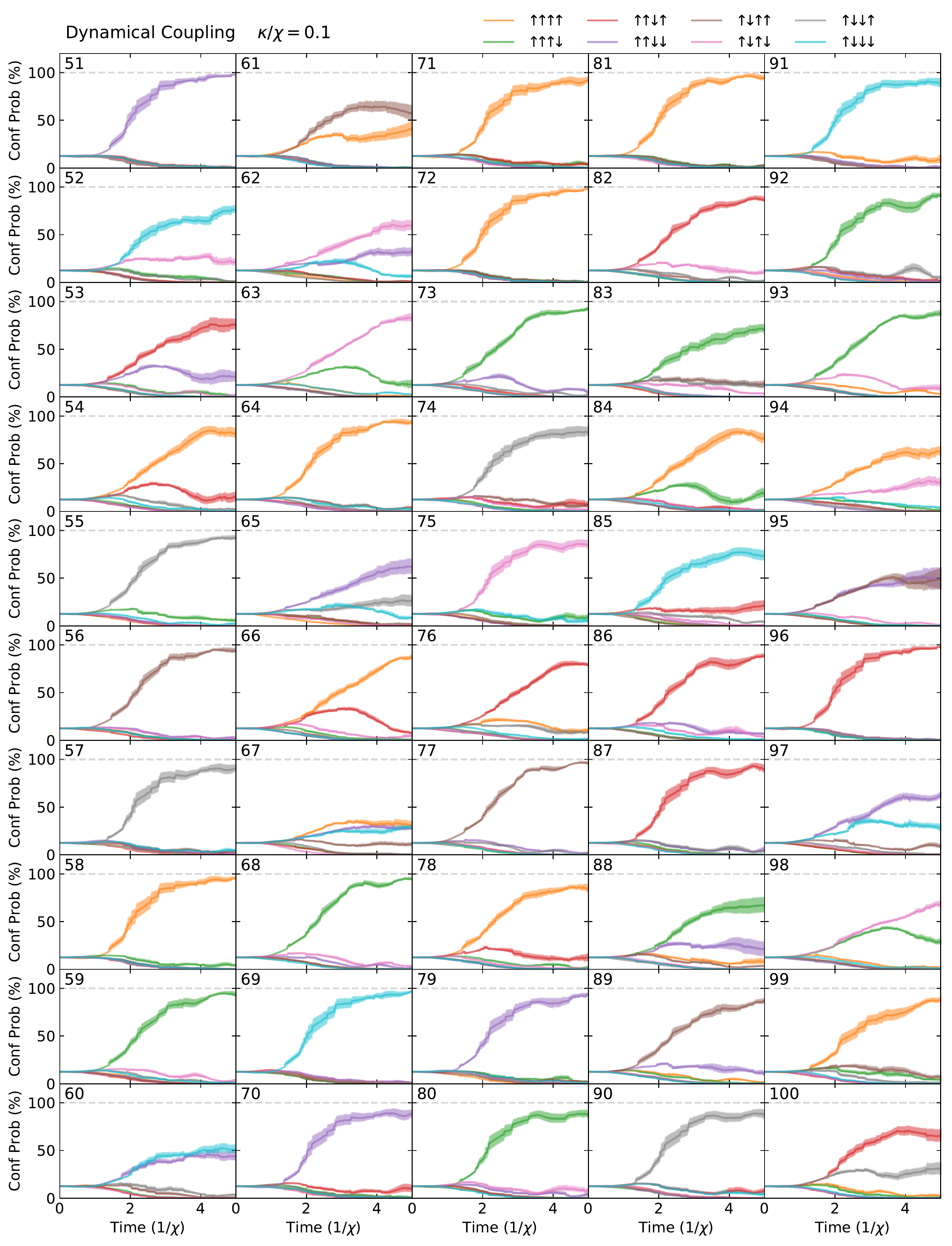}
    \caption{ \textbf{Configuration probabilities of dynamically-coupled system, high-loss case (part 2/2)}. Evolution of the probabilities $\text{P}_{\sigma}(t)$ of obtaining each of the possible spin configurations in the dynamically-coupled system as a function of the evolution time, for the problem instances depicted inset in the corresponding sub-axis (see upper-left corner) of Figure~\ref{fig:gs_noloss_2}. The configuration probability is defined according to \eqref{eq:prob-conf} in Section~\ref{sec:quantum}. Note that due to the symmetries of the system, the configuration probability is exactly the same upon flipping every spin, so we take the convention that each curve represents the sum of the probabilities for the two degenerate configurations. The system parameters are chosen according to the prescription described in Section~\ref{sec:parameter-summary} for each Ising problem instance. See Section~\ref{sec:quantum} for additional details of our quantum simulation methodology. Due to the stochastic nature of the simulation, $N_\text{traj} = 10$ trajectories are run for each problem instance; the solid line indicates the mean over the trajectories, while the fill indicates the standard error of the mean, as described in Section~\ref{sec:quantum}. }
    \label{fig:dynconf_highloss_2}
\end{figure}

\begin{figure}
    \includegraphics[width=1.0\linewidth]{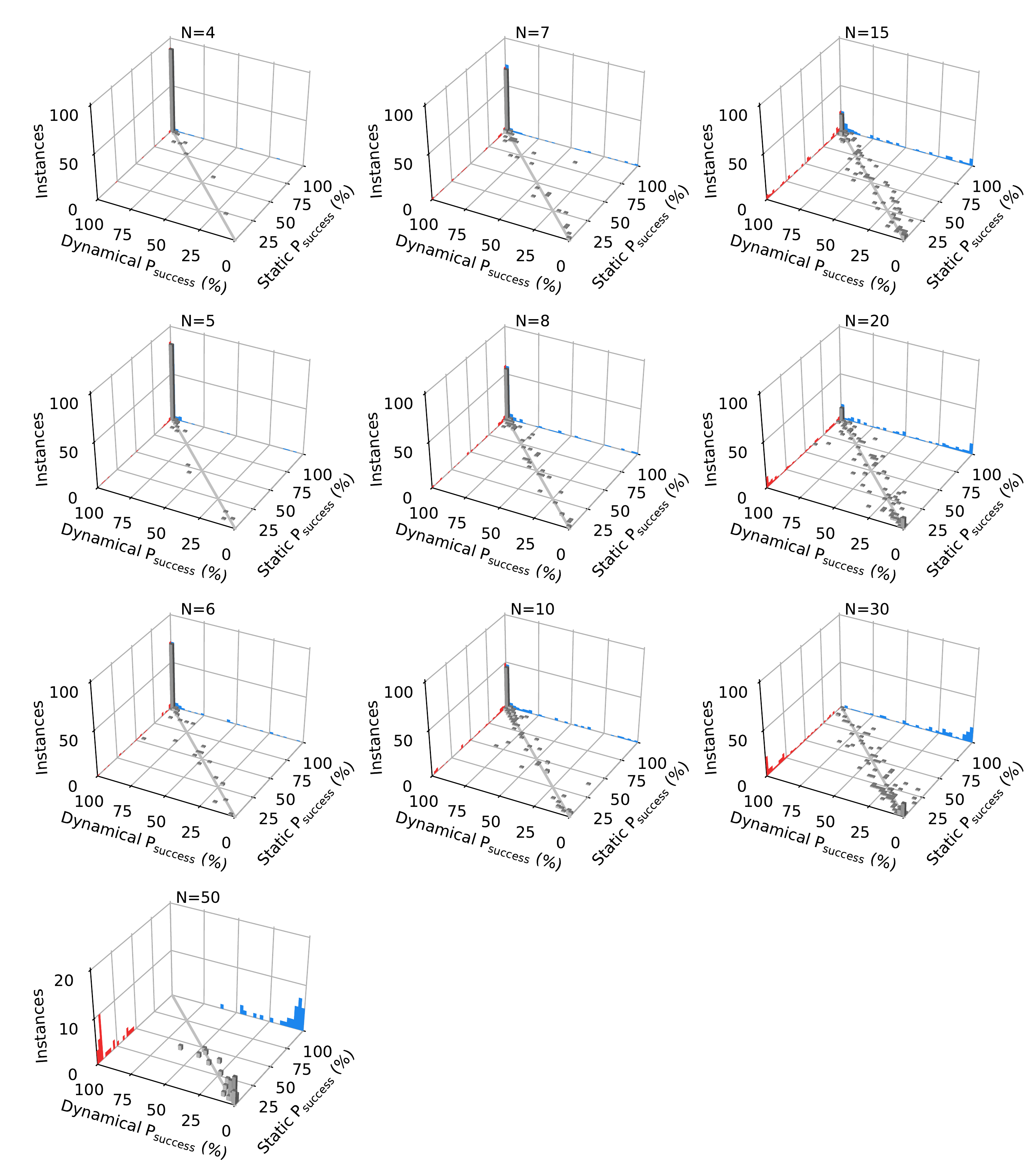}
    \caption{\textbf{Correlation of success probabilities between statically- and dynamically-coupled classical systems}. For various problem sizes $N$, we show the correlation matrix of success probabilities $P_\text{success}$ (evaluated at the end of the trajectory) between the statically-coupled and dynamically-coupled systems, using the classical EOMs for each respective system. Projected to the sides in blue and red are the corresponding histograms for the dynamically-coupled and statically-coupled systems, respectively. The system parameters are chosen according to the prescription described in Section~\ref{sec:parameter-summary} for each Ising problem instance. See Section~\ref{sec:classical} for additional details on our classical simulation methodology.}
    \label{fig:classical-hist}
\end{figure}